\documentclass[useAMS,usenatbib]{mn2e}
\usepackage{amssymb}
\usepackage{graphicx}
\usepackage{amsmath}
\usepackage{natbib}
\usepackage{epsfig}
\usepackage{epstopdf}

\usepackage{mathrsfs}
\usepackage{amsfonts}

\title[SEDs of Type 1 AGN in XMM-COSMOS II]{Spectral Energy Distributions of Type 1 AGN in XMM-COSMOS Survey II - Shape Evolution}

\author[Heng Hao et al.]{Heng Hao$^{1,2}$\thanks{E-mail:henghao@post.harvard.edu},
Martin Elvis$^{2}$, Francesca Civano$^{2,3}$, Gianni Zamorani$^{4}$,
Luis C. Ho$^{5}$,
\newauthor
Andrea Comastri$^{4}$, Marcella Brusa$^{6,7}$, Angela Bongiorno$^{6,
8}$, Andrea Merloni$^{6}$,
\newauthor
Jonathan R. Trump$^{9}$, Mara Salvato$^{10,11}$, Chris D.
Impey$^{12}$, Anton M. Koekemoer$^{13}$,
\newauthor
Giorgio Lanzuisi$^{6}$, Annalisa Celotti$^{1,14,15}$, Knud
Jahnke$^{16}$, Cristian Vignali$^{4,7}$,
\newauthor
John D. Silverman$^{17}$, C. Megan Urry$^{18}$, Kevin Schawinski$^{19}$, Peter Capak$^{20}$\\
$^{1}$SISSA, Via Bonomea 265, I-34136 Trieste, Italy\\
$^{2}$Harvard-Smithsonian Center for Astrophysics, 60 Garden Street,
Cambridge, MA 02138, USA\\
$^{3}$Dartmouth College, Department of Physics and Astronomy, 6127
Wilder Lab, Hanover, NH 03755\\
$^{4}$INAF-Osservatorio Astronomico di Bologna, via Ranzani 1,
I-40127 Bologna, Italy\\
$^{5}$The Observatories of the Carnegie Institute for Science, Santa
Barbara Street, Pasadena, CA 91101, USA\\
$^{6}$Max Planck Institute f\"ur Extraterrestrische Physik, Postfach
1312, 85741, Garching bei M\"{u}nchen, Germany\\
$^{7}$Dipartimento di Fisica e Astronomia, Universit\`{a} degli
studi di Bologna, viale Berti Pichat 6/2 40127 Bologna Italy\\
$^{8}$INAF-Osservatorio Astronomico di Roma, Via di Frascati 33,
00040, Monteporzio Catone, Rome, Italy\\
$^{9}$UCO/Lick Observatory, University of California, Santa Cruz, CA
95064, USA\\
$^{10}$IPP - Max-Planck-Institute for Plasma Physics, Boltzmann
Strasse 2, D-85748, Garching bei M\"{u}nchen, Germany\\
$^{11}$Excellence Cluster, Boltzmann Strasse 2, D-85748,
Garching bei M\"{u}nchen, Germany\\
$^{12}$Steward Observatory, University of Arizona, 933 North Cherry
Avenue, Tucson, AZ 85721, USA\\
$^{13}$Space Telescope Science Institute, 3700 San Martin Drive,
Baltimore, MD 21218, USA\\
$^{14}$INAF - Osservatorio Astronomico di Brera, via E. Bianchi 46,
I-23807 Merate, Italy\\
$^{15}$INFN - Sezione di Trieste, via Valerio 2, 34127, Trieste,
Italy\\
$^{16}$Max-Planck-Institut f\"ur Astronomie, K\"onigstuhl 17,
Heidelberg, D-69117, Germany\\
$^{17}$Kavli Institute for the Physics and Mathematics of the
Universe, Todai Institutes for Advanced Study, the University of
Tokyo,\\
Kashiwa, Japan 277-8583 (Kavli IPMU, WPI)\\
$^{18}$Physics Department and Yale Center for Astronomy and
Astrophysics, Yale University, New Haven, CT 06511, USA\\
$^{19}$Institute for Astronomy, Department of Physics, ETH
Zurich, Wolfgang-Pauli-Strasse 16, CH-8093 Zurich, Switzerland\\
$^{20}$California Institute of Technology, MC 105-24, 1200 East
California Boulevard, Pasadena, CA 91125, USA}

\begin{document}

\date{Version Nov 18th, 2013.}

\pagerange{\pageref{firstpage}--\pageref{lastpage}}\pubyear{2012}

\maketitle

\label{firstpage}

\begin{abstract}
The mid-infrared to ultraviolet (0.1 -- 10 $\mu m$) spectral energy
distribution (SED) shapes of 407 X-ray-selected radio-quiet type 1
AGN in the wide-field ``Cosmic Evolution Survey" (COSMOS) have been
studied for signs of evolution. For a sub-sample of 200 radio-quiet
quasars with black hole mass estimates and host galaxy corrections,
we studied their mean SEDs as a function of a broad range of
redshift, bolometric luminosity, black hole mass and Eddington
ratio, and compared them with the Elvis et al. (1994, E94) type 1
AGN mean SED. We found that the mean SEDs in each bin are closely
similar to each other, showing no statistical significant evidence
of dependence on any of the analyzed parameters. We also measured
the SED dispersion as a function of these four parameters, and found
no significant dependencies. The dispersion of the XMM-COSMOS SEDs
is generally larger than E94 SED dispersion in the ultraviolet,
which might be due to the broader ``window function'' for COSMOS
quasars, and their X-ray based selection.

\end{abstract}


\section{Introduction}
The physical details of the continuum emission of Active Galactic
Nuclei (AGN) remain unsettled after several decades of study. Yet,
understanding the continuum emission of AGN, from X-rays to radio,
is essential to unlocking the physics of accretion onto super
massive black holes (SMBHs). The continuum in each spectral region
can be ascribed to distinct energy generation mechanisms: jets in
the radio (see e.g. the review by Harris \& Krawczynski 2006), dust
in the infrared (IR, McAlary \& Rieke 1988; Sanders et al. 1989),
accretion disks in the optical-ultraviolet (UV) and soft X-rays
(Shakura \& Sunyaev 1973; Rees 1984; Czerny \& Elvis 1987), and
Compton up-scattering by hot coronae in the hard X-rays (e.g.,
Zamorani et al. 1981; Laor et al. 1990; Haardt \& Maraschi 1991;
Williams et al. 1992; Zdziarski et al. 2000; Kawaguchi et al. 2001;
Mateos et al. 2005; Mainieri et al. 2007).

Continuum changes with redshift, luminosity or Eddington ratio might
be expected. Most SMBH growth occurs during the active `AGN' phases
(the `Soltan argument', Soltan 1982), implying that most galaxy
bulges went through an AGN phase (e.g. Magorrian et al. 1998). Rapid
growth of central SMBHs happens in high-redshift and high-luminosity
quasars emitting near the Eddington limit (Barth et al. 2003;
Vestergaard 2004; Jiang et al. 2006; Kurk et al. 2007; but see also
Steinhardt \& Elvis 2010). The space density of X-ray-selected,
highly luminous AGN peaks at around $z=2.5$, and declines at $z>3$
(Silverman et al. 2005, Brusa et al. 2009, Civano et al. 2011).
Low-luminosity AGN are more prevalent at $z<1$ than higher
luminosity ones (Cowie et al. 2003; Fiore et al. 2003; Ueda et al.
2003; Silverman et al. 2005). As the central SMBH is the driver of
the emission, one might expect the quasar SED to evolve as the black
hole grows due to accretion.

Many parameters -- the black hole mass, the AGN luminosity relative
to the host galaxy, the accretion rate, the physical properties of
the accretion disk and the properties of the absorbing dust -- might
affect the shape of the AGN SED (Wilkes 2003). For instance, the
optical to X-ray spectral index
[$\alpha_{OX}$=0.384log$(F_{2keV}/F_{2500\AA})$], correlates with
luminosity but not with redshift (e.g. Vignali et al. 2003a; Steffen
et al. 2006; Just et al. 2007; Young et al. 2010; Lusso et al.
2010). It is possible that similar dependency of luminosity in the
SED shape exists at other wavelengths.

Observations indicate a tight link between SMBH growth and galaxy
evolution (e.g. Magorrian et al. 1998; Marconi \& Hunt 2003;
Tremaine et al. 2002; Menci et al. 2008). Locally, SMBHs appear to
reside at the center of most galaxies and the SMBH masses are
tightly correlated with their masses (e.g. Kormendy \& Richstone
1995; Marconi \& Hunt 2003) and velocity dispersions (i.e.
$M_{BH}-\sigma$ relations; Ferrarese \& Merritt 2000; Gebhardt et
al. 2000; Tremaine et al. 2002). Some evidence for evolution of this
relationship has been reported using several methods (e.g. Peng et
al. 2006; Shields et al. 2006; Ho 2007; Merloni et al. 2010). This
evolution would imply that the feedback of the SMBH to the host
galaxy evolves. Similar evolution in the innermost regions (within
the torus) is possible too. All of these processes could lead to
different SED shapes.

However, no evolution of the AGN SED has yet been demonstrated.
There is no convincing evidence for any change of SED with redshift
(Silverman et al. 2002; Mathur et al. 2002; Brandt et al. 2002;
Vignali et al. 2003b). High redshift quasars (up to redshift 7) show
optical spectra similar to low redshift quasars from the SDSS (Jiang
et al. 2007, Mortlock et al. 2011). There is evidence, though, that
the SEDs of extremely low-luminosity
($L_{bol}\lesssim10^{42}~erg/s$) nuclei are remarkably different
from those of luminous ($L_{bol}\gtrsim10^{44}~erg/s$) AGN (Ho 1999,
2008).

So far, the systematic study of the dependency of the SED shape with
physical parameters has been limited by difficulty in obtaining a
large sample size with good multi-wavelength coverage. The
Cosmological Evolution Survey (COSMOS, Scoville et al. 2007) has the
appropriate combination of depth, area and extensive
multi-wavelength data that allows for a sensitive survey of AGN to
address this question.

The COSMOS field has been imaged with XMM-Newton for a total of
$\sim1.5$~Ms (Hasinger et al. 2007; Cappelluti et al. 2007, 2009).
Optical identifications were made by Brusa et al. (2010) for the
entire XMM-COSMOS sample. Photometric properties and redshifts were
produced for each point source. This extensive data set allows us to
make a systematic study of the evolution of the SED shape, which is
the main purpose of this paper.

From this complete sample, we extracted a sample of 413 type 1 AGN
(broad emission line FWHM$>$2000~km s$^{-1}$). The type 1 AGN SED
sample catalog is described in detail in Elvis et al. (2012,
hereafter Paper I). It includes quasars with redshifts $0.1\leq z
\leq4.3$ and magnitudes $16.9\leq i_{AB}\leq24.8$, with 98\% of the
sources being radio-quiet (Hao et al. 2013b). This sample is twenty
times larger than the Elvis et al. (1994, hereafter E94) radio-quiet
type 1 AGN SED sample, and has full wavelength coverage from radio
to X-rays (for a total of 43 photometric bands, Paper I) and high
confidence level spectroscopic redshifts (Trump et al. 2009a;
Schneider et al. 2007; Lilly et al. 2007, 2009). The mean SED of the
XMM-COSMOS type 1 AGN was calculated and compared to previous
studies: E94, Richards et al. (2006, hereafter R06), Hopkins et al.
(2007) and Shang et al. (2011) in paper I Figure 21. In this figure
we can see, in the near-IR to optical-UV range, all recent studies
have very similar shapes to E94, while the XMM-COSMOS mean
host-corrected quasar SED has a less prominent `big-blue bump',
possibly due to remnant host contributions, not corrected because of
the dispersion in the black hole mass and host luminosity scaling
relationship itself. In this paper, we compare the Paper I sample
with E94 as a representative.

Paper I presented the selection and properties of the XMM-COSMOS
type 1 AGN sample of 413 quasars (XC413 hereinafter). We used
various radio-loud criteria ($R_L = \log(f_{5GHz}/f_{B})>1$, Wilkes
\& Elvis 1987; $q_{24} = \log(f_{24\mu m}/f_{1.4GHz})<0$, Appleton
et al. 2004; $R_{1.4}=\log(f_{1.4GHz}/f_J)$, in the observed frame;
$q_{24,obs}$, the $q_{24}$ in the observed frame;
$R_{uv}=\log(f_{5GHz}/f_{2500\AA})>1$, Stocke et al. 1992;
$P_{5GHz}=\log[P_{5 GHz}(W/Hz/Sr)]>24$, Goldschmidt et al. 1999; and
$R_X=\log(\nu L_{\nu} (5GHz)/L_X)>-3$, Terashima \& Wilson 2003) to
define a radio-loud quasar. We find that the radio-loud fraction is
1.5\%--4.5\% using any criterion, except $R_{uv}$, which is subject
to reddening and host contamination issues (Hao et al. 2013b). Using
two criteria at the same time, the radio-loud fraction is $\lesssim
8/413=2\%$. Only 6 XC413 quasars satisfy all the seven criteria. We
define these 6 quasars in this catalog as radio-loud (Paper I; Hao
et al. 2013). We refer to the radio-quiet sub-sample of XC413 as
XCRQ407 hereinafter.

\begin{table*}
\begin{minipage}{\textwidth}\centering
\caption{\label{t:ps}Source Properties in Parameter Space$^1$}
\begin{tabular}{@{}cccccccccccc@{}}
\hline XID & z\footnote{The spectroscopic redshifts are from Trump
et al. (2009a), Schneider et al. (2007), Lilly et al. (2007, 2009)}
& log$M_{BH}$\footnote{The black hole mass estimates are from Trump
et al. (2009b) and Merloni et al. (2010)} &
log$L_{bol}$\footnote{Calculated by integrating the rest frame SED
from $1~\mu m$ to 40~keV.} &
$\lambda_{E}$\footnote{$\lambda_E=L_{bol}/L_{Edd}$, details in
\S~\ref{s:mbhedd}.} & log$L_{all}$\footnote{Calculated by
integrating the rest frame SED from 1.4~GHz to 40~keV.} &
$L_{ir}/L_{all}$\footnote{$L_{ir}$ is calculated by integrating the
rest frame SED from $24~\mu m$ to $1~\mu m$.}&
$L_{opt}/L_{all}$\footnote{$L_{opt}$ is calculated by integrating
the rest frame SED from $1~\mu m$ to 912\AA.}
  & $L_{X}/L_{all}$\footnote{$L_{X}$ is calculated by integrating the rest
frame SED from 0.5~keV to 40~keV.} & $L_{bol}/L_{opt}$ &
log$L_{bol,hc}$\footnote{Calculated by integrating the rest frame
host-corrected SED from $1~\mu m$ to 912\AA\ for 203 quasars in the
XMM-COSMOS sample.}  & $L_{bol,hc}/$ \\ & & [M$_{\bigodot}$] &
[erg~s$^{-1}$] &  &
[erg~s$^{-1}$] & \% & \%  & \% & & [erg~s$^{-1}$]& $L_{opt,hc}$ \\
\hline
      1 & 0.373 &    8.58 &   45.36 &   0.048 &   45.46 &    19.2 &    37.1 &     5.6 &    2.11 &$\cdots$ &$\cdots$   \\
      2 & 1.024 &    8.96 &   45.82 &   0.056 &   46.09 &    32.5 &    17.8 &    27.6 &    2.98 &   45.79 &    3.35   \\
      3 & 0.345 &    8.66 &   45.18 &   0.026 &   45.39 &    33.8 &    24.1 &    17.2 &    2.56 &   45.11 &    3.44   \\
      4 & 0.132 &    7.31 &   44.25 &   0.068 &   44.55 &    40.8 &    35.5 &     3.0 &    1.41 &   44.19 &    1.50   \\
      5 & 1.157 &$\cdots$ &   45.95 &$\cdots$ &   46.21 &    34.2 &    25.2 &    12.8 &    2.18 &$\cdots$ &$\cdots$   \\
      6 & 0.360 &    8.64 &   44.89 &   0.014 &   45.20 &    31.1 &    34.7 &     4.7 &    1.42 &   44.74 &    1.71   \\
      7 & 0.519 &    8.38 &   45.21 &   0.053 &   45.44 &    37.6 &    27.5 &    17.9 &    2.13 &   45.16 &    2.46   \\
      8 & 0.699 &    7.96 &   45.78 &   0.518 &   45.97 &    33.2 &    32.2 &     7.3 &    2.00 &   45.78 &    1.98   \\
      9 & 1.459 &    8.86 &   45.90 &   0.088 &   46.28 &    40.1 &    18.4 &    11.7 &    2.30 &   45.89 &    2.43   \\
$\cdots$ & $\cdots$ & $\cdots$ & $\cdots$ & $\cdots$ & $\cdots$ &
$\cdots$ & $\cdots$ & $\cdots$ & $\cdots$ & $\cdots$ & $\cdots$ \\
\hline
\end{tabular}\\
$^1$ A portion of the table is shown here for guidance. The complete
table will be available online.\end{minipage}
\end{table*}

Estimates of black hole mass ($M_{BH}$) for 206 of XC413 have been
made by Merloni et al. (2010) and Trump et al. (2009b). Both papers
used single-epoch spectra and applied the scaling relations from
reverberation mapping found by Vestergaard \& Peterson (2006). This
method requires high S/N spectra with the broad emission line not
near the ends of the spectra. For the quasars with only zCOSMOS
spectra, the black hole mass was estimated only for those with MgII
lines in the spectra (Merloni et al. 2010),  using the calibration
of McLure \& Jarvis (2002). 206 quasars out of XC413, and 203
quasars out of XCRQ407 have black hole mass estimates. Paper I
estimated the host galaxy contribution for 203 of these 206 quasars
using the scaling relationship between black hole mass and host
luminosity (Marconi \& Hunt 2003) adding an evolutionary term
(Bennert et al. 2010, 2011), excluding the 3 which had
over-subtraction problems (the estimated host galaxy luminosity is
larger than the observed luminosity). We define this sub-sample (SS)
as SS203. In SS203, 200 quasars are radio-quiet. We refer to this
sub-sample as SSRQ200.

In Paper I, the rest frame SEDs of XC413 were constructed on a
uniform frequency grid ($\Delta \log\nu=0.02$) from radio to X-rays.
As there is limited data in the far-IR ($>10\mu m$ in rest frame)
and longer wavelengths, and because the UV flux is strongly affected
by both variability and strong broad emission lines, we defer the
analysis of these regions to a later paper. In this paper, we focus
on the SED shape in the optical to mid-IR range (0.1 -- 10 $\mu m$),
while all the plots are shown in the rest frame $0.1-10\mu m$ range.
Note that all the sources discussed in this paper are broad line AGN
with typical luminosity ($\sim10^{44}-10^{47}$erg/s), typical black
hole mass ($\sim 10^7-10^9 M_{\bigodot}$) and typical accretion
rates (Eddington ratio larger than 0.01, see in \S~\ref{s:mbhedd}).

All the wavelengths discussed in this paper are in the rest frame.
We adopt the WMAP 5-year cosmology (Komatsu et al. 2009), with H$_0$
=71~km~s$^{-1}$~Mpc$^{-1}$, $\Omega_M$ = 0.26 and
$\Omega_{\Lambda}$= 0.74.

\section{Parameter Space}
\subsection{Bolometric Luminosity} \label{s:Lbol}

The bolometric luminosity is the total energy per second radiated by
the quasar at all wavelengths in all directions. This luminosity is,
in principle, simply calculated by directly integrating the
rest-frame SED over the whole wavelength range. In practice, this is
difficult and observationally expensive. For the sample discussed in
this paper, it is possible to approximate the bolometric luminosity
with the SEDs available (Paper I).

As described in Paper I, the SEDs are produced by linearly
interpolating between the data points in $\log\nu L_{\nu}$ versus
$\log\nu$ space (i.e. connecting the individual points with power
laws in linear space). The COSMOS photometry for the XC413 sample is
$>$90\% complete from $u$ (CFHT) to MIPS 24$\mu$m, that is over the
1.8 dex wide 0.35$\mu$m -- 24$\mu$m ($\sim$0.14$\mu$m -- 10$\mu$m
for the typical z = 1.5 of XC413) observed frame interpolation is
unproblematic (Paper I). In the mid-infrared range, for quasars with
70$\mu m$ or 160$\mu m$ detections, we joined the 24$\mu$m data to
the longer wavelength points with a power-law in log~$\nu$f$_{\nu}$
vs. log~$\nu$ space; for the others we extrapolated from the rest
frame 24$\mu$m to 8$\mu$m slope and checked that the extrapolation
generally works. As the far-IR to radio photometry data are sparse,
for each source with a $>3\sigma$ VLA detection, we assumed a power
law $f_{\nu}\propto \nu^{-0.5}$ (e.g., Ivezi\'{c} 2004) in the rest
frame 1.4 GHz (21 cm) to 100 GHz (3 mm) range. In the 100 GHz (3 mm)
to $160\mu m$ part, the SED can be approximated by the red end of
the grey body $f_{\nu}\propto \nu^{3+\beta}/(e^{h\nu/kT}-1)$, when
$h\nu\ll kT$, $f_{\nu}\propto \nu^{2+\beta}$ (e.g. Lapi et al.
2011). $\beta$ is generally chosen in the range 1--2 (Dunne \& Eales
2001). We thus choose $\beta=1$, i.e. we assume a power-law
$f_{\nu}\propto \nu^{3}$ in this wavelength range. We directly
linearly interpolate from Lyman break (1216\AA) and 0.5~keV SED in
$\log\nu L_{\nu}$ versus $\log\nu$ space (Laor et al. 1997). In the
X-rays, we use the measured photon spectral index ($\Gamma$,
$f_{\nu}\propto \nu^{(1-\Gamma)}$, Mainieri et al. 2007) and the
observed 2~keV luminosity to get the SED in the 0.5~keV-40~keV
(rest-frame), which is the range of the XMM data for XC413.

We used the SEDs from Paper I to calculate two approximations to the
bolometric luminosities by integrating the rest frame SED over
different wavelength ranges: (1) from $1~\mu m$ to 40~keV as
$L_{bol}$; (2) from 21~cm/1.4~GHz to 40~keV as $L_{all}$. $L_{all}$
was integrated over all the wavelengths for which we have data.

As the great majority of the XC413 quasars are radio-quiet and the
photometric coverage from $24~\mu m$ to 1.4~GHz is currently sparse
(Paper I), using $L_{all}$ requires unwarranted extrapolation. The
contribution of the radio emission to the bolometric luminosity is
less than 3\% even for the radio-loud sources. Most of the far-IR
(at $\gtrsim100\mu m$) luminosity is probably due to star-forming
activity not the AGN (Netzer et al. 2007, Mullaney et al. 2012; but
see Ho 2005). From the optical short-ward the primary emission is
from the innermost region of the quasar (SMBH and accretion disk).
The near and mid-IR continuum is the result of reprocessing shorter
wavelength radiation of the quasar by dust (Sanders et al. 1989;
Suganuma et al. 2006). The reprocessed dust emission thereby
includes reprocessed primary radiation emitted in directions
different from our line-of-sight. In a non-spherical geometry, as is
likely present in quasars, this makes us count more radiation that
we would in a $4\pi$ averaged calculation. In this sense, we are
double-counting the primary emission (e.g. Lusso et al. 2010).
Hence, $L_{bol}$ which integrates from $1~\mu m$ to 40~keV is a good
approximation of the true bolometric quasar luminosity defined over
the entire wavelength range.

We note that directly integrating the SED in the rest frame from
$1\mu m$ to 40~keV overestimates the quasar emission as the
contribution from the host galaxy is not excluded. In practice, we
do not have good estimates of the host contribution. For SSRQ200, we
calculated the bolometric luminosity by integrating the host
corrected SED from 1$\mu m$ to 912\AA\ as $L_{bol, hc}$ (last column
in Table~\ref{t:ps}). As discussed in Paper I, we applied the
scaling relationship between black hole mass and host luminosity
reported in Marconi \& Hunt (2003), and added an evolutionary term
(Bennert et al. 2010, 2011), to estimate the contribution of host
galaxy.

We note that $L_{bol}$ will underestimate the total emission for
three reasons: (1) excluding the IR to radio emission; (2) excluding
the hard X-ray emission, where we do not have data; (3) ignoring the
reddening of the primary emission in the optical/UV. Note that most,
if not all, of these factors are not significant underestimate. For
example, as we stated above, the contribution of the radio emission
to the bolometric luminosity s less than 3\% even for the radio-loud
quasars.

To investigate the contribution to bolometric luminosity in
different wavelength ranges, we calculated $L_{ir}$ by integrating
from $24~\mu m$ to $1~\mu m$, $L_{opt}$ from $1~\mu m$ to $912~\AA$
and $L_{X}$ from $0.5~keV$ to $40~keV$. The fractions of the
luminosity in different wavelength ranges are reported in
Table~\ref{t:ps} (The full table is available on line).
Table~\ref{t:pr} shows the median values and the ranges for
$L_{bol}$, the fractions in three wavelength ranges, and the median
values and the ranges for $M_{BH}$ and Eddington ratio (which will
be discussed in \S~\ref{s:mbhedd}). In general, the IR component
($1-24~\mu m$) provides an equal or greater contribution compared to
the optical/UV `big blue bump' component ($0.1-1~\mu m$). We also
calculate the ratio of $L_{bol}$ over $L_{opt}$ as listed in
Table~\ref{t:ps} \& \ref{t:pr}. The mean $L_{bol}/L_{opt}$ is
$2.04\pm0.75$ (for XCRQ407) and $2.00\pm0.71$ (for SSRQ200)
respectively. The mean $L_{bol,~hc}/L_{opt,~hc}$ is $2.11\pm0.91$.
These values are larger than the ratio for E94 template (1.86) and
for R06 template (1.61). This is probably caused by the selection
effect, that E94 and R06 are optical selected samples that select
more quasars with large optical contribution.

\begin{table}
\begin{minipage}{\columnwidth}
\centering \caption{Parameter Range \label{t:pr}}
\begin{tabular}{@{}lc@{\hspace{1mm}}c@{\hspace{1mm}}cc@{\hspace{1mm}}c@{\hspace{1mm}}c@{}} \hline
Parameter &
\multicolumn{3}{c}{XCRQ407} &  \multicolumn{3}{c}{SSRQ200}\\
  & min & med & max & min & med & max \\ \hline
z & 0.10 & 1.57 & 4.26 & 0.13 & 1.50 & 4.26\\
log($M_{BH}/M_{\bigodot}$) & $\cdots$& $\cdots$& $\cdots$ & 7.18 & 8.39 & 9.34\\
log$L_{bol}$ (erg/s) & 44.01 & 45.49 & 46.91 & 44.25 & 45.56 & 46.91 \\
log$L_{all}$ (erg/s) & 44.29 & 45.71 & 47.40 & 44.55 & 45.77 & 47.40 \\
$L_{ir}/L_{all}$ & 12.7\% & 33.2\% & 76.7\% & 12.7\% & 33.4\% & 76.7\% \\
$L_{opt}/L_{all}$ & 2.1\% & 32.0\% & 60.5\% & 9.0\% & 31.8\% & 58.9\% \\
$L_{X}/L_{all}$ & 0.33\% & 8.3\% & 43.9\% & 0.47\% & 8.1\% & 43.9\% \\
$L_{bol}/L_{opt}$ & 1.10 & 1.90 & 8.30 & 1.10 &
1.84 & 8.30\\
$\lambda_E$ & $\cdots$ &$\cdots$& $\cdots$ & 0.008 & 0.114 & 2.505\\
$\log L_{bol,~hc}$ (erg/s) & $\cdots$ &$\cdots$& $\cdots$ & 44.19 & 45.53 & 46.86\\
$L_{bol,~hc}/L_{opt,~hc}$ & $\cdots$ & $\cdots$ &
$\cdots$ & 1.11 & 1.94 & 11.48\\
\hline
\end{tabular}
\end{minipage}
\end{table}

\subsection{Bolometric Correction}
\begin{figure}
\epsfig{file=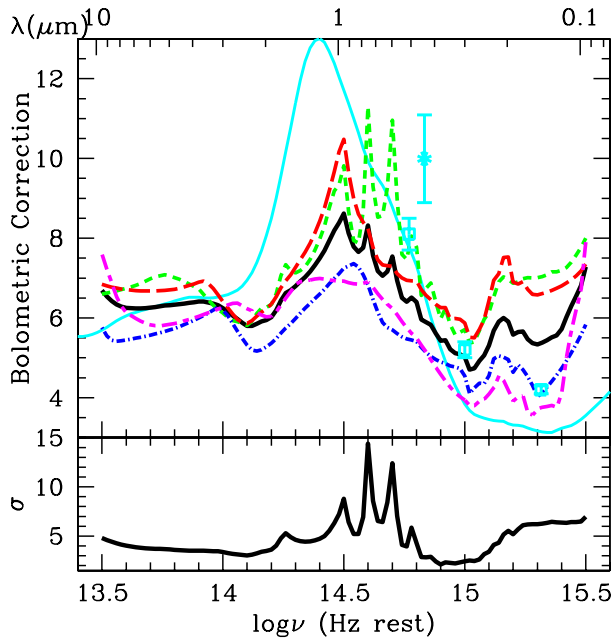, angle=0,width=\linewidth} \caption{The
frequency dependent bolometric correction for the SSRQ200 sample
(black solid). The bolometric corrections in different redshift bins
are also shown in the same plot (redshift: 0.1--1.2= green short
dashed line, 50 quasars; 1.2--1.5= red long dashed line, 50 quasars;
1.5--1.8= blue dot dashed line, 51 quasars; and 1.8--4.3= magenta
short long dashed line, 49 quasars).  The cyan solid line show the
bolometric correction for E94 RQ mean SED. The cyan star point at B
band (4400\AA) show the bolometric correction calculate from the
Hopkins et al. (2007), assuming $\log L_{bol}=45.5~erg/s$, which is
the median $L_{bol}$ of SSRQ200. The cyan square show the bolometric
correction at 3000\AA\ and 5100\AA\ respectively from Runnoe et al.
(2012). In the bottom panel we show the dispersion of the bolometric
correction for the SSRQ200 sample. \label{bc}}
\end{figure}

We then calculated the bolometric correction at various frequencies.
The bolometric correction ($BC_\nu=L_{bol}/\nu L_\nu$) is the factor
which transforms the luminosity in one band to the bolometric
luminosity. As the host galaxy contribution is prominent for
X-ray-selected quasars, we calculated the frequency-dependent $BC$
in the rest frame 0.1--10$\mu m$ only for SSRQ200 after performing
the host correction. The bolometric correction was calculated for a
$\Delta\log\nu=0.02$ grid. The bolometric luminosity used in the
calculation is $L_{bol,hc}$, which is the integration of
host-corrected SEDs, listed as the last column in Table~\ref{t:ps}.

The mean and dispersion of the $BC_\nu$ are listed in Table
\ref{t:bc} and shown in Figure \ref{bc}, where we also show the mean
and dispersion of the $BC_\nu$ for quasars at different redshift
bins. The mean $BC_\nu$ curves for different redshift bins are
consistent with each other given the large dispersion at each
wavelength (Figure~\ref{bc}). The dispersion is largest at around
1~$\mu m$, where the host contribution is the highest, and in the
extreme UV, where quasar variability is likely to contribute
significantly to the observed dispersion.

For comparison, we plot the E94 mean SED bolometric correction as
the cyan solid line in Figure~\ref{bc}. Hopkins et al. (2007) used a
double power law to approximate the B band bolometric correction.
$$\frac{L_{bol}}{L_B}=6.25\left(\frac{L_{bol}}{10^{10}L_{\bigodot}}\right)^{-0.37}+
9.00\left(\frac{L_{bol}}{10^{10}L_{\bigodot}}\right)^{-0.012}$$ To
compare with the bolometric correction we got, we use the median
value of $L_{bol}$ ($10^{45.5}~erg/s$) in SSRQ200 to apply to the
formula. The B band bolometric correction from Hopkins et al. (2007)
is thus 9.99 (red star in Figure~\ref{bc}). More recently, Runnoe et
al. (2012) studied the SEDs of 63 bright quasars at low redshift and
found a linear bolometric correction of $4.2\pm0.1$, $5.2\pm0.2$ and
$8.1\pm0.4$ at 1450, 3000 and 5100\AA, respectively (red square in
Figure~\ref{bc}). These results generally agree with
bolometric correction from SSRQ200. 

\begin{table*}
\begin{minipage}{\textwidth}
\centering \caption{Bolometric Corrections for the SSRQ200 sample
$^1$ \label{t:bc}}
\begin{tabular}{c|c|c@{\hspace{1mm}}c|c@{\hspace{1mm}}c|c@{\hspace{1mm}}c|c@{\hspace{1mm}}c|c@{\hspace{1mm}}c}
\hline $\lambda$ & log$\nu$ & \multicolumn{2}{c}{All z} &
\multicolumn{2}{c}{$0.1<z<1.2$} & \multicolumn{2}{c}{$1.2<z<1.5$} &
\multicolumn{2}{c}{$1.5<z<1.8$} & \multicolumn{2}{c}{$1.8<z<4.3$}\\
($\mu m$)& (Hz) & BC & $\sigma$ & BC & $\sigma$ & BC & $\sigma$ & BC
& $\sigma$ &
BC & $\sigma$ \\
\hline
 9.48 &    13.5 &  6.69 &  4.80 &  6.60 &  4.11 &  6.84 &  4.95 &  5.76 &  3.45 &  7.59 &  6.23 \\
 7.53 &    13.6 &  6.25 &  4.02 &  6.74 &  3.97 &  6.71 &  4.18 &  5.45 &  2.99 &  6.13 &  4.76 \\
 5.98 &    13.7 &  6.26 &  3.73 &  6.98 &  3.96 &  6.67 &  3.81 &  5.57 &  2.83 &  5.82 &  4.11 \\
 4.75 &    13.8 &  6.33 &  3.60 &  7.01 &  3.80 &  6.72 &  3.64 &  5.74 &  2.82 &  5.85 &  3.98 \\
 3.77 &    13.9 &  6.41 &  3.50 &  6.73 &  3.34 &  6.89 &  3.69 &  6.00 &  3.03 &  6.04 &  3.90 \\
 3.00 &    14.0 &  6.25 &  3.38 &  6.12 &  3.20 &  6.41 &  3.20 &  6.19 &  3.29 &  6.27 &  3.88 \\
 2.38 &    14.1 &  5.80 &  3.04 &  5.82 &  3.21 &  5.86 &  2.86 &  5.32 &  2.19 &  6.22 &  3.74 \\
 1.89 &    14.2 &  6.02 &  3.62 &  6.34 &  5.05 &  6.35 &  3.25 &  5.37 &  2.23 &  6.03 &  3.41 \\
 1.50 &    14.3 &  6.80 &  4.61 &  7.13 &  7.16 &  7.27 &  3.84 &  5.98 &  2.57 &  6.83 &  3.55 \\
 1.19 &    14.4 &  7.49 &  4.73 &  7.97 &  7.16 &  8.43 &  4.55 &  6.60 &  2.75 &  6.98 &  3.01 \\
 0.95 &    14.5 &  8.62 &  8.80 &  9.81 & 14.01 & 10.48 &  9.58 &  7.25 &  2.94 &  6.92 &  2.89 \\
 0.75 &    14.6 &  8.32 & 14.40 & 11.30 & 28.08 &  8.25 &  4.90 &  6.83 &  2.44 &  6.90 &  3.30 \\
 0.60 &    14.7 &  7.54 & 12.40 & 10.96 & 24.12 &  7.29 &  3.54 &  5.70 &  2.10 &  6.23 &  2.86 \\
 0.48 &    14.8 &  6.40 &  4.29 &  7.77 &  7.07 &  6.93 &  3.12 &  5.41 &  1.96 &  5.49 &  2.71 \\
 0.38 &    14.9 &  5.43 &  2.11 &  5.98 &  1.89 &  6.26 &  2.63 &  4.90 &  1.60 &  4.58 &  1.73 \\
 0.30 &    15.0 &  5.02 &  2.43 &  5.71 &  1.89 &  5.87 &  3.56 &  4.54 &  1.76 &  3.93 &  1.38 \\
 0.24 &    15.1 &  5.31 &  3.41 &  6.14 &  3.24 &  6.44 &  5.08 &  4.55 &  1.89 &  4.10 &  1.79 \\
 0.19 &    15.2 &  5.58 &  5.19 &  6.87 &  4.79 &  6.90 &  8.47 &  4.64 &  2.12 &  3.91 &  1.74 \\
 0.15 &    15.3 &  5.33 &  6.22 &  7.01 &  6.54 &  6.58 &  9.69 &  4.03 &  2.63 &  3.71 &  2.13 \\
 0.12 &    15.4 &  5.72 &  6.36 &  7.20 &  7.75 &  6.81 &  9.18 &  4.74 &  2.62 &  4.14 &  2.49 \\
 0.09 &    15.5 &  7.28 &  6.97 &  8.00 &  7.57 &  7.35 &  8.77 &  5.83 &  3.00 &  7.99 &  7.20 \\
\hline
\end{tabular}\\
$^1$ After host galaxy subtraction.
\end{minipage}
\end{table*}

\subsection{Eddington Ratio}
\label{s:mbhedd}

For SS200, the accretion rate relative to the Eddington rate (the
Eddington ratio, $\lambda_E$) can be calculated given the bolometric
luminosities derived in \S~\ref{s:Lbol}, i.e.

$$\lambda_E=\frac{L_{bol}}{L_{Edd}}=\frac{L_{bol}}{\frac{4\pi
Gcm_{p}}{\sigma_e}M_{BH}}
=\frac{L_{bol}}{1.26\times10^{38}(M_{BH}/M_{\odot})}$$

The black hole mass ($M_{BH}$) and the corresponding Eddington ratio
($\lambda_{E}$) of the XC413 are listed in Table~\ref{t:ps} (Full
table available on-line). The median and ranges of these parameters
are reported in Table~\ref{t:pr}.

\subsection{Parameter Space Classification}
\begin{table*}
\begin{minipage}{\textwidth}
\centering \caption{Parameter Space Bins \label{t:psc}}
\begin{tabular}{@{}c|ccc|ccc|cc|cc@{}}\hline
bin & redshift & \multicolumn{2}{c}{N} & $L_{bol}$ &
\multicolumn{2}{c}{N} & $M_{BH}$ & N & log$\lambda_{E}$ & N\\
& & XCRQ407 & SSRQ200 &  & XCRQ407 & SSRQ200 & & SSRQ200 & & SSRQ200
\\ \hline
      1 & 0.103--1.166 & 107 & 50 & 44.00--45.23 & 126 & 50 & 7.18--8.10 & 50 & 0.008-- 0.057 &  50 \\
      2 & 1.166--1.483 &  72 & 50 & 45.23--45.56 &  99 & 50 & 8.10--8.39 & 50 & 0.057-- 0.114 &  50 \\
      3 & 1.483--1.848 &  84 & 51 & 45.56--45.87 &  93 & 50 & 8.39--8.66 & 50 & 0.114-- 0.252 &  50 \\
      4 & 1.848--4.256 & 144 & 49 & 45.87--46.91 &  89 & 50 & 8.66--9.34 & 50 & 0.252-- 2.506 &  50 \\
\hline
\end{tabular}
\end{minipage}
\end{table*}

\begin{figure*}
\includegraphics[angle=0,width=0.33\textwidth]{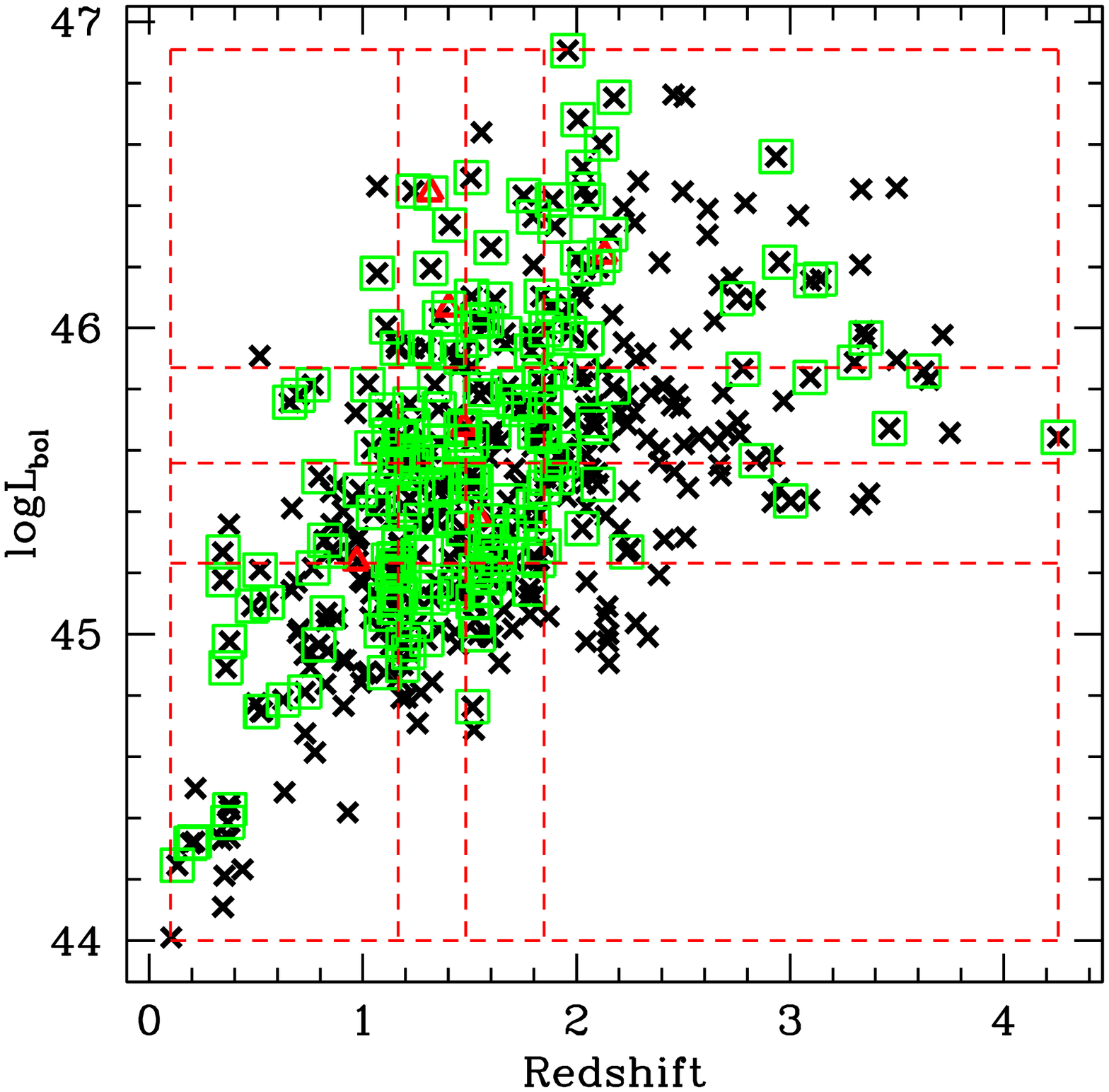}
\includegraphics[angle=0,width=0.33\textwidth]{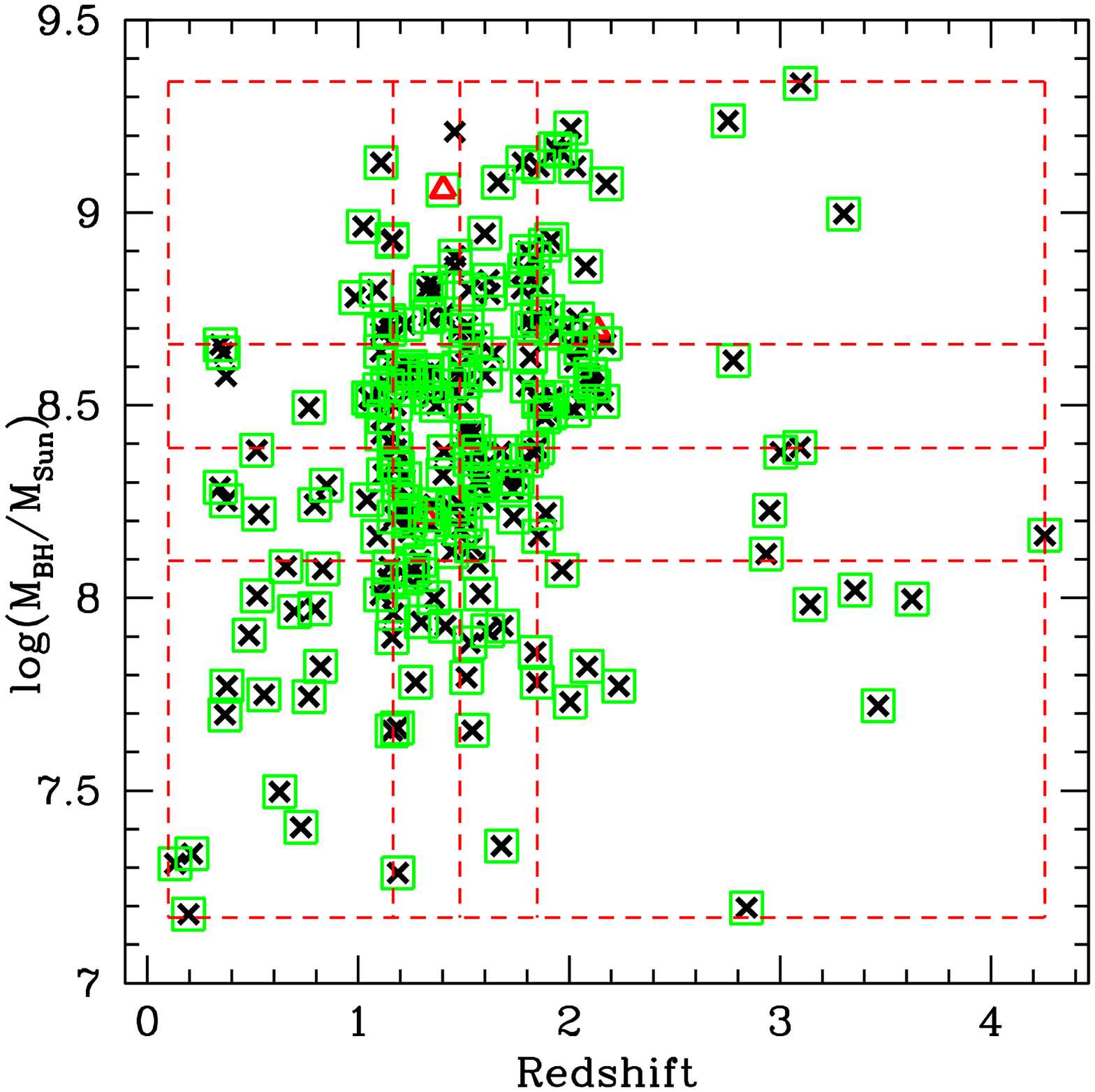}
\includegraphics[angle=0,width=0.33\textwidth]{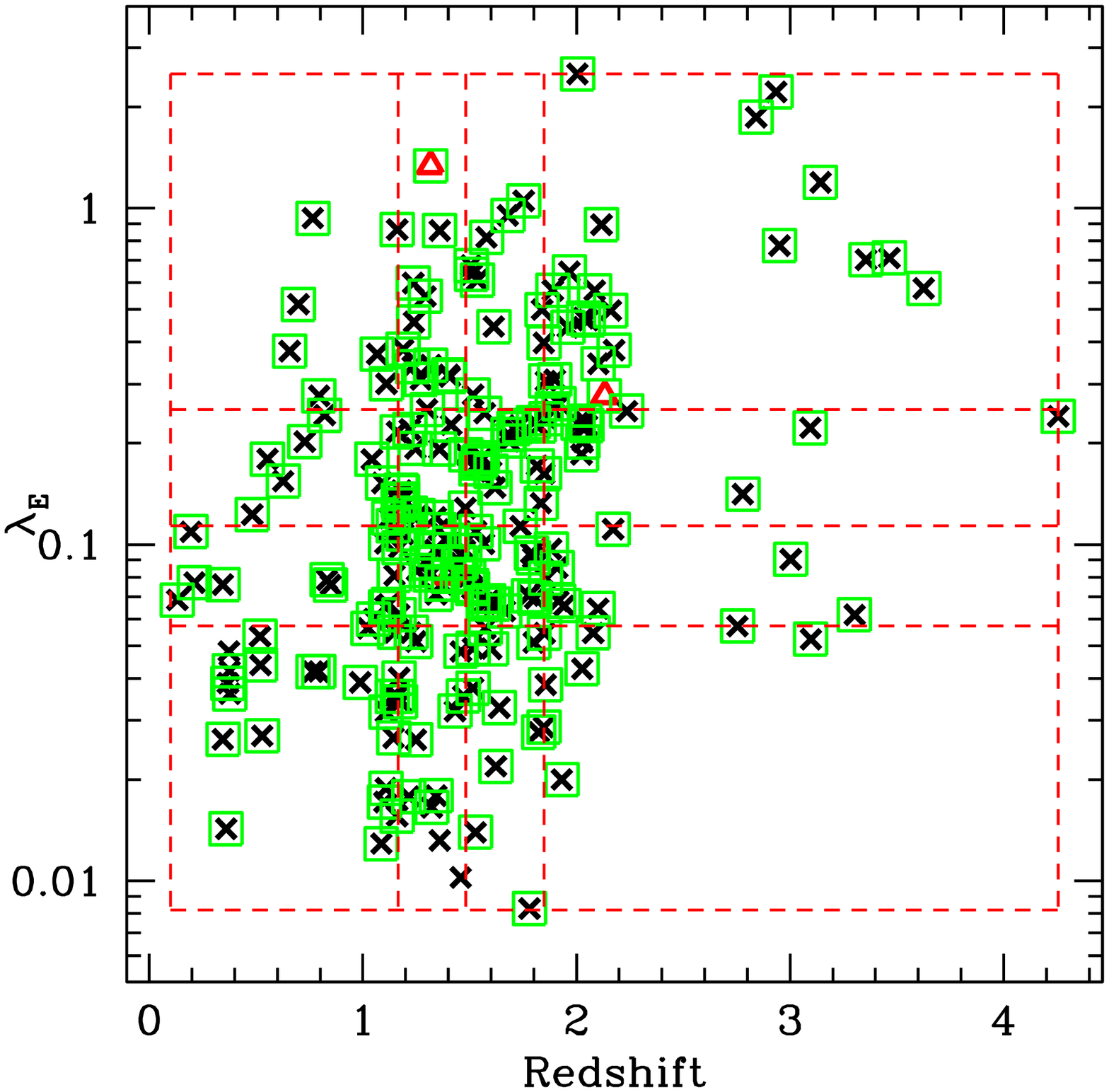}
\caption{The classification of $z$, $L_{bol}$, $M_{BH}$ and
log$\lambda_{E}$ bins of the XMM-COSMOS type 1 AGN sample. The
dashed lines show the borders of quartile bins: the $z$ bins (0.1
$\sim$ 1.2 $\sim$ 1.5 $\sim$ 1.8 $\sim$ 4.3), $L_{bol}$ bins (44.0
$\sim$ 45.2 $\sim$ 45.6 $\sim$ 45.9 $\sim$ 46.9), $M_{BH}$ bins (7.2
$\sim$ 8.1 $\sim$ 8.4 $\sim$ 8.7 $\sim$ 9.4) and $\lambda_{E}$ bins
(0.008 $\sim$ 0.057 $\sim$ 0.114 $\sim$ 0.252 $\sim$ 2.506). The
black crosses show all the radio-quiet XMM-COSMOS type 1 AGN. The
red triangles show the 6 radio-loud quasars. The green squares show
the sub-sample SS203. \label{Lzarb}}
\end{figure*}

The XCRQ407 sample spans a wide range of $z$, $L_{bol}$, $M_{BH}$
and $\lambda_E$ (Table~\ref{t:pr}, Figure~\ref{Lzarb}). The redshift
range is comparable to that of Spitzer-SDSS sample (R06), and the
luminosity, black hole mass and Eddington ratio ranges are
comparable to that of PG quasars (Sikora et al. 2007). The size of
the XCRQ407 sample is more than double that of previous samples, and
is an order of magnitude larger than the E94 sample. This large
sample size spanning a wide range of the parameter spaces is useful
to understand whether and how the properties of the SMBH affect the
SED shape.

XCRQ407 is an X-ray-selected sample and so includes sources with
large host galaxy contribution (Paper I). For this reason, we focus
on the host-corrected SEDs of the SSRQ200. For comparison we also
study the SEDs of the XCRQ407 before the host correction. The median
values and ranges in $z$, $L_{bol}$, $M_{BH}$ and $\lambda_{E}$ are
similar for the two sub-samples (Table~\ref{t:pr}).

We can now check the SED shape dependence on each physical parameter
by dividing the samples into quartiles of quasar $z$, $L_{bol}$,
$M_{BH}$ and $\lambda_{E}$, as shown by the dashed red lines in
Figure~\ref{Lzarb}. The bin boundaries and number of sources in
different bins are given in Table~\ref{t:psc}. With these divisions,
different bins have a similar number of quasars to calculate the
mean SED, and therefore the possible statistical differences between
different bins are minimized.

\section{Mean Quasar SED Dependency on Physical Parameters}
\label{s:msedevl}
\begin{figure*}
\includegraphics[angle=0,width=0.35\textwidth]{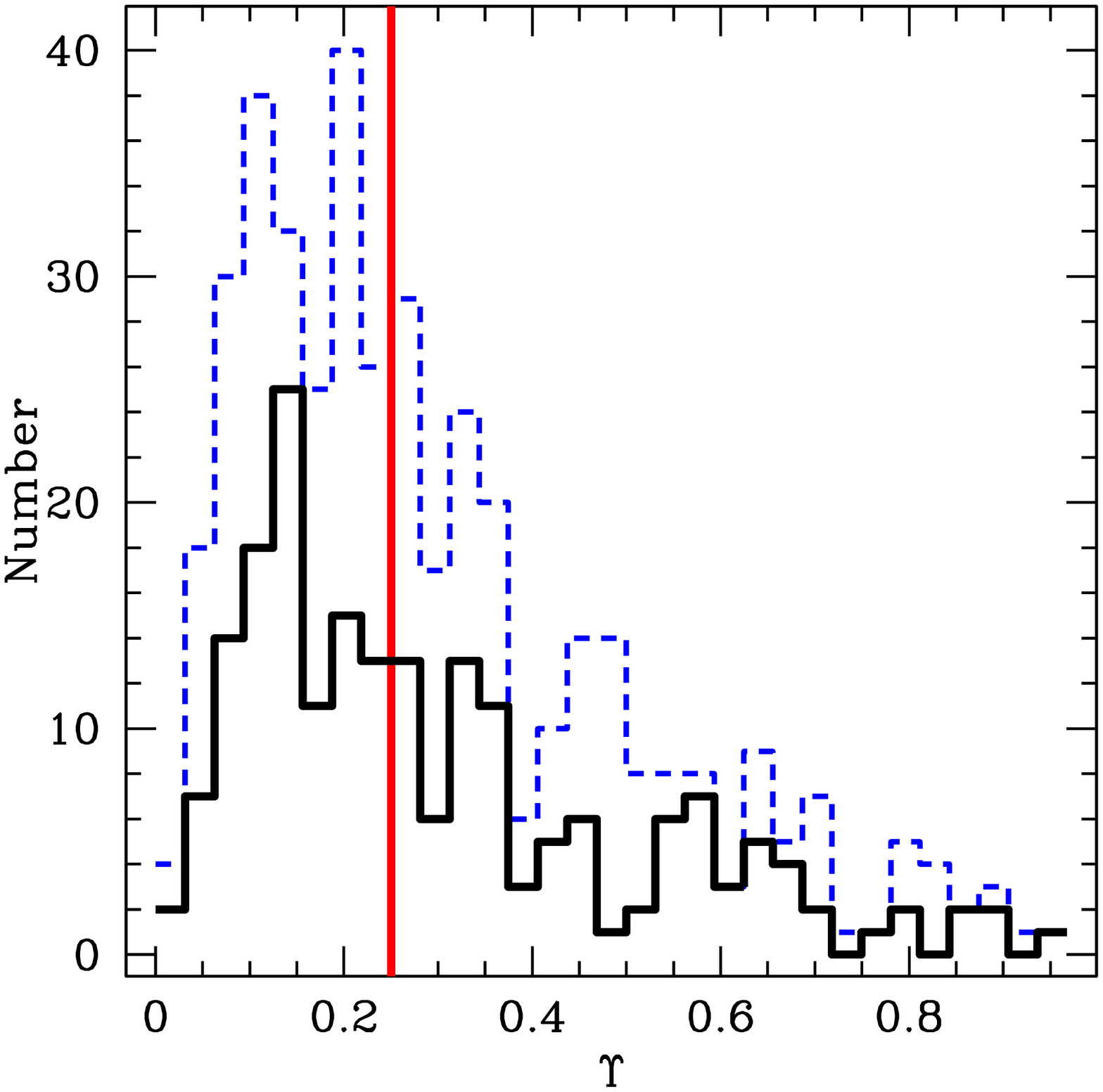}
\includegraphics[angle=0,width=0.35\textwidth]{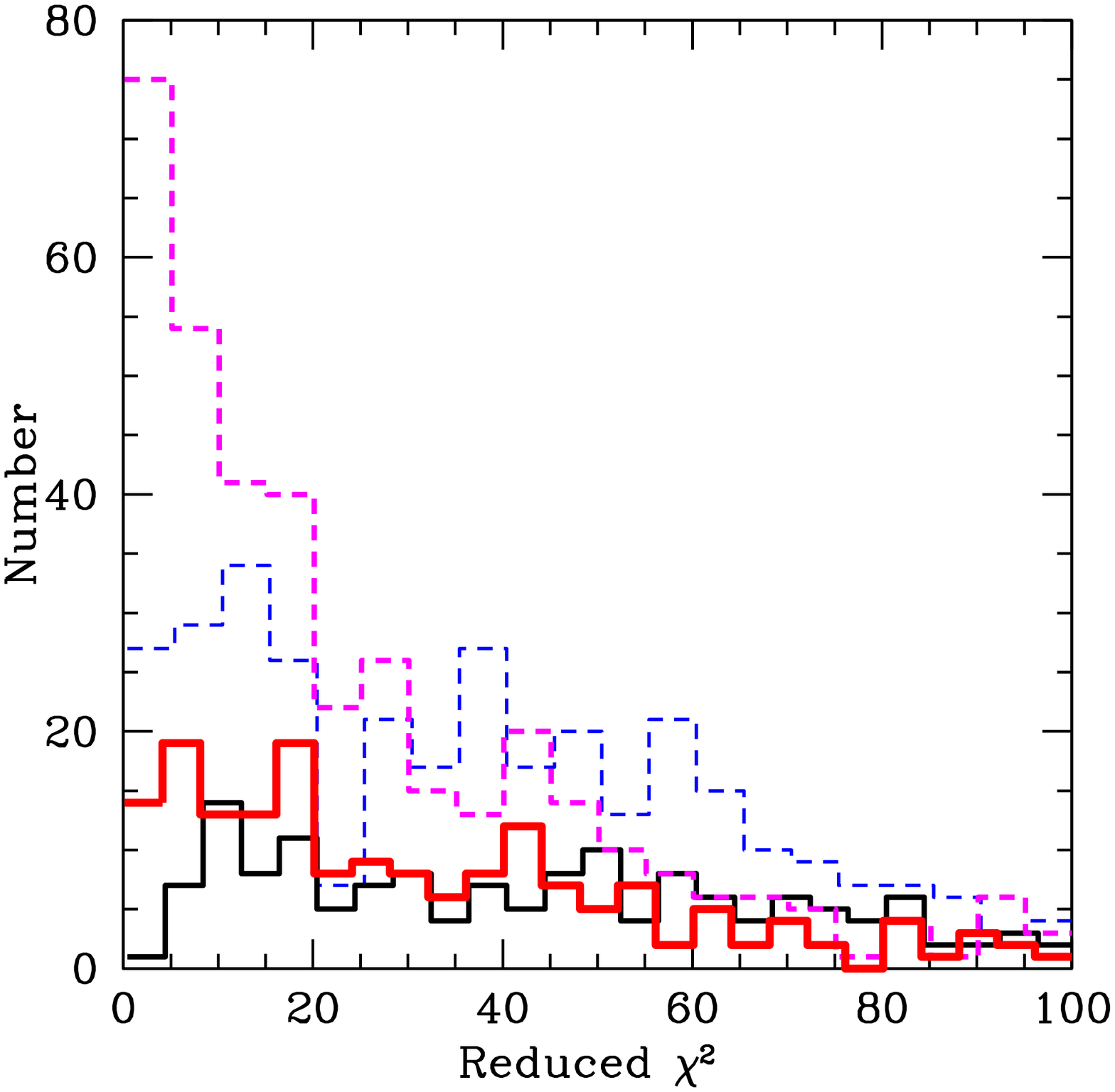}
\caption{{\em Left:} Distribution of the variability parameter
$\Upsilon$ (Salvato et al. 2009) of the SSRQ200 AGN (black solid
line) and XC413 AGN (blue dashed line). The vertical red line
divides the sources into variable ($>$0.25) and not-variable
($<$0.25). {\em Right:} The reduced $\chi^2$ of the SED-fitting (see
\S~\ref{s:msedevl} for details) both before (black solid line for
SSRQ200 and blue dashed line for XC413) and after (red solid line
for SSRQ200 and magenta dashed line for XC413) the restriction of
the data to the 2004-2007 interval.\label{varsel}}
\end{figure*}

\begin{figure*}
\includegraphics[angle=0,width=0.45\textwidth]{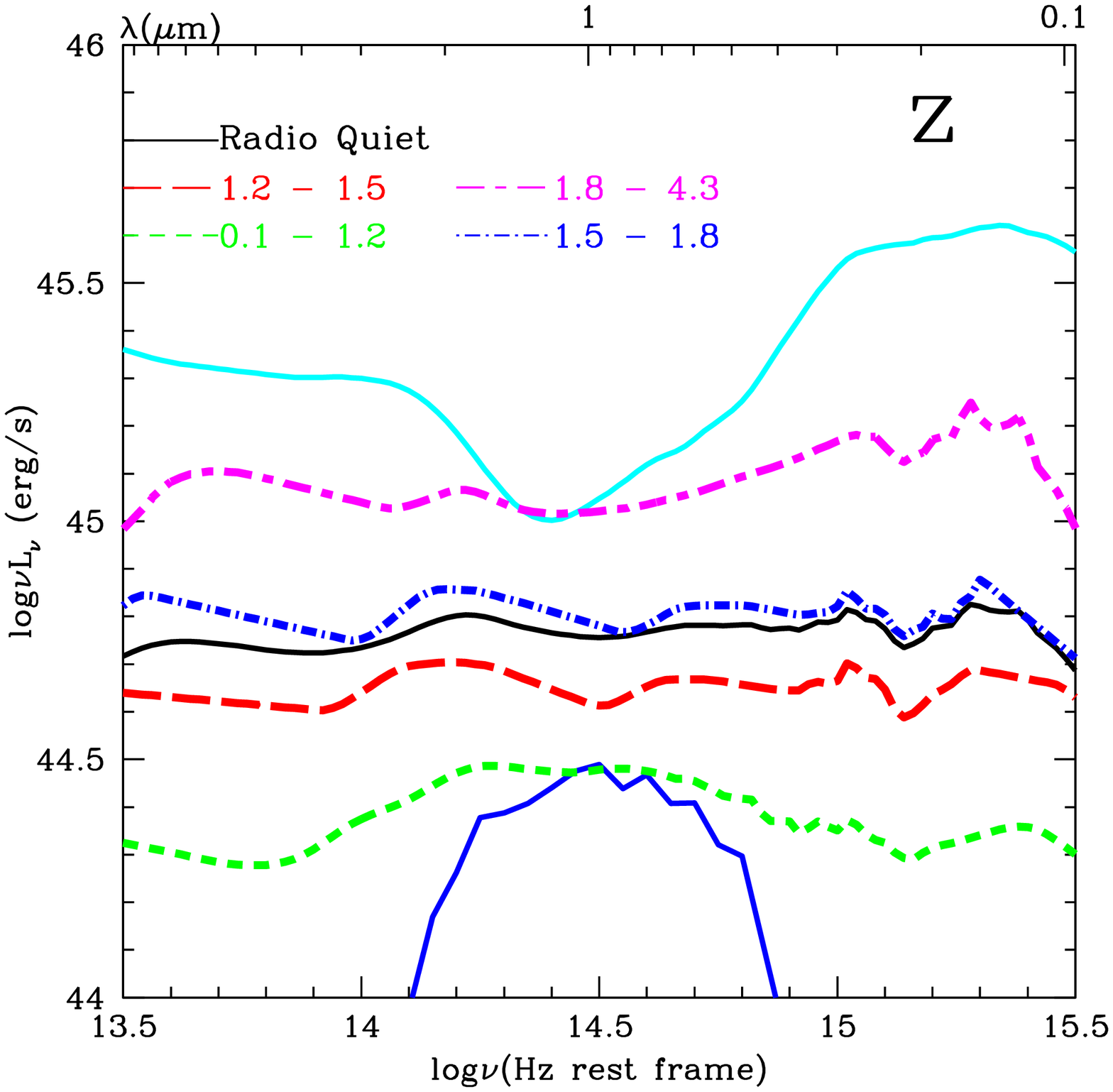}
\includegraphics[angle=0,width=0.45\textwidth]{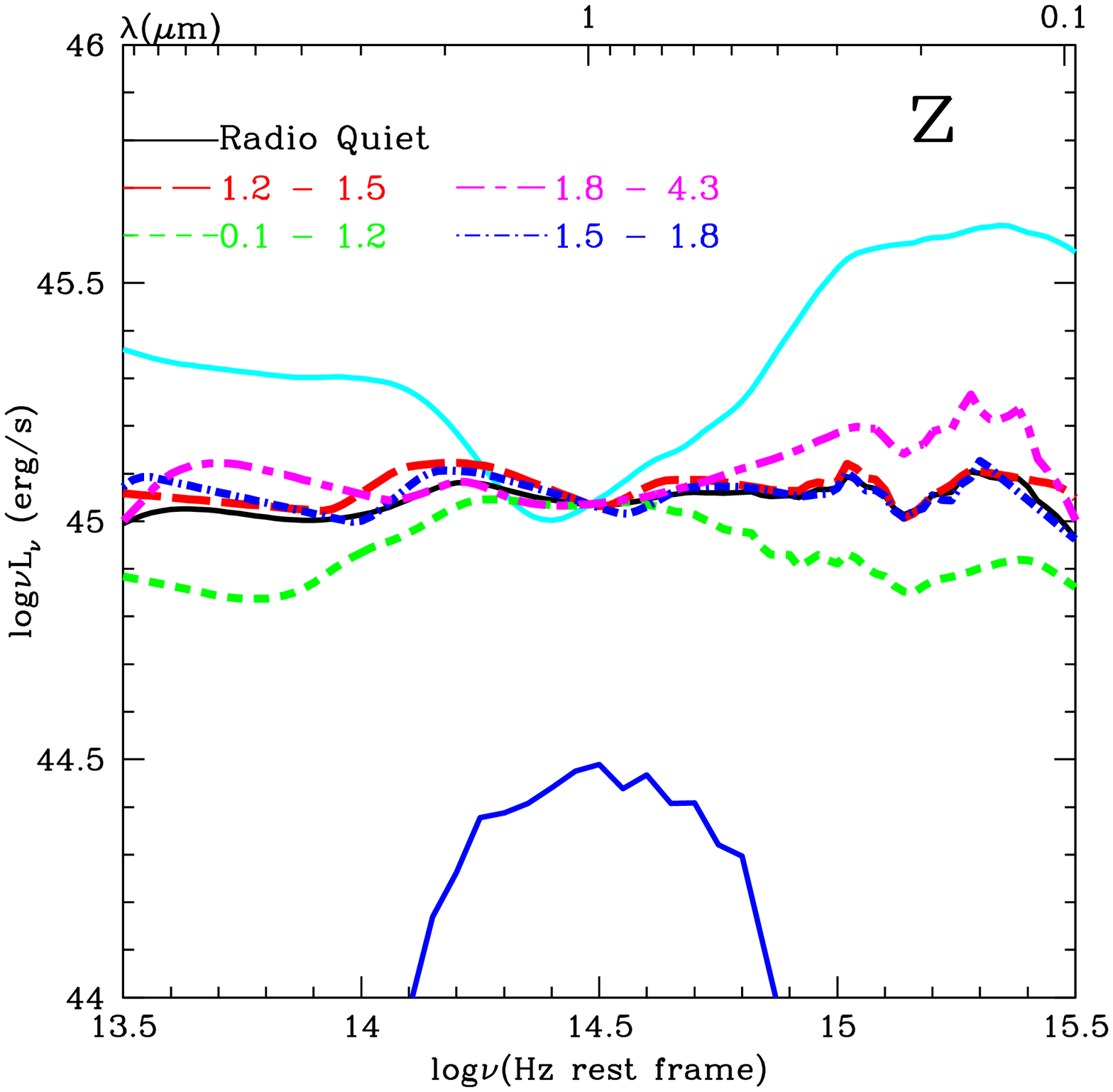}
\includegraphics[angle=0,width=0.45\textwidth]{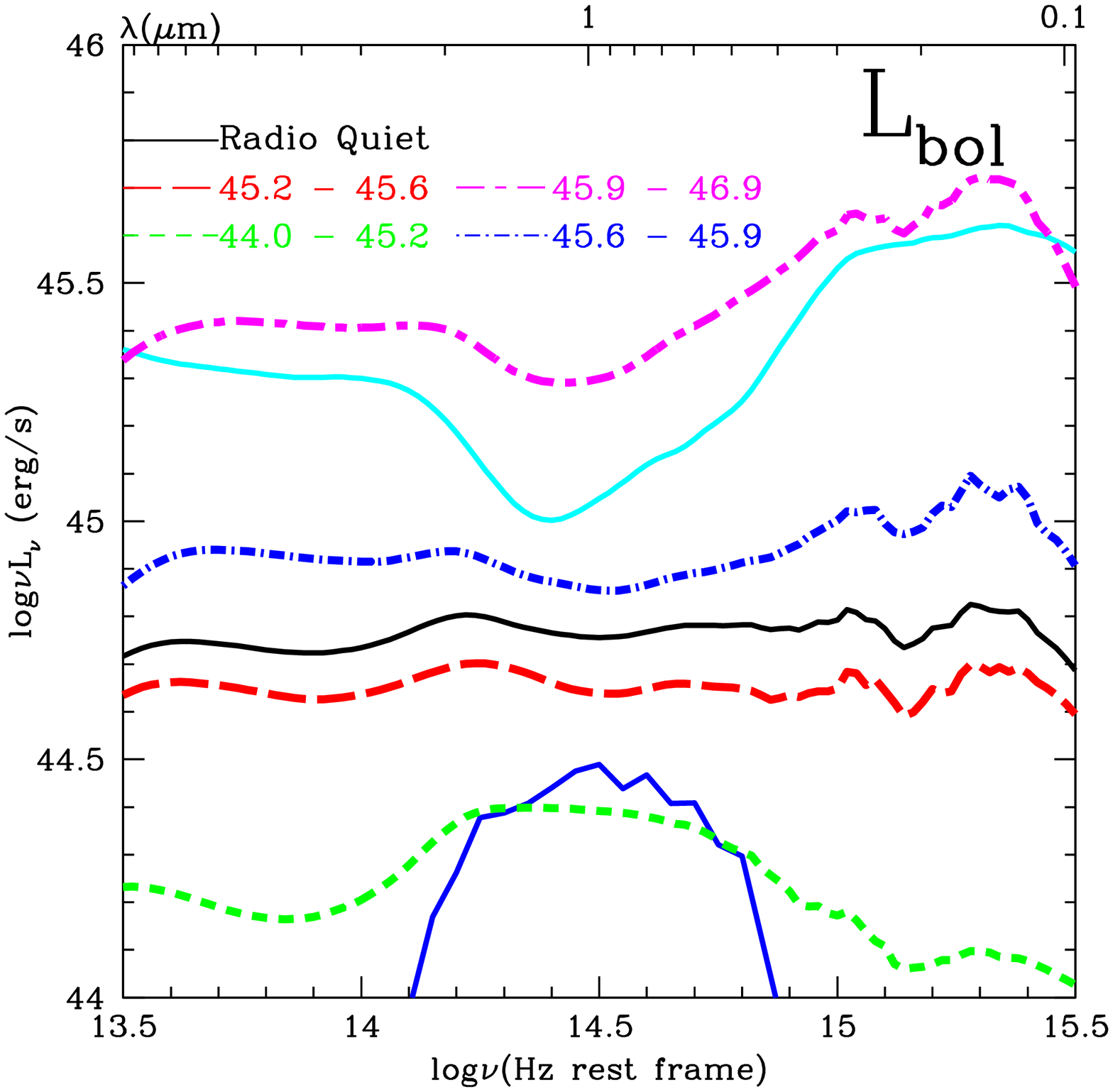}
\includegraphics[angle=0,width=0.45\textwidth]{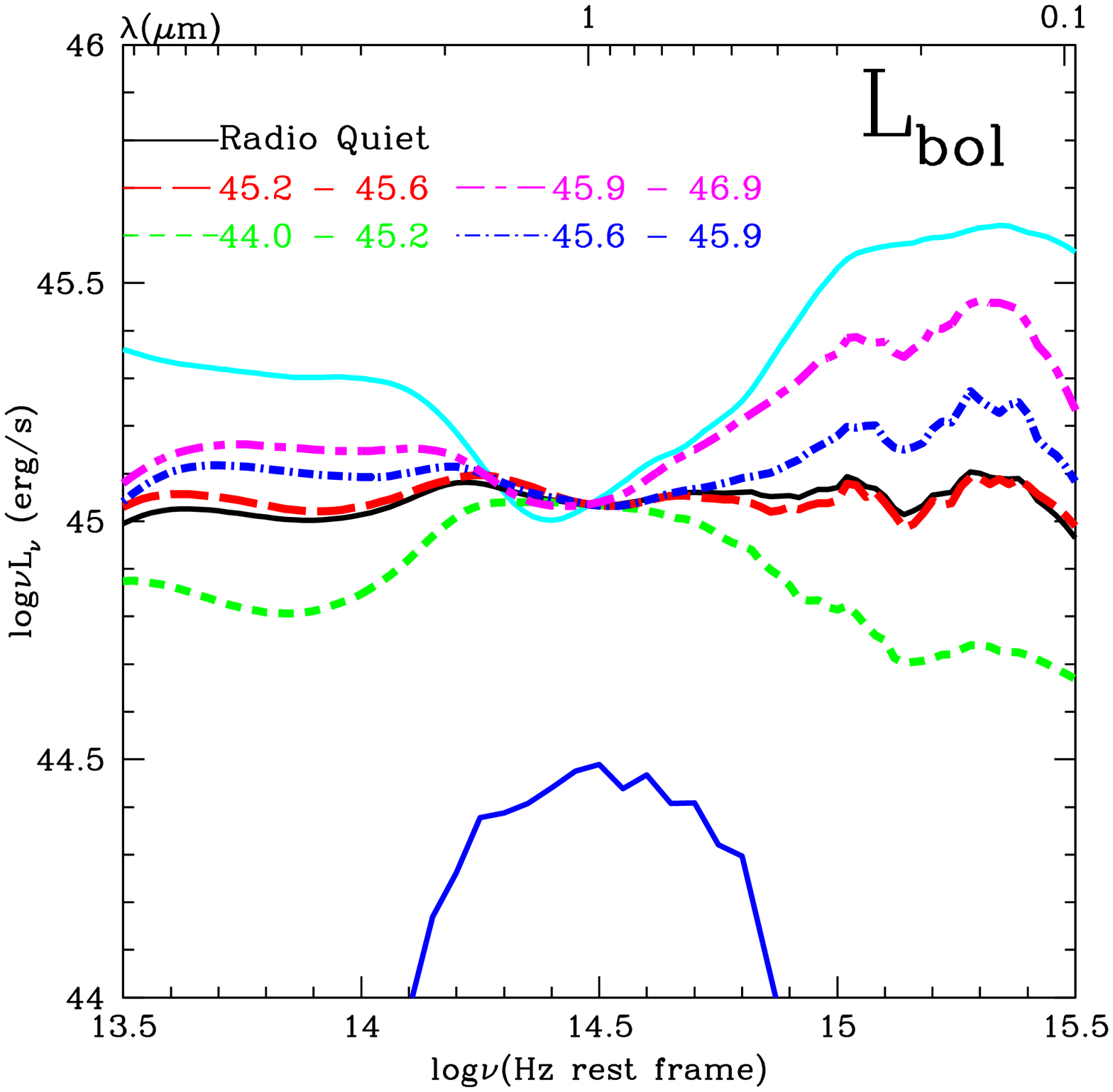}
\caption{The mean SED for XCRQ407 in bins of $z$, $L_{bol}$ before
(left) and after (right) normalization at 1$\mu m$ compared to E94
mean radio-quiet SED (cyan solid line). The black lines show the
mean SED for all the radio-quiet type 1 AGN in the XMM-COSMOS sample
(XCRQ407). A host galaxy template (an Elliptical (5~Gyr), E5, blue
solid line) is normalized to $L_{*}$ from UKIDSS Ultra Deep Survey
(Cirasuolo et al. 2007). \label{rqmsedbin}}
\end{figure*}

\begin{figure*}
\includegraphics[angle=0,width=0.45\textwidth]{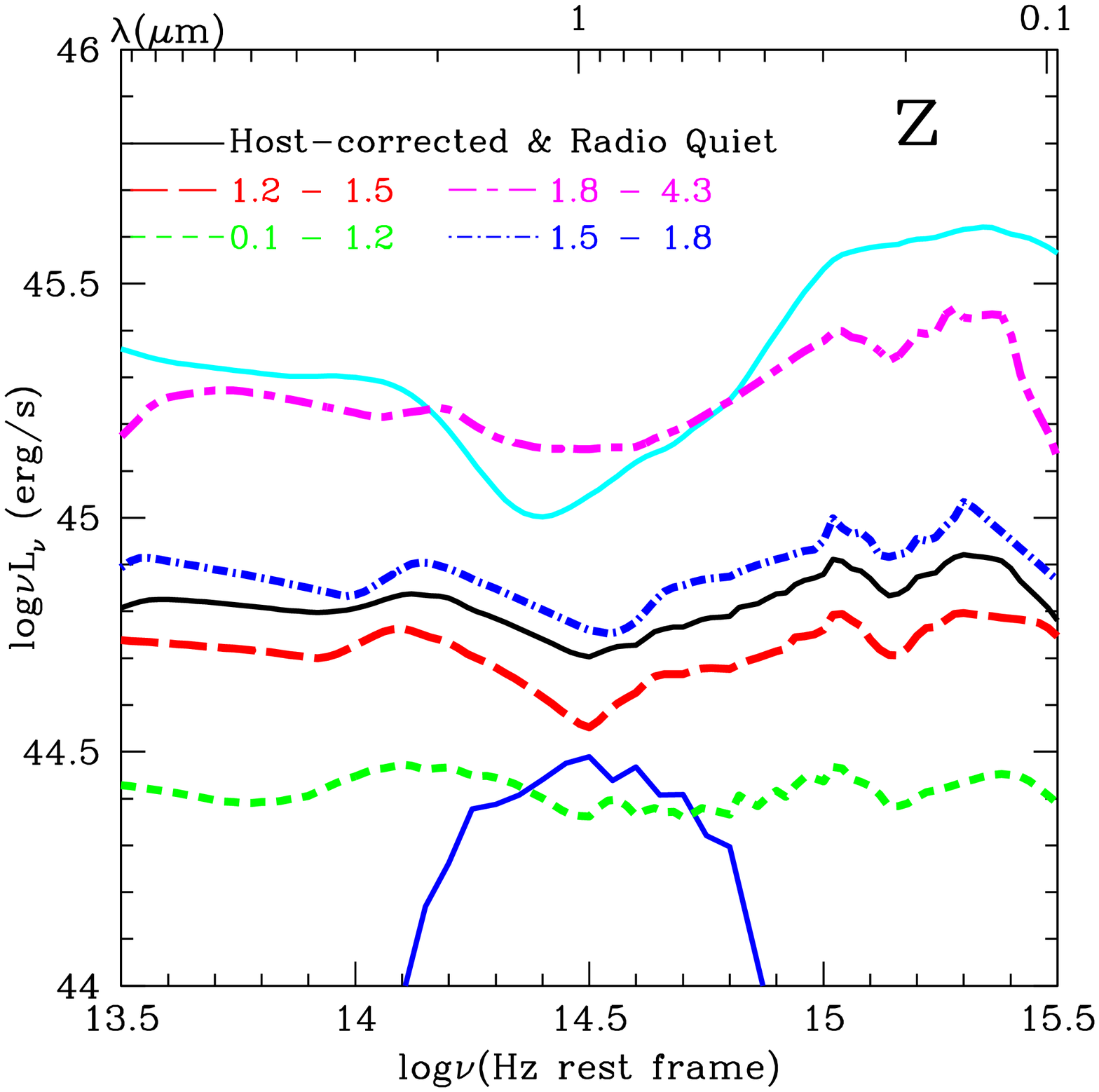}
\includegraphics[angle=0,width=0.45\textwidth]{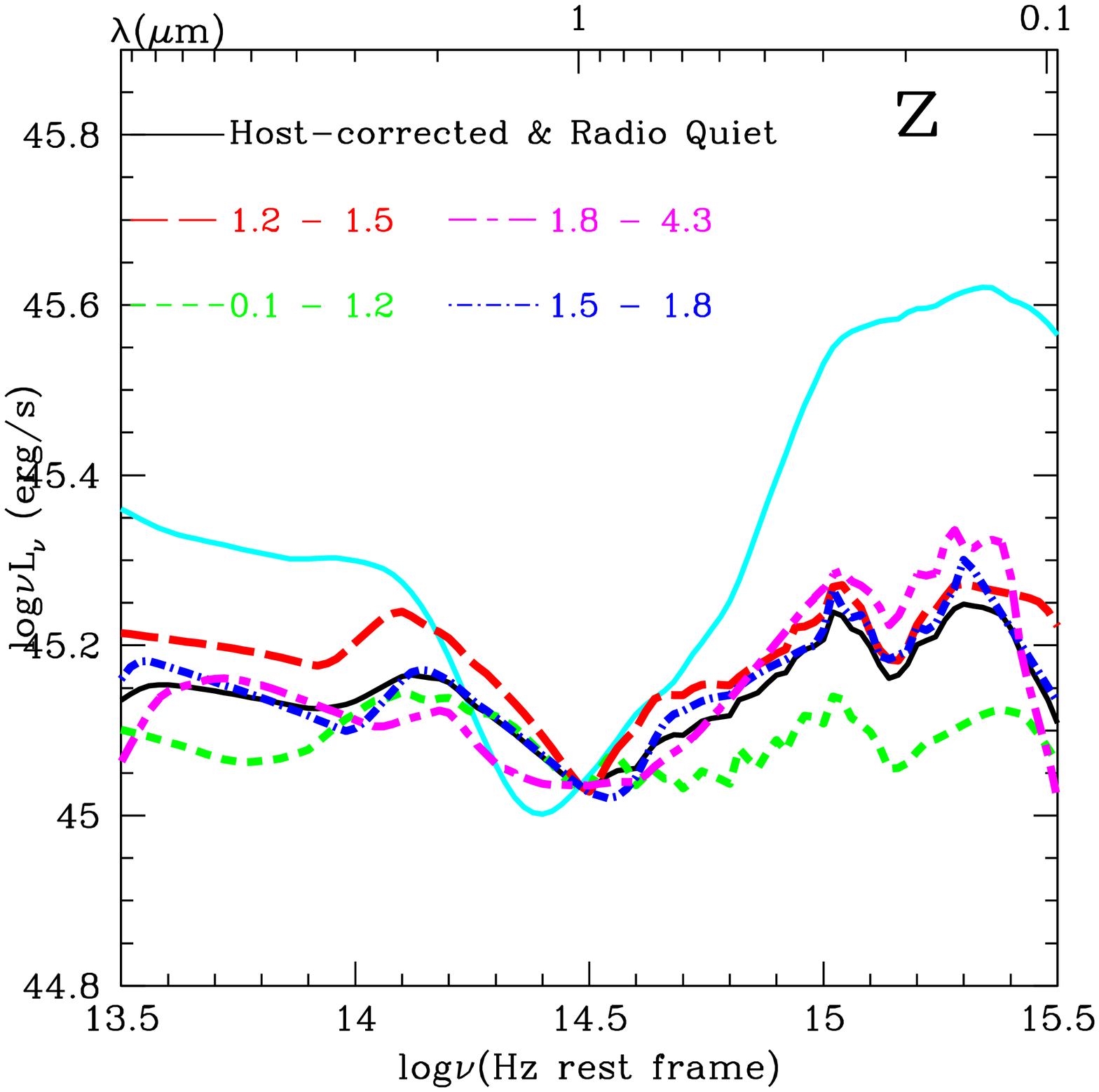}
\includegraphics[angle=0,width=0.45\textwidth]{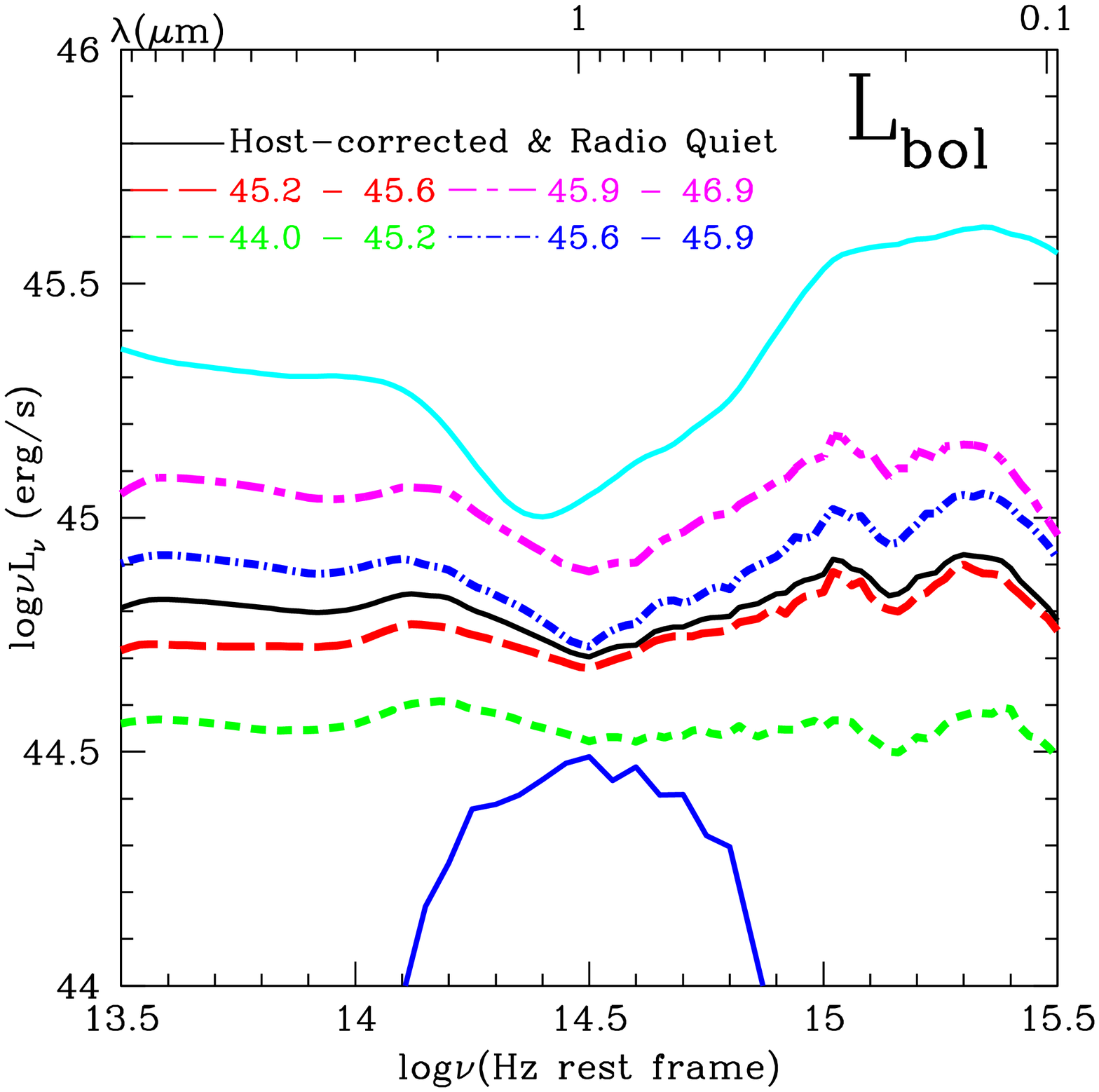}
\includegraphics[angle=0,width=0.45\textwidth]{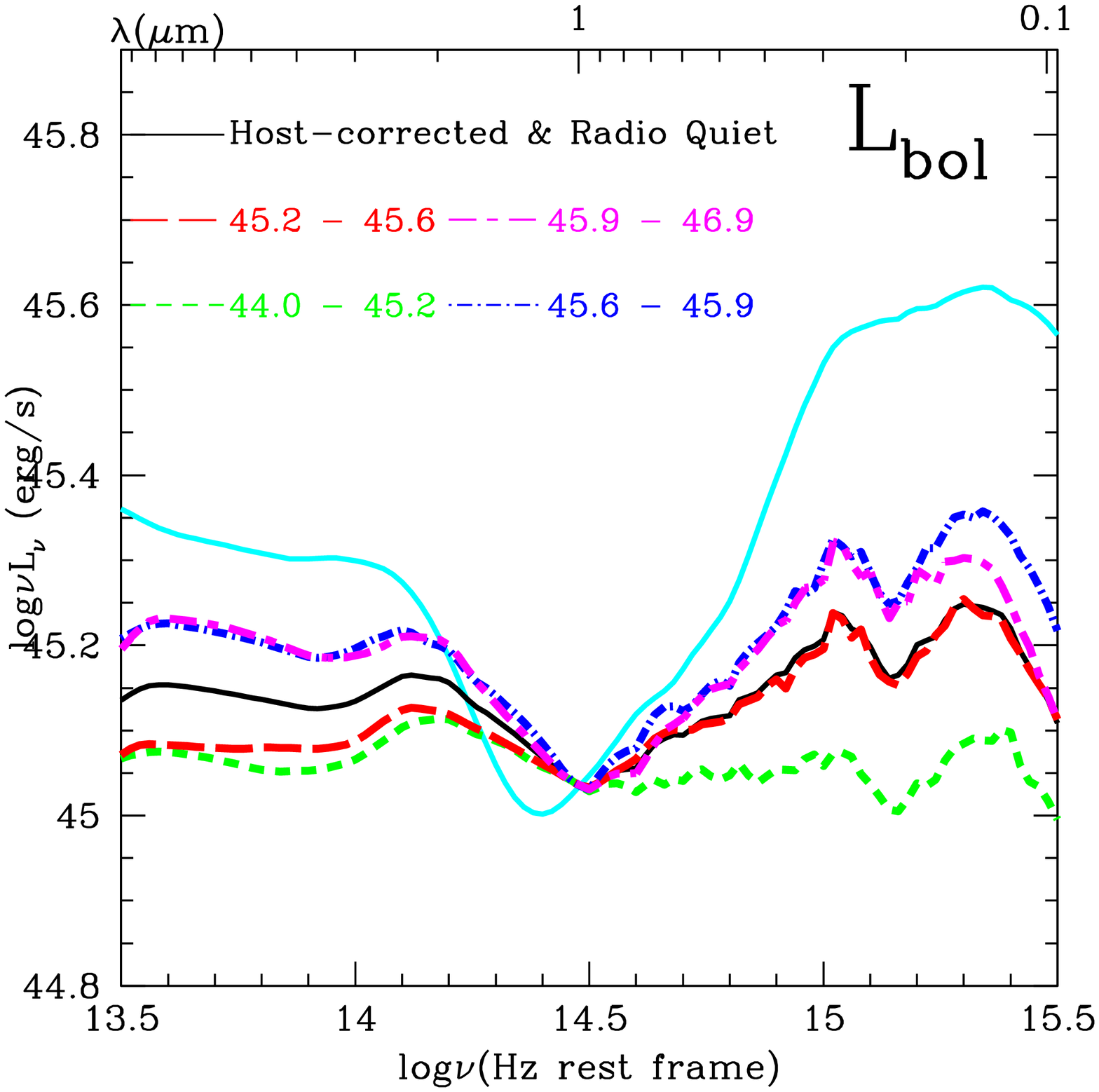}
\caption{The host-corrected mean SED for SSRQ200 in bins of $z$ and
$L_{bol}$ compared to E94 mean radio-quiet SED (cyan solid
line):{\em left:} before normalization; {\em right:} normalized at
1$\mu m$. The black solid lines show the host-corrected mean SED of
all quasars in SSRQ200. A host galaxy template (an Elliptical
(5~Gyr), blue solid line) is normalized to $L_{*}$ from UKIDSS Ultra
Deep Survey (Cirasuolo et al. 2007). Note that, to show the
difference of the mean SEDs in different bins clearly, we expanded
the y axis in all the plots of the right panel.\label{msedbin1}}
\end{figure*}

\begin{figure*}
\includegraphics[angle=0,width=0.45\textwidth]{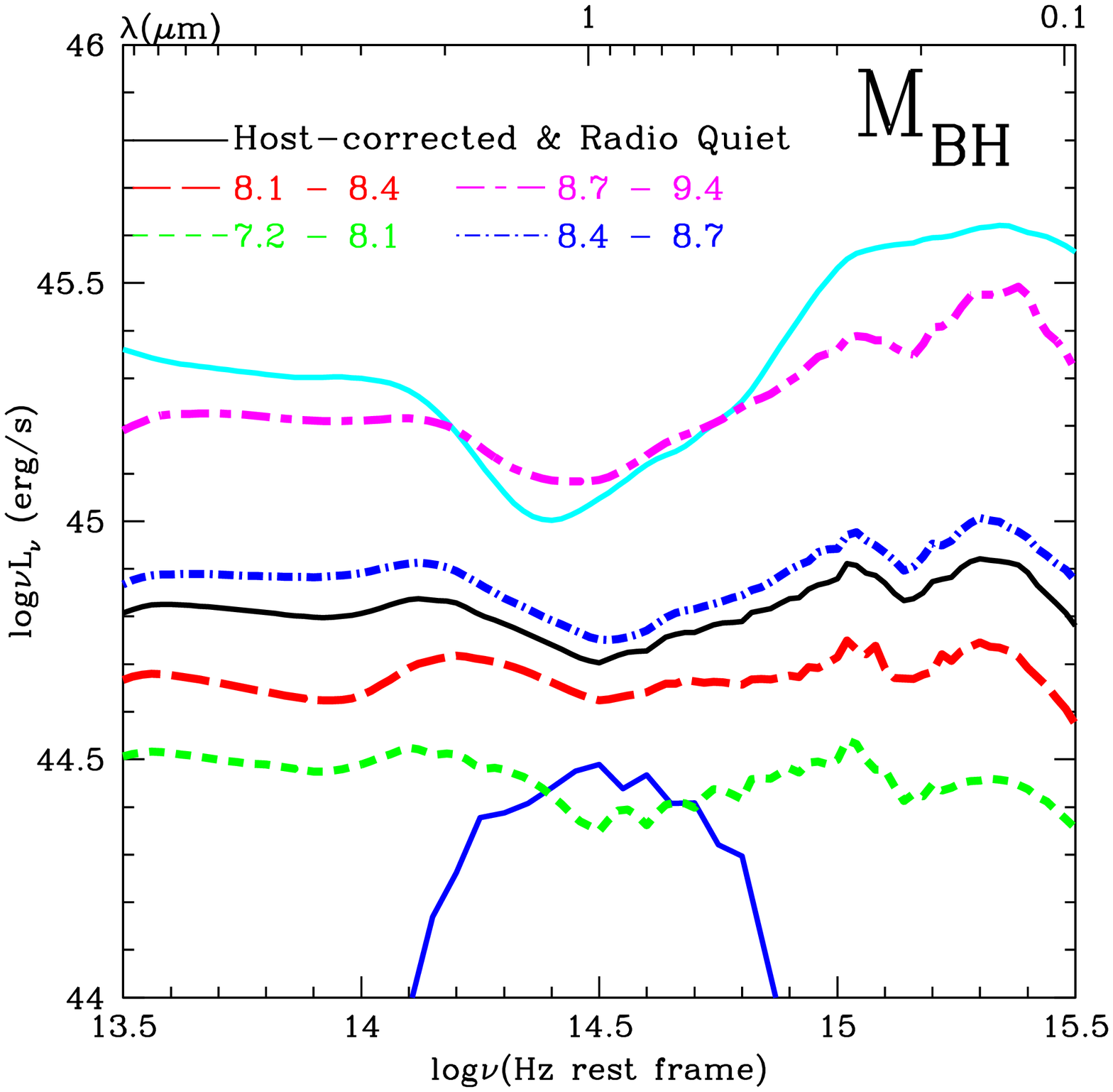}
\includegraphics[angle=0,width=0.45\textwidth]{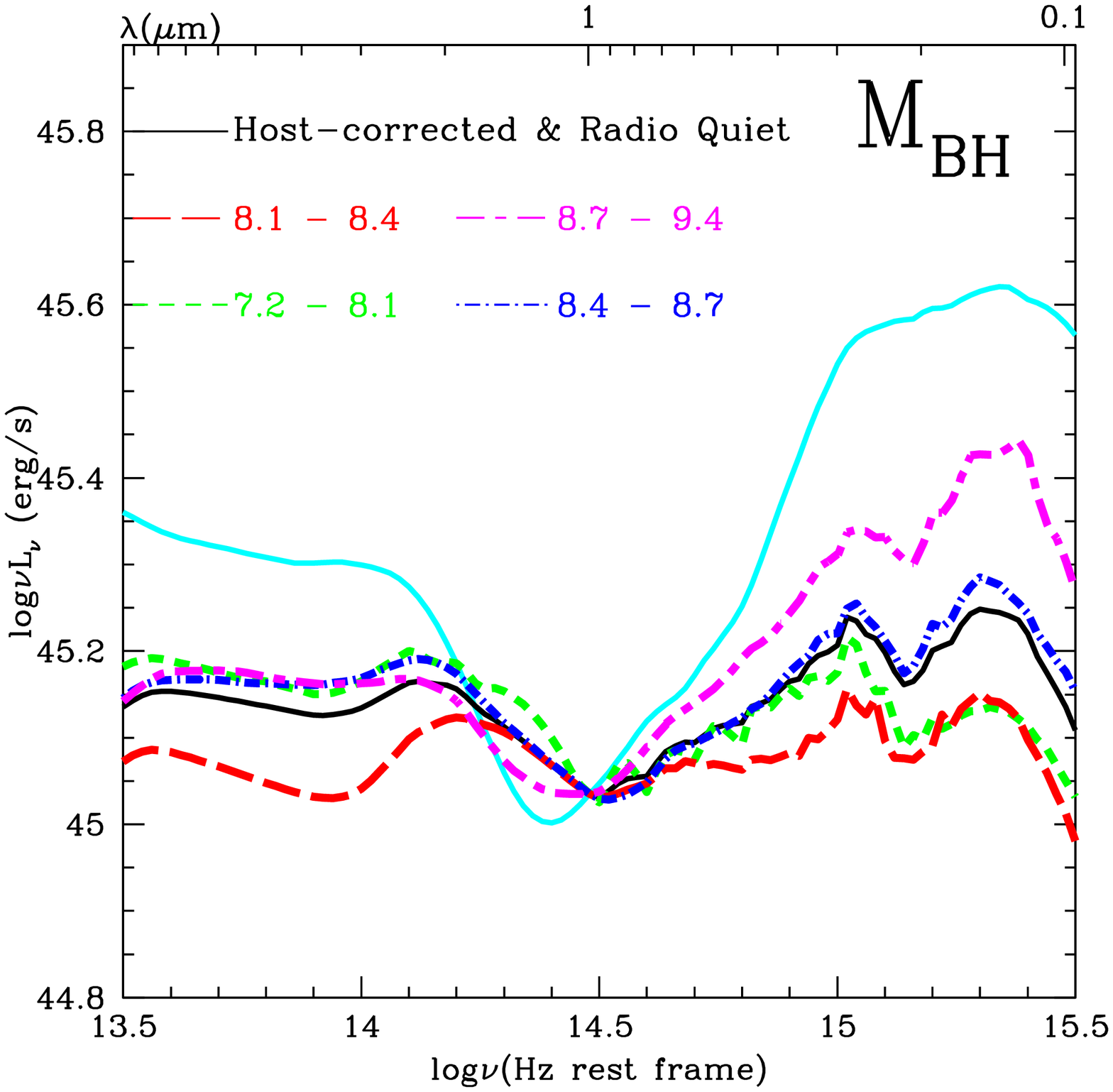}
\includegraphics[angle=0,width=0.45\textwidth]{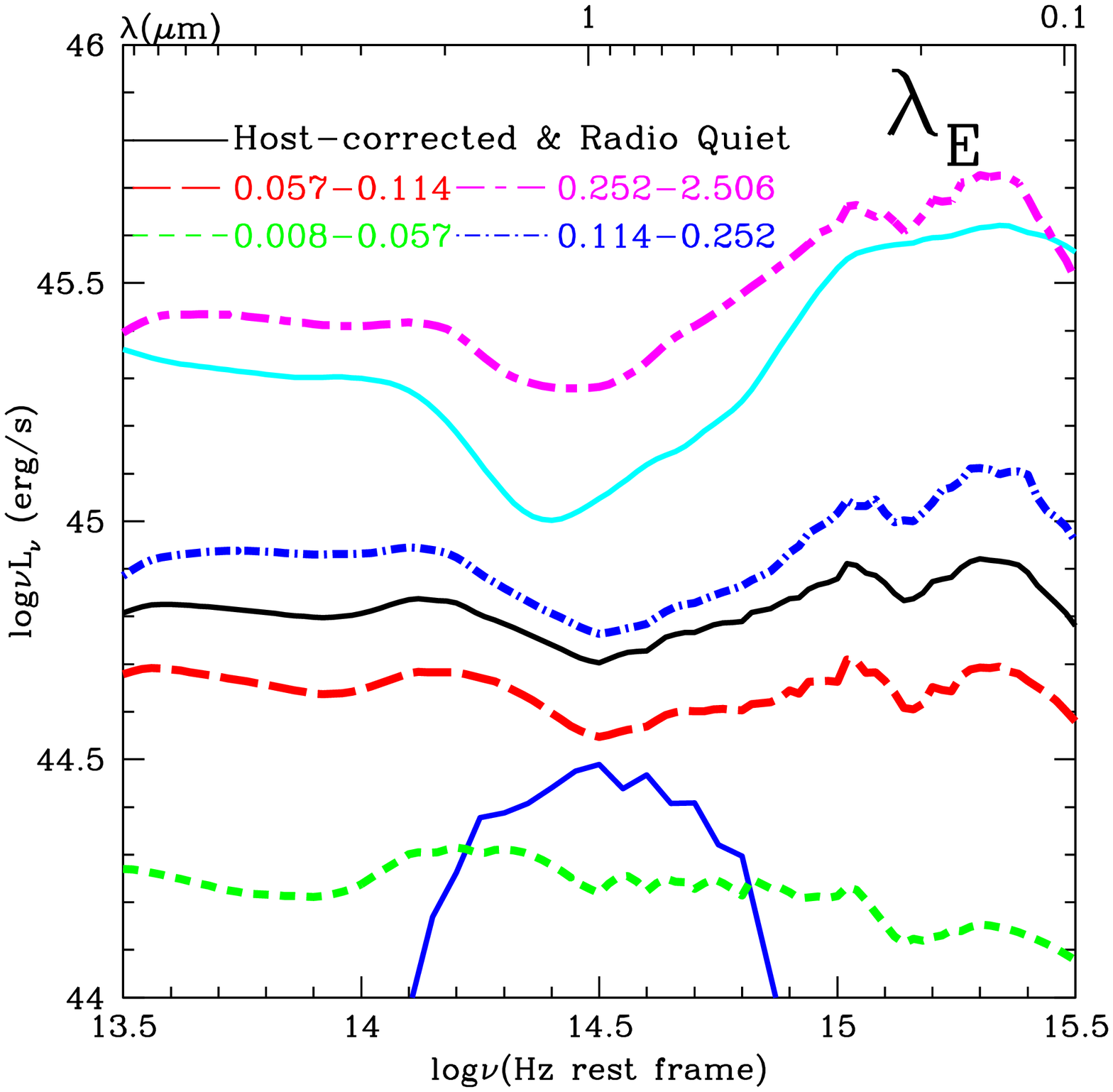}
\includegraphics[angle=0,width=0.45\textwidth]{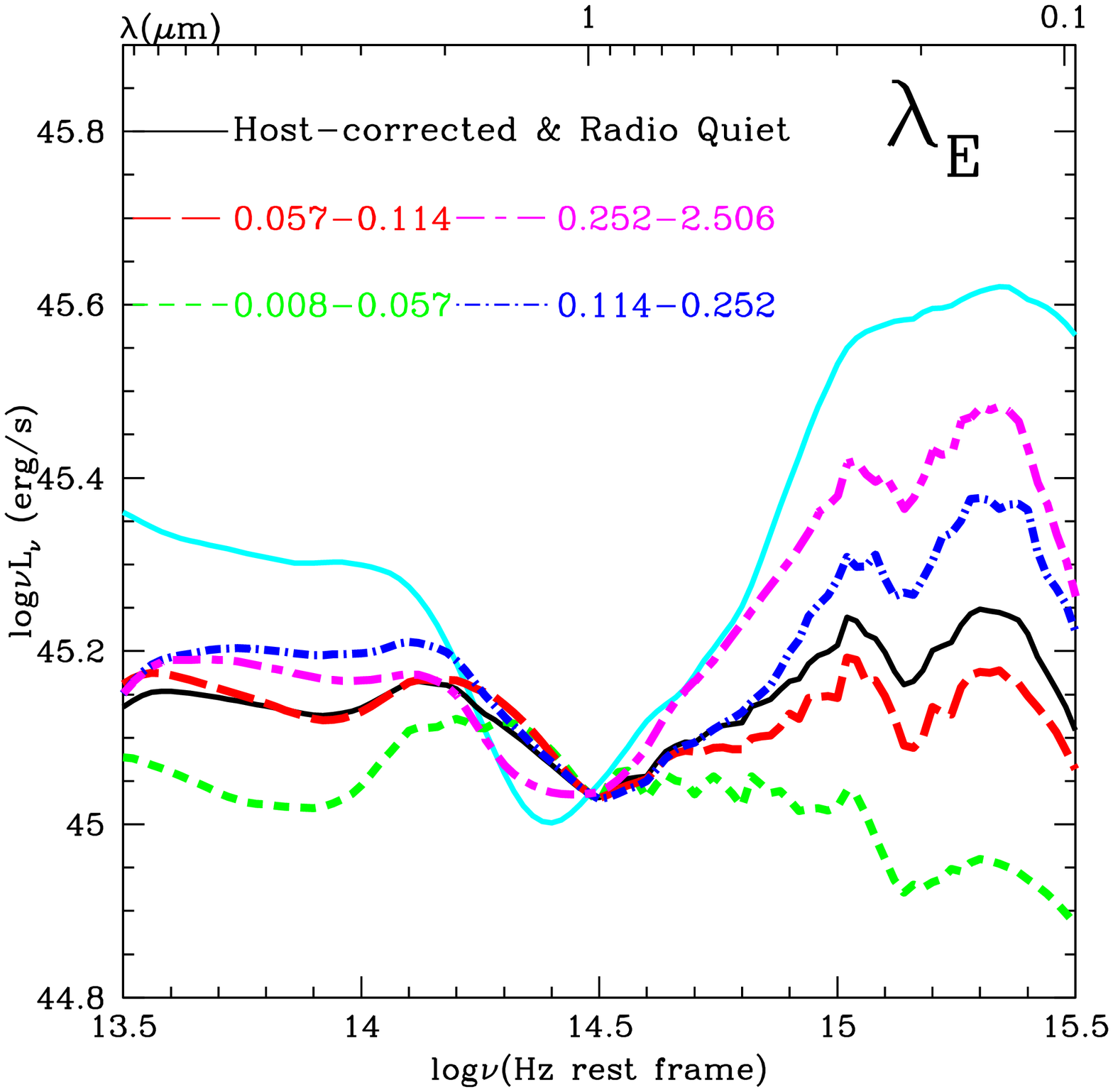}
\caption{The host-corrected mean SED for SSRQ200 in bins of $M_{BH}$
and $\lambda_{E}$ compared to E94 mean radio-quiet SED (cyan solid
line):{\em left:} before normalization; {\em right:} normalized at
1$\mu m$. The black solid lines show the host-corrected mean SED of
all quasars in SSRQ200. A host galaxy template (an Elliptical
(5~Gyr), blue solid line) is normalized to $L_{*}$ from UKIDSS Ultra
Deep Survey (Cirasuolo et al. 2007). Note that, to show the
difference of the mean SEDs in different bins clearly, we expanded
the y axis in all the plots of the right panel.\label{msedbin2}}
\end{figure*}

\begin{table}
\begin{minipage}{\columnwidth}
\centering \caption{Number of SSRQ200 quasars in Sub-bins for
Partial Evolution\label{t:npart}}
\begin{tabular}{@{}cc|cccc@{}}
\hline\multicolumn{2}{c}{bin} & \multicolumn{4}{c}{N in
sub-bin}\\
& sub-bin & $z$ & $L_{bol}$ & $M_{BH}$ & $\lambda_{E}$ \\
\hline
                    & 1 &    &  9 & 11 & 11 \\
$z$(2)              & 2 &    & 19 & 15 & 17 \\
                    & 3 &    & 14 & 13 & 14 \\
                    & 4 &    &  8 & 11 &  8 \\
                    \hline
                    & 1 &    & 10 & 11 & 11 \\
$z$(3)              & 2 &    & 16 & 15 & 15 \\
                    & 3 &    & 10 & 15 & 11 \\
                    & 4 &    & 15 & 10 & 14 \\
                    \hline
                    & 1 &  9 &    &  6 &  9 \\
$\log L_{bol}$(2)   & 2 & 19 &    & 17 & 19 \\
                    & 3 & 16 &    & 17 & 13 \\
                    & 4 &  6 &    & 10 &  9 \\
                    \hline
                    & 1 & 10 &    & 15 & 10 \\
$\log L_{bol}$(3)   & 2 & 14 &    & 14 & 14 \\
                    & 3 & 10 &    & 10 & 12 \\
                    & 4 & 16 &    & 11 & 14 \\
                    \hline
                    & 1 & 11 &  9 &    & 12 \\
$\log M_{BH}$(2)    & 2 & 15 & 17 &    & 23 \\
                    & 3 & 15 & 14 &    & 11 \\
                    & 4 &  9 & 10 &    &  4 \\
                    \hline
                    & 1 & 11 & 15 &    & 10 \\
$\log M_{BH}$(3)    & 2 & 13 & 17 &    & 10 \\
                    & 3 & 15 & 10 &    & 19 \\
                    & 4 & 11 &  8 &    & 11 \\
                    \hline
                    & 1 & 11 &  9 & 13 &    \\
$\log\lambda_{E}$(2)& 2 & 17 & 19 & 23 &    \\
                    & 3 & 15 & 14 & 10 &    \\
                    & 4 &  7 &  8 &  4 &    \\
                    \hline
                    & 1 &  8 & 11 &  8 &    \\
$\log\lambda_{E}$(3)& 2 & 14 & 13 & 11 &    \\
                    & 3 & 11 & 12 & 19 &    \\
                    & 4 & 17 & 14 & 12 &    \\
\hline
\end{tabular}
\end{minipage}
\end{table}

\begin{figure*}
\includegraphics[angle=0,width=0.32\textwidth]{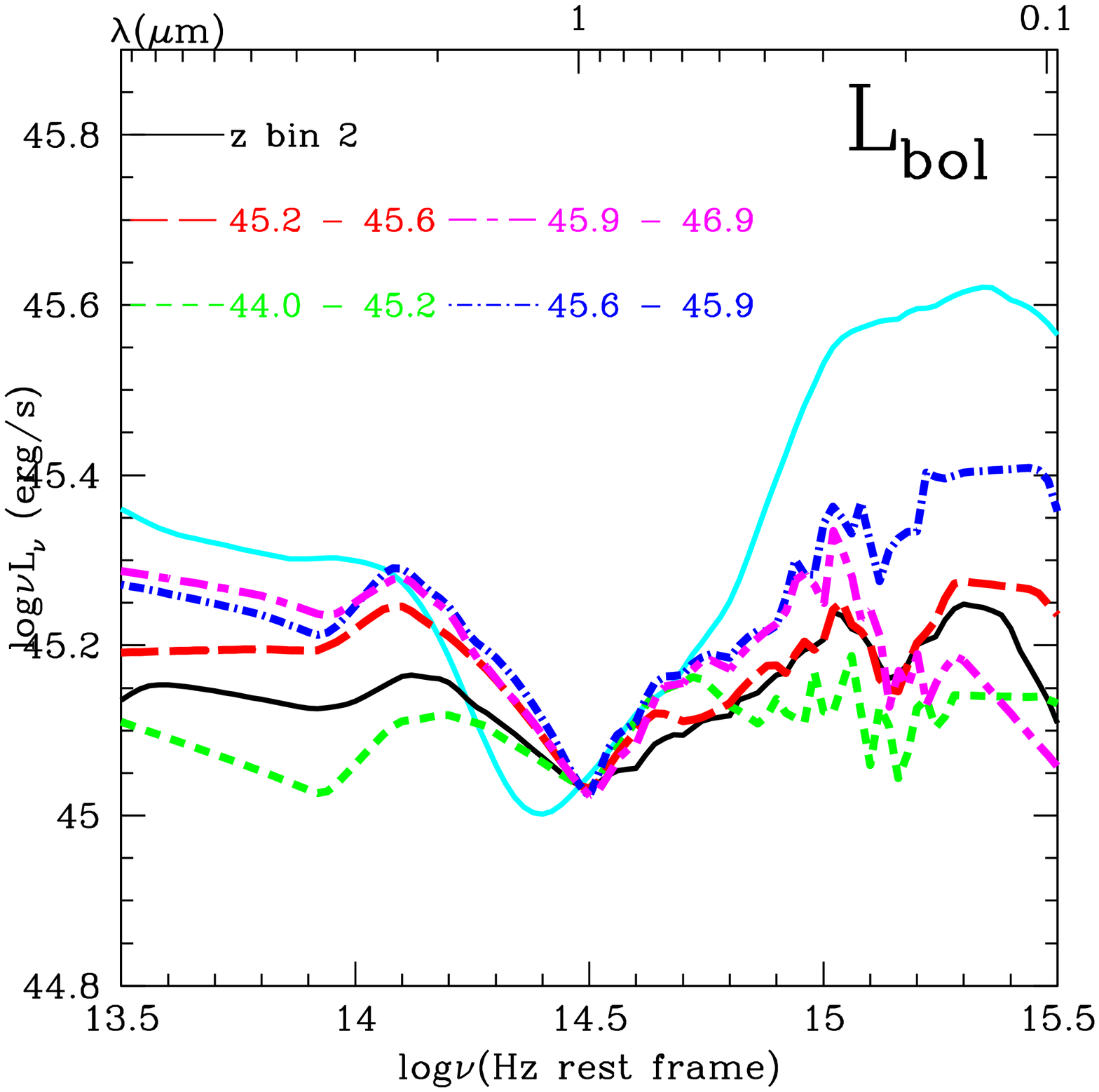}
\includegraphics[angle=0,width=0.32\textwidth]{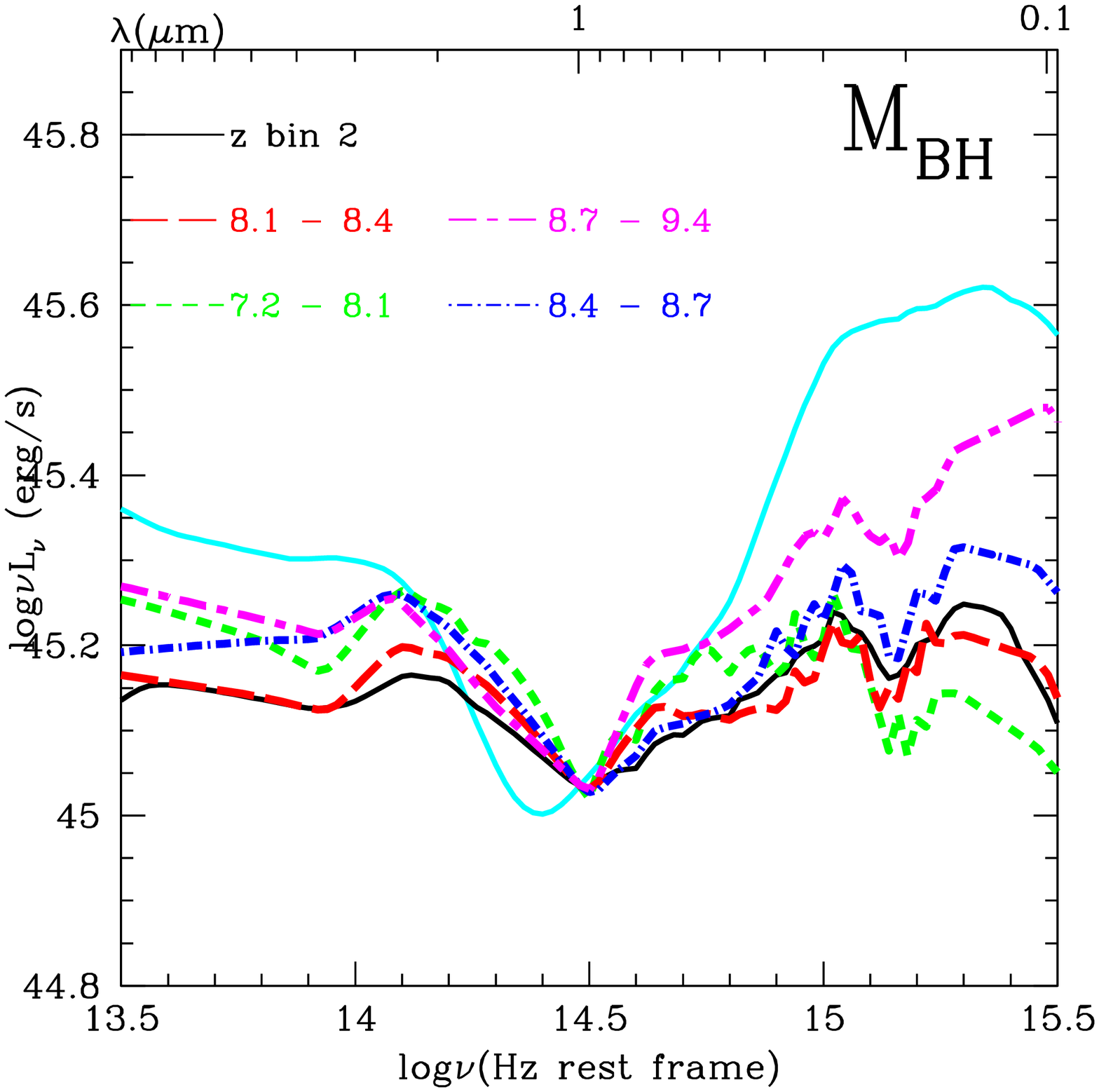}
\includegraphics[angle=0,width=0.32\textwidth]{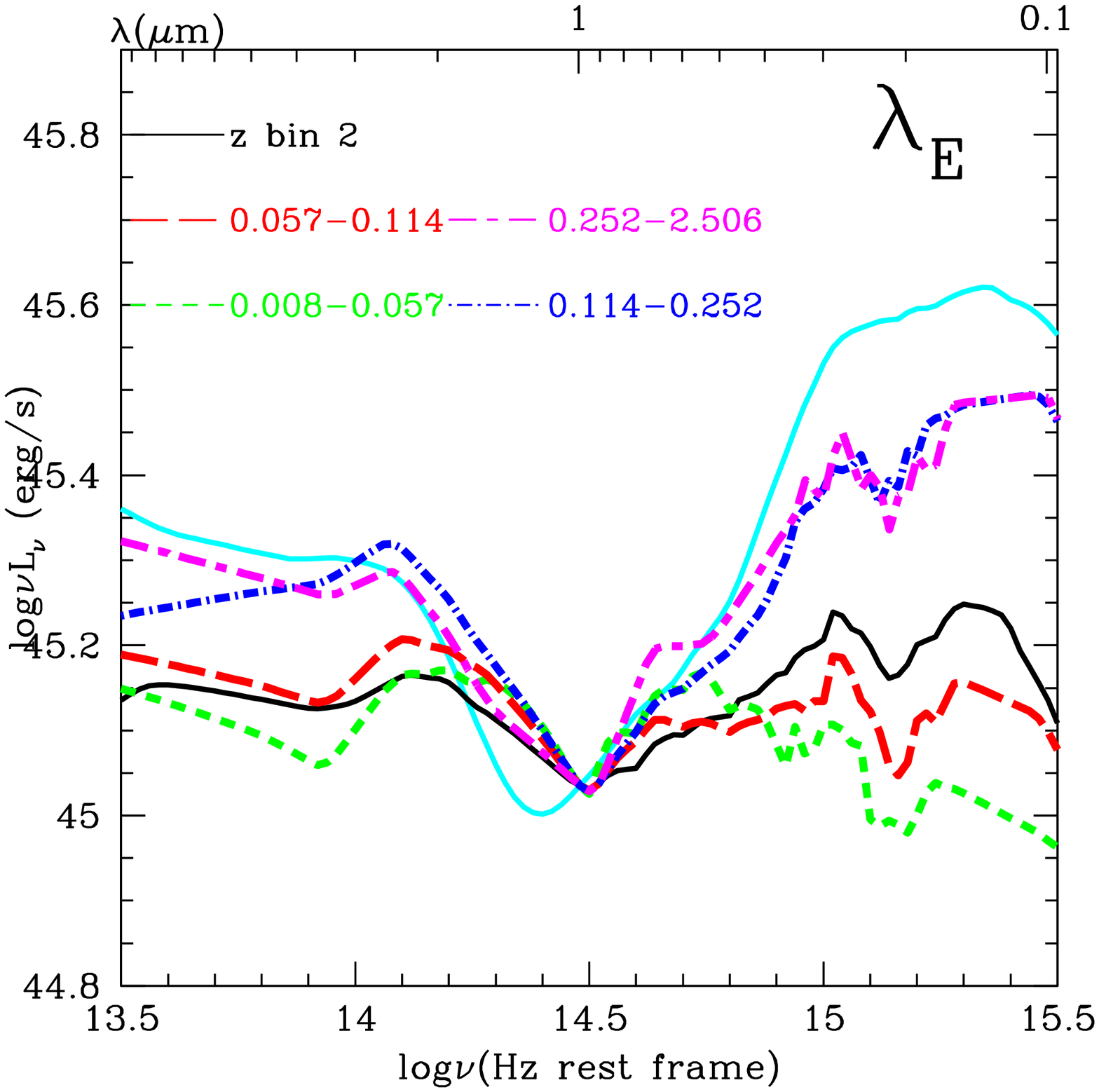}
\includegraphics[angle=0,width=0.32\textwidth]{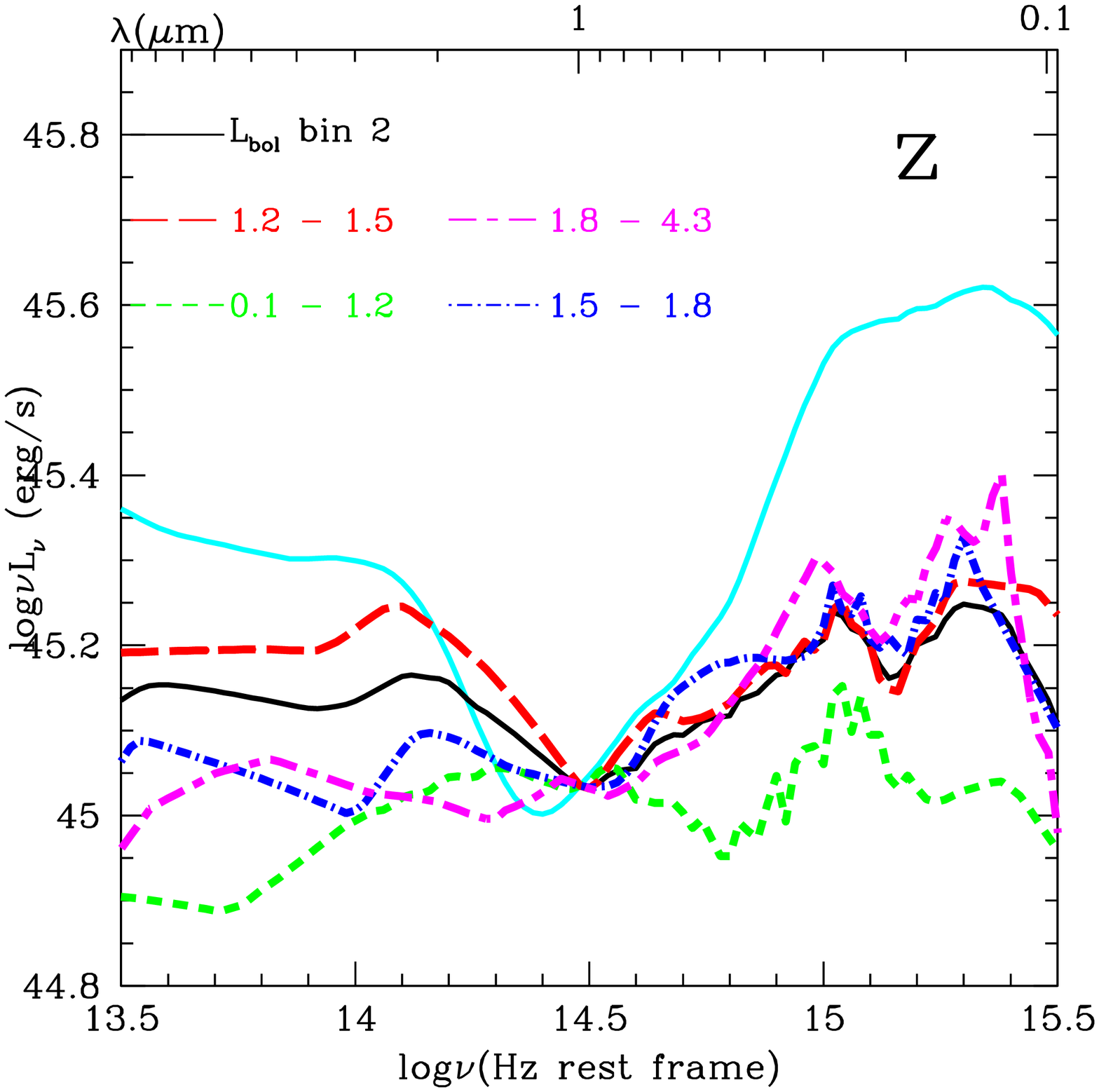}
\includegraphics[angle=0,width=0.32\textwidth]{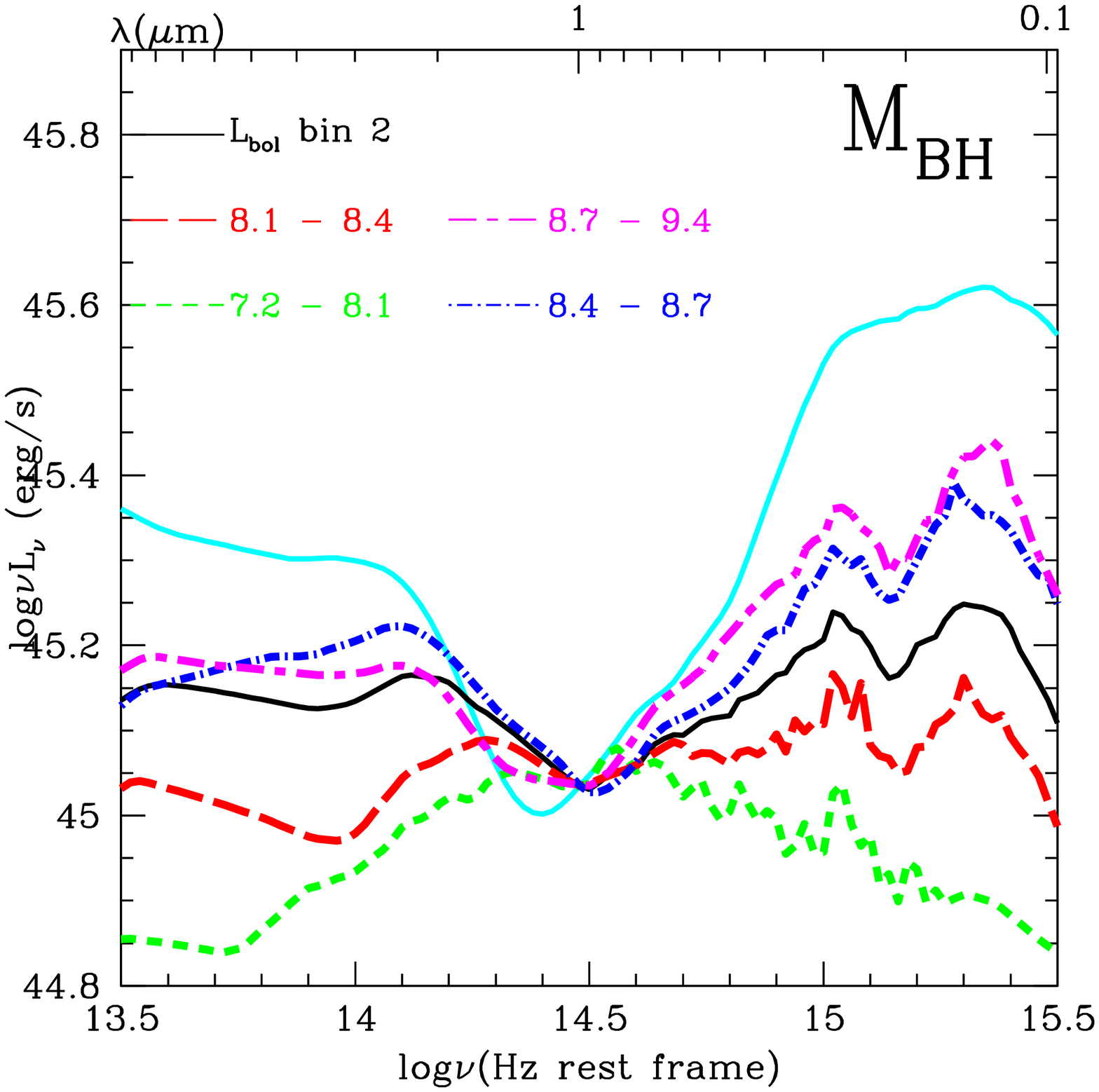}
\includegraphics[angle=0,width=0.32\textwidth]{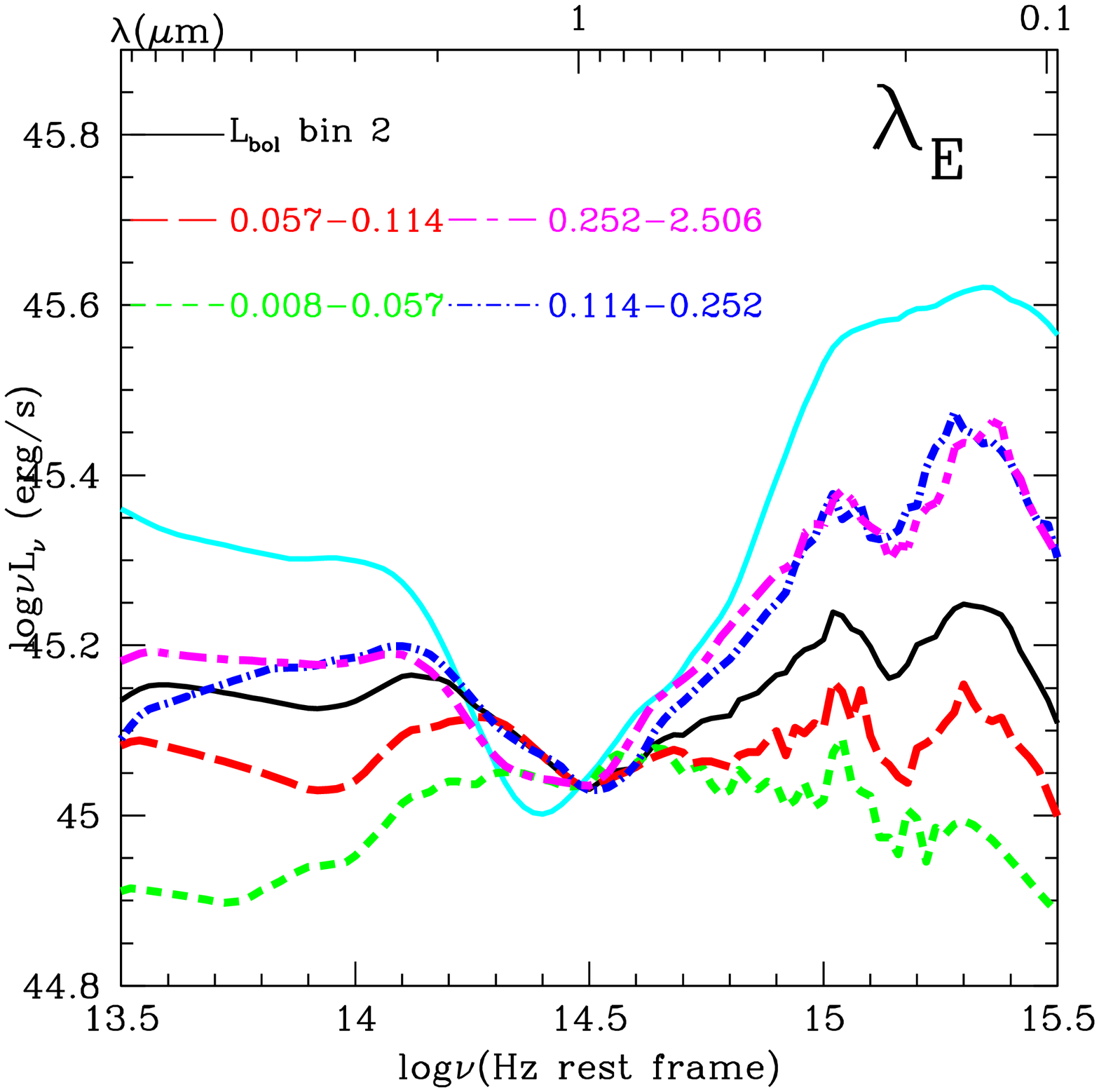}
\includegraphics[angle=0,width=0.32\textwidth]{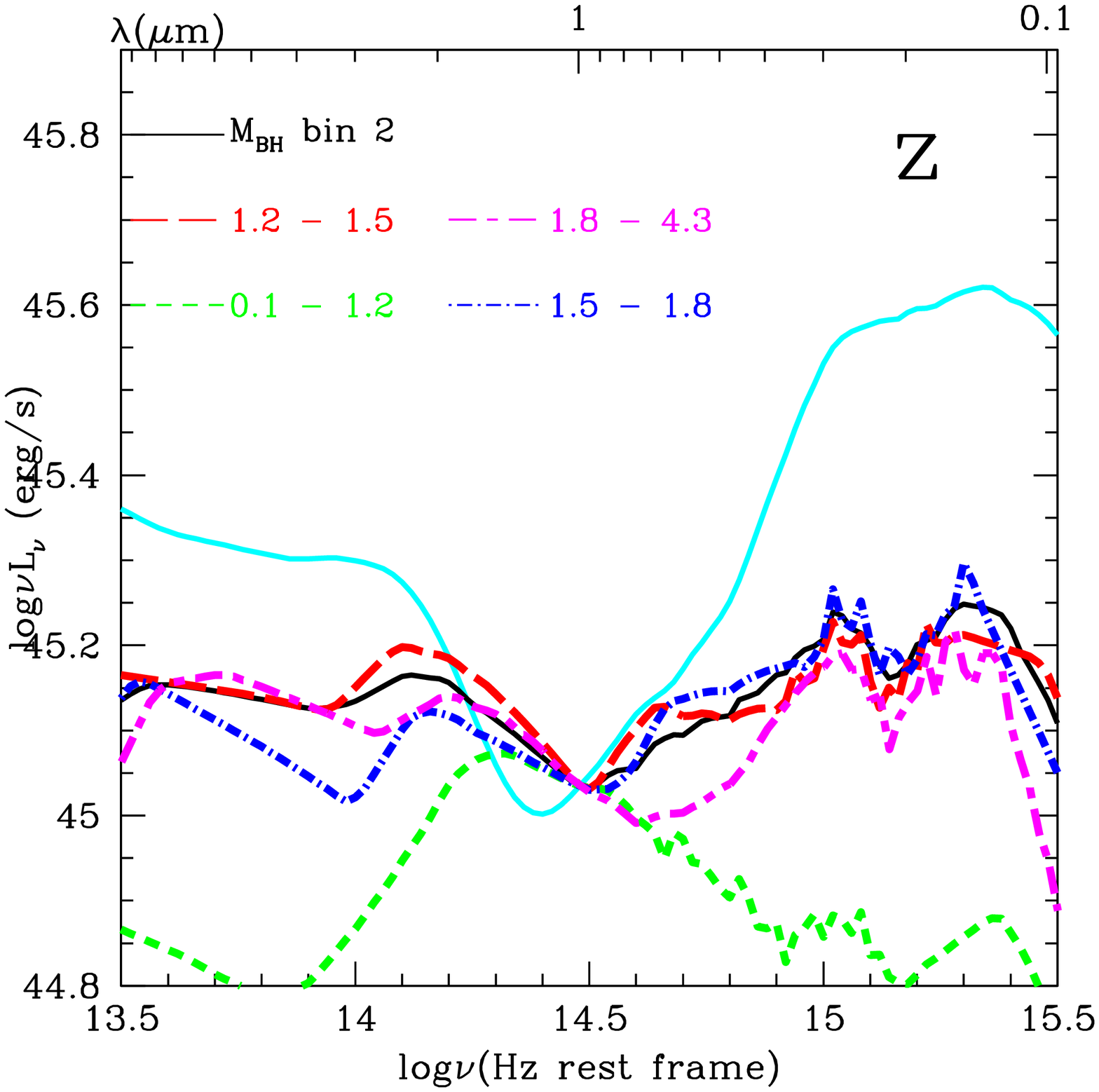}
\includegraphics[angle=0,width=0.32\textwidth]{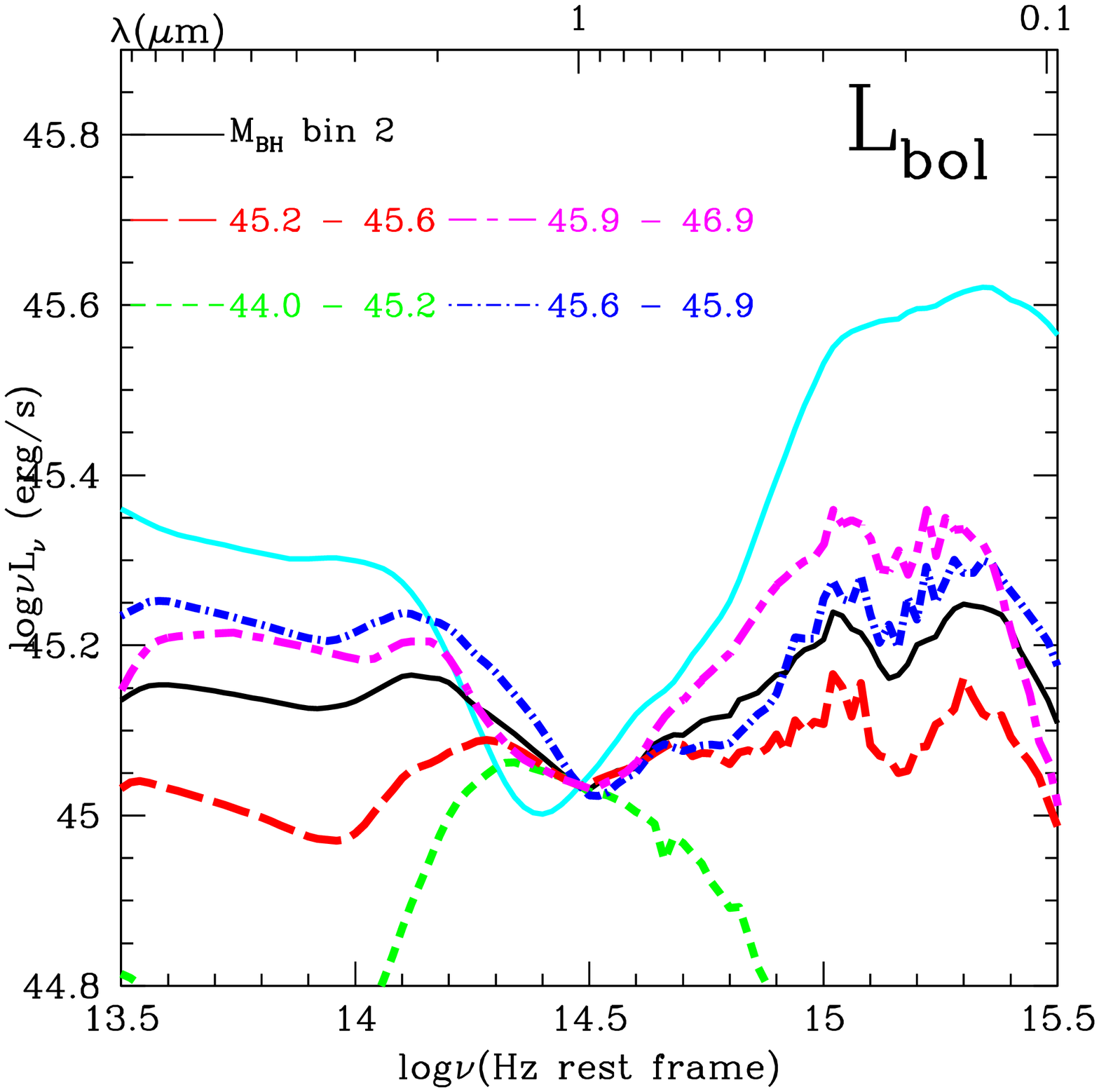}
\includegraphics[angle=0,width=0.32\textwidth]{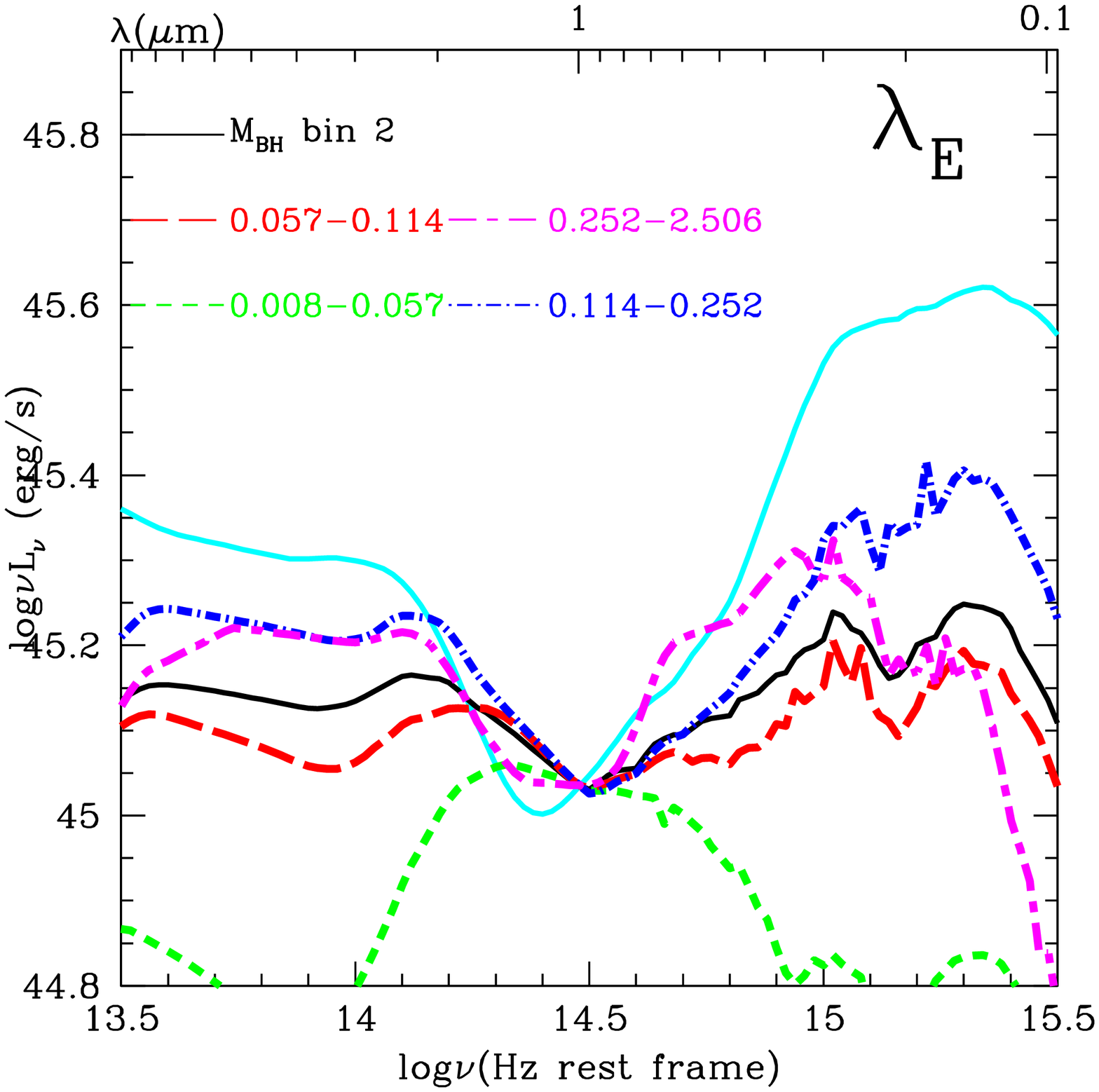}
\includegraphics[angle=0,width=0.32\textwidth]{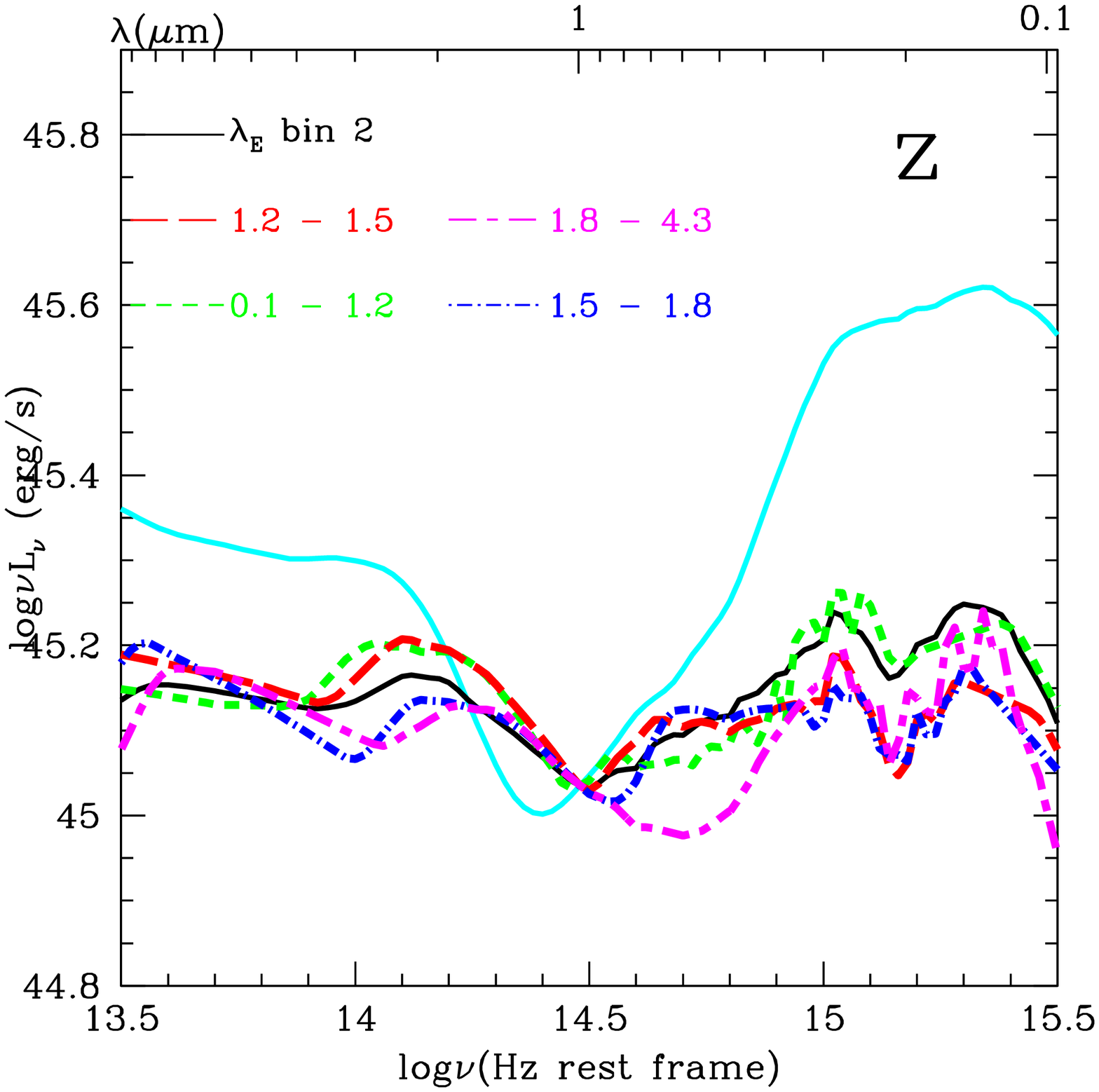}
\includegraphics[angle=0,width=0.32\textwidth]{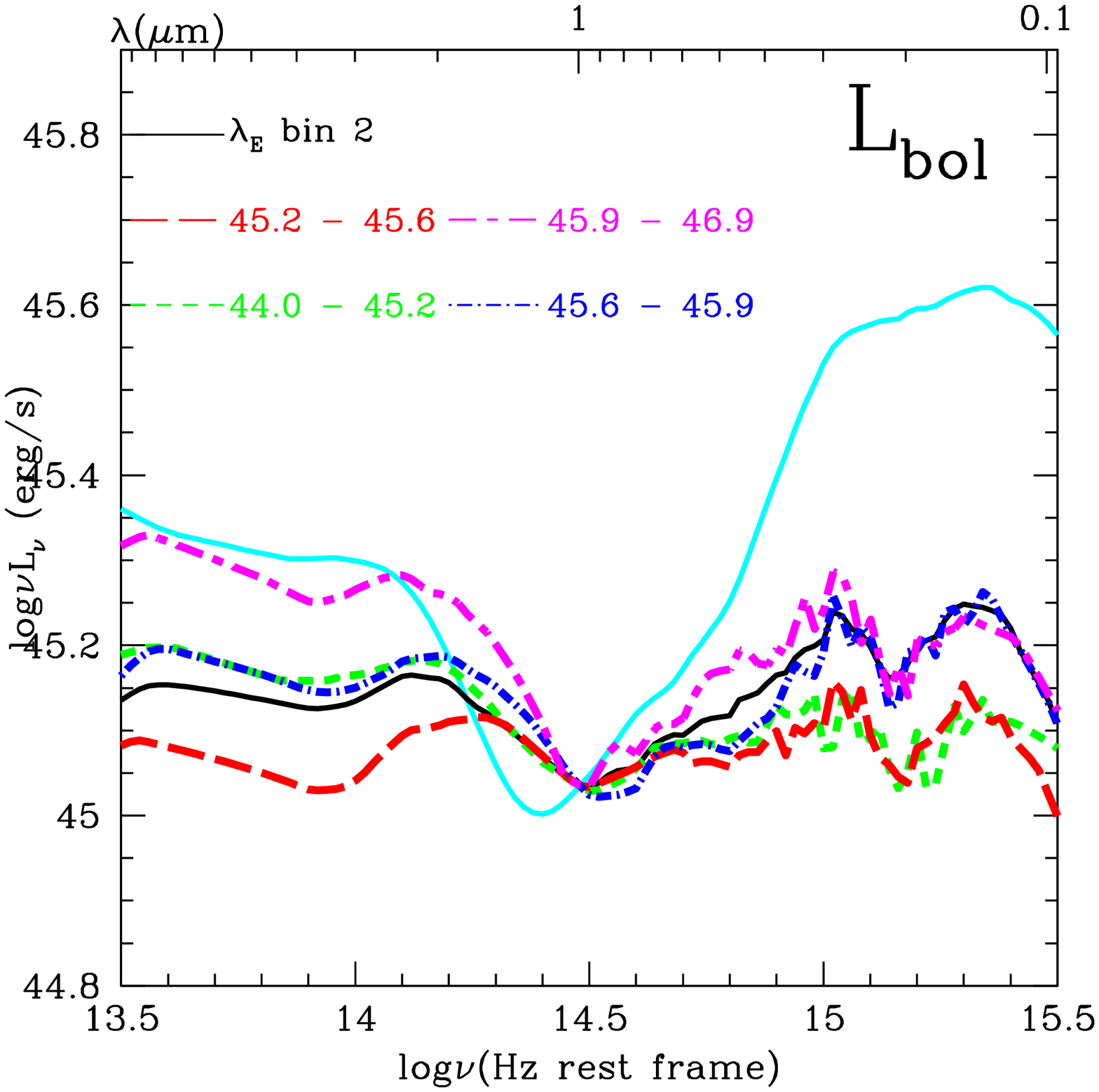}
\includegraphics[angle=0,width=0.32\textwidth]{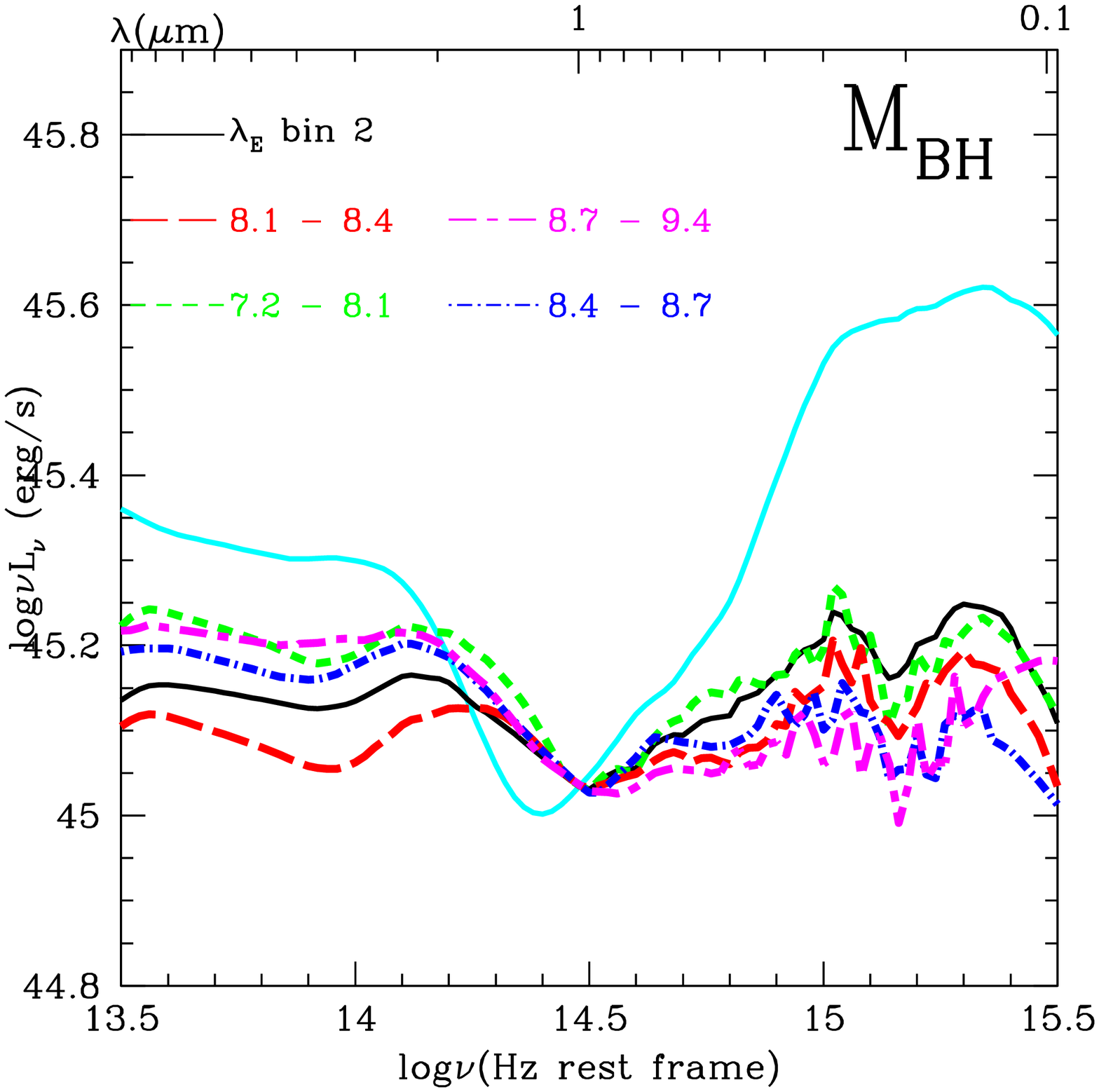}
\caption{The mean host-subtracted SED normalized at 1$\mu m$
compared to E94 mean radio-quiet SED (cyan solid line). Different
rows are for quasars in specific bins: (1) $z$ bin 2 ($1.2<z<1.5$);
(2) $\log L_{bol}$ bin 2 ($45.2<\log L_{bol}<45.6$); (3)
$\log(M_{BH}/M_{\bigodot})$ bin 2
($8.1<\log(M_{BH}/M_{\bigodot})<8.4$); (4) $\log\lambda_E$ bin 2
($0.057<\lambda_E<0.114$). The symbol on the upper right corner of
the each plot shows which sub-bin is considered. The lines are color
coded as in Figure~\ref{msedbin1}~\&~\ref{msedbin2}.
\label{msedbinpart2}}
\end{figure*}

\begin{figure*}
\includegraphics[angle=0,width=0.32\textwidth]{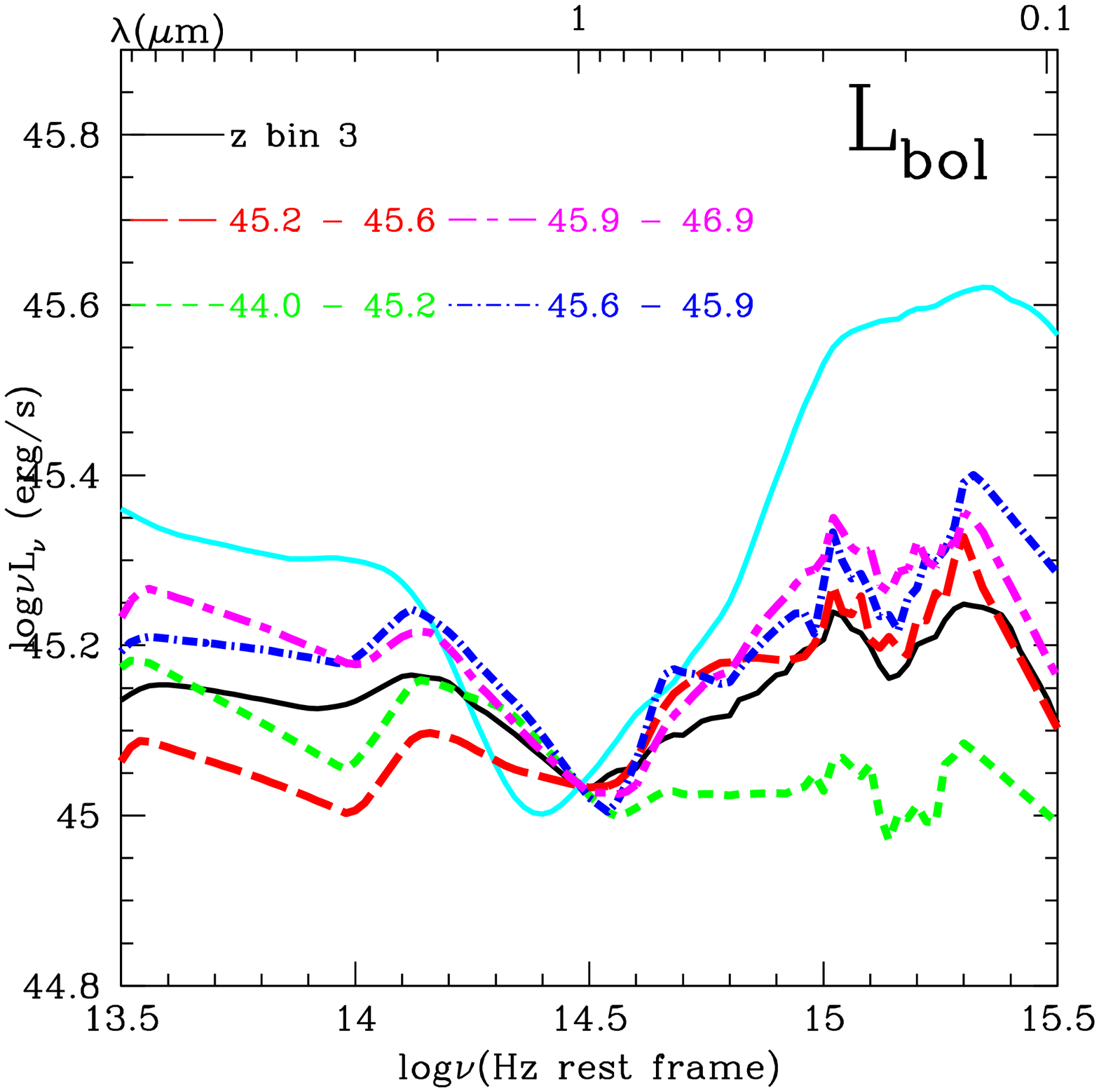}
\includegraphics[angle=0,width=0.32\textwidth]{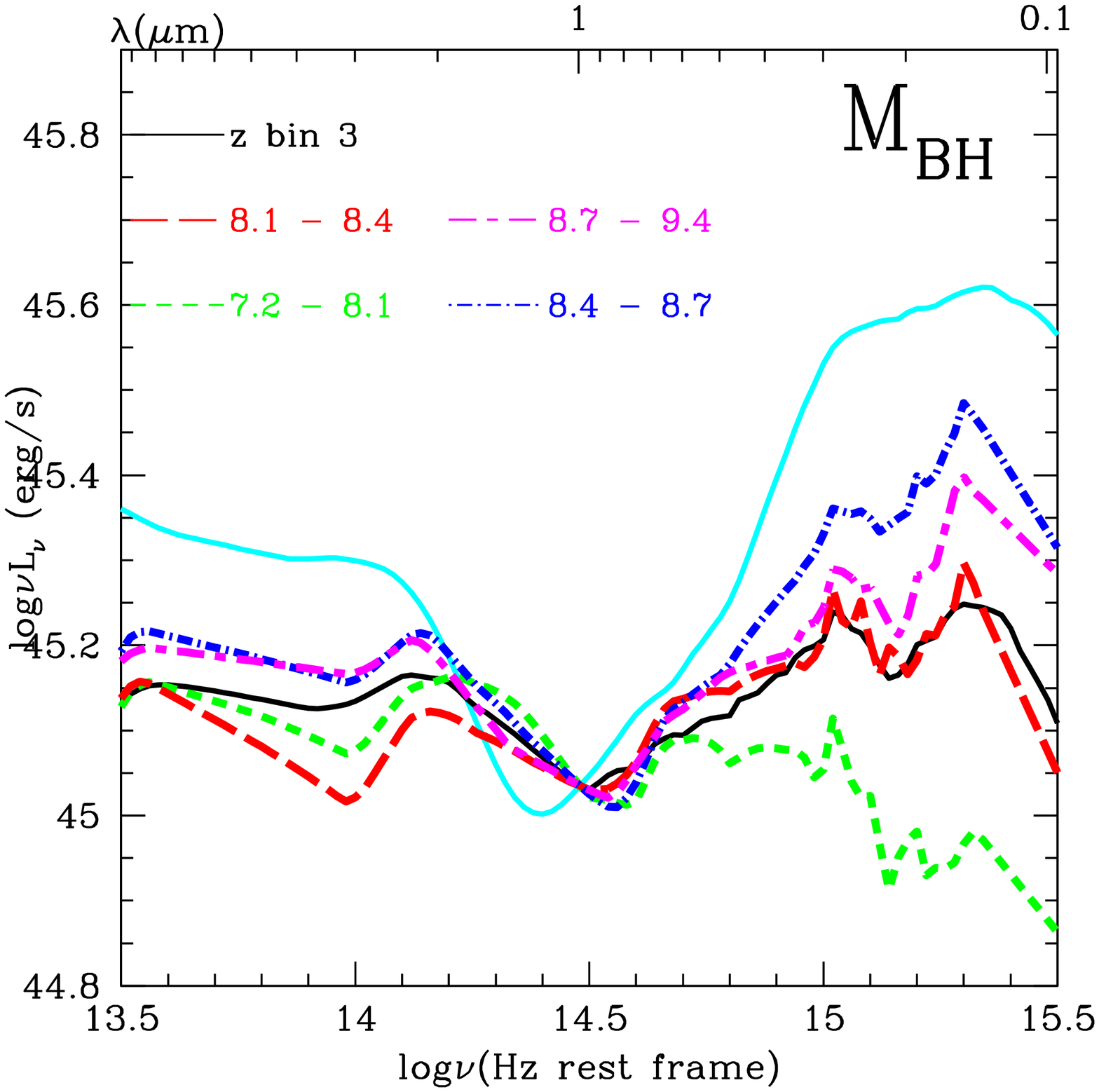}
\includegraphics[angle=0,width=0.32\textwidth]{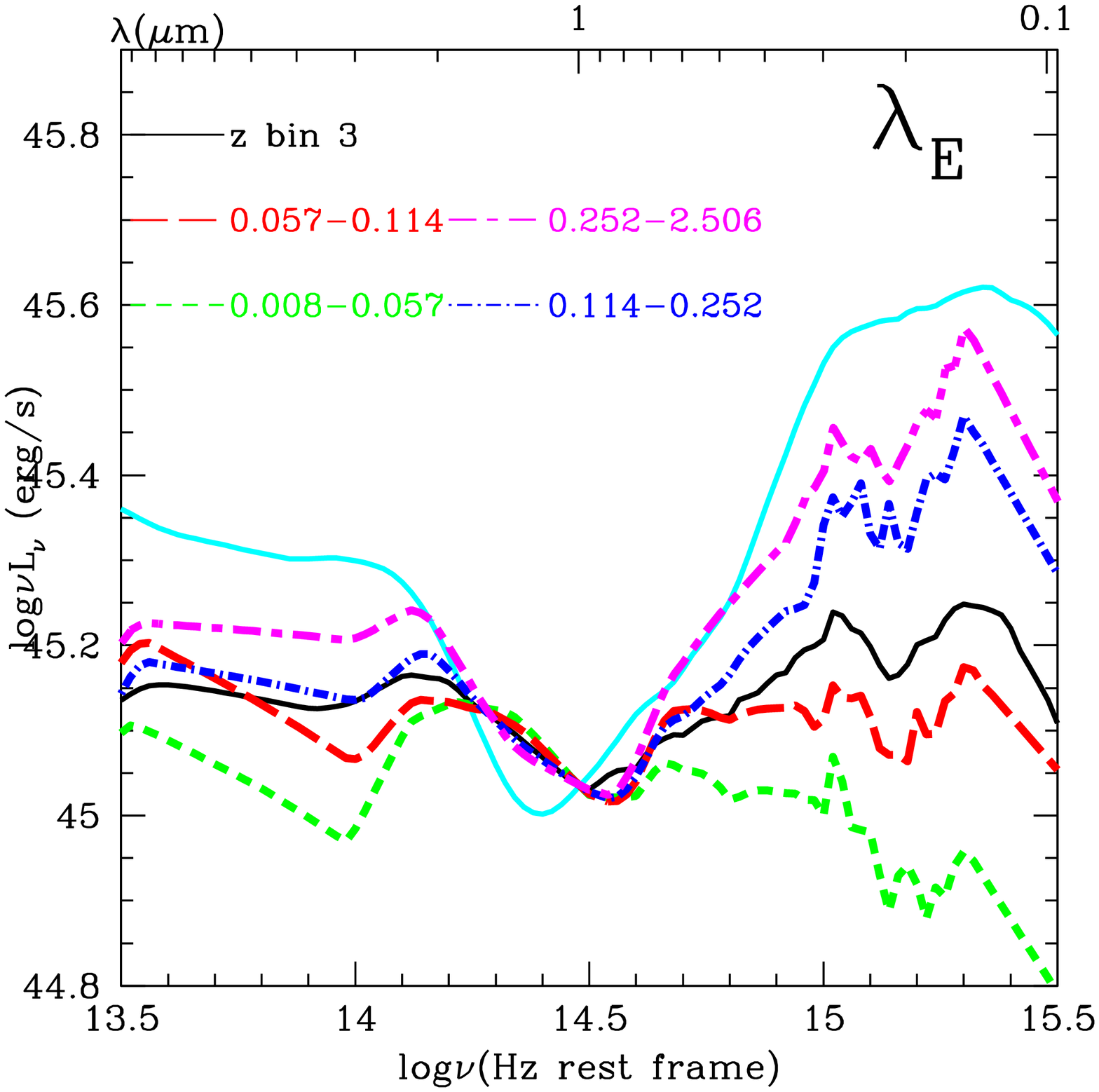}
\includegraphics[angle=0,width=0.32\textwidth]{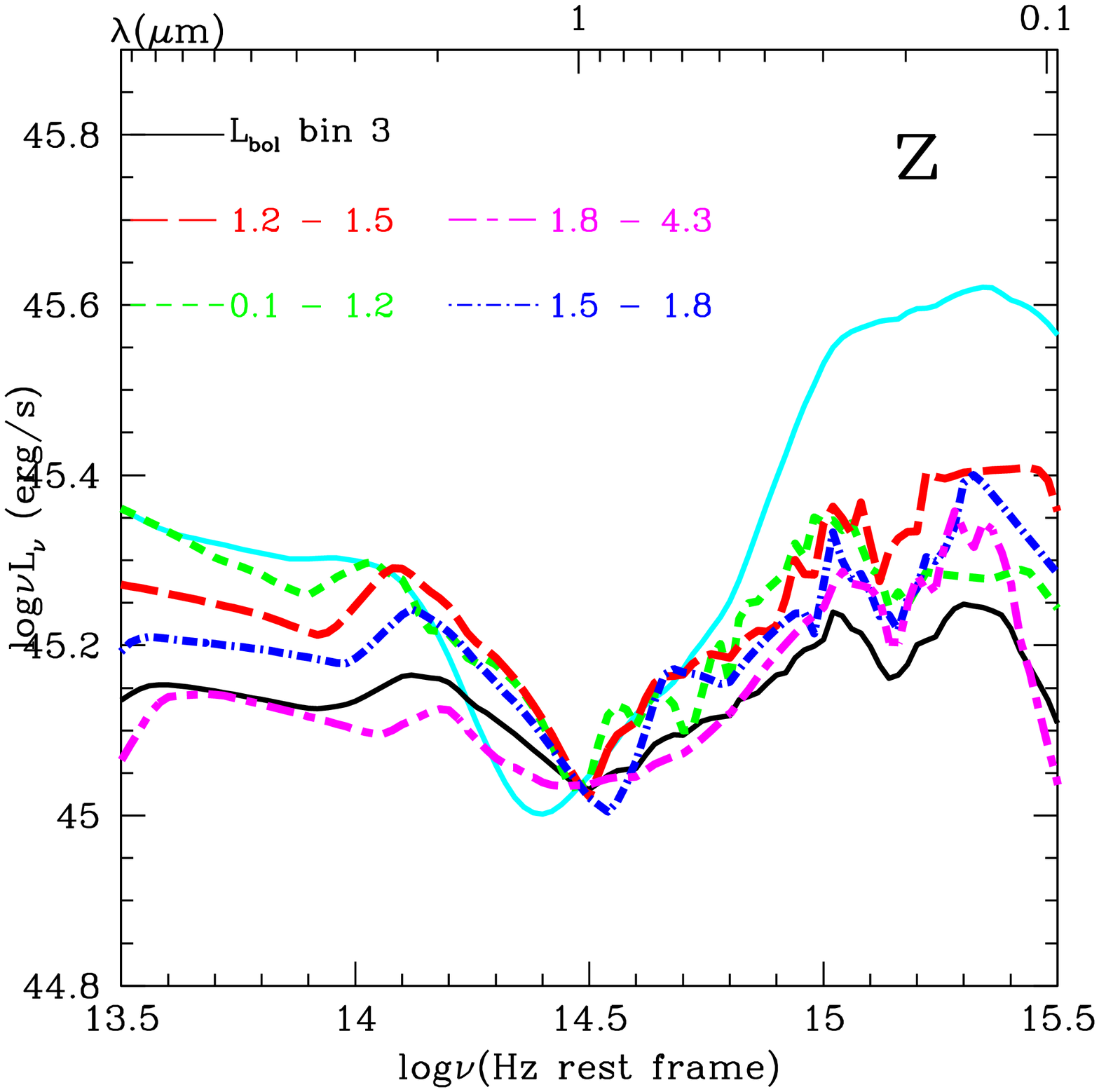}
\includegraphics[angle=0,width=0.32\textwidth]{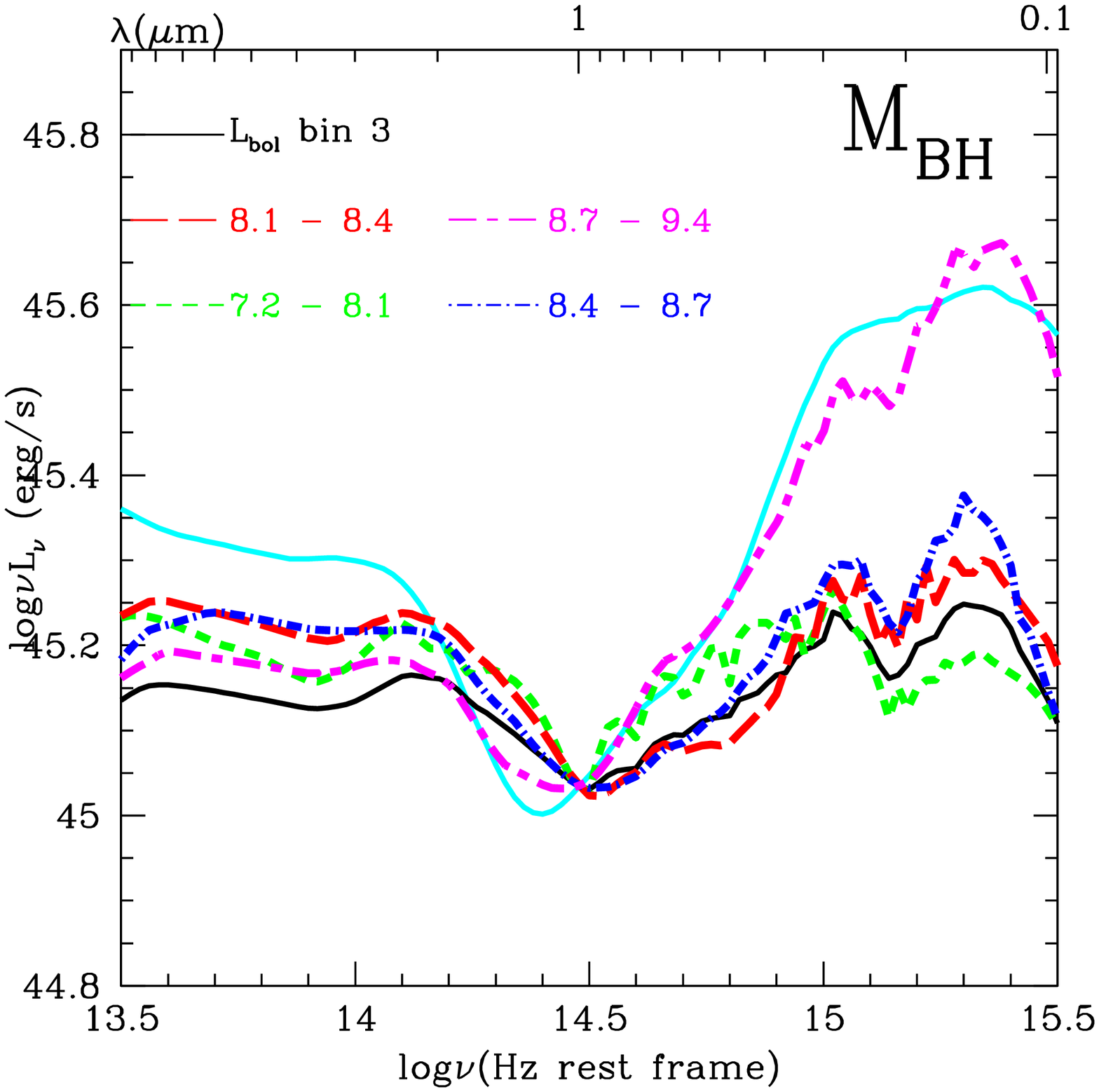}
\includegraphics[angle=0,width=0.32\textwidth]{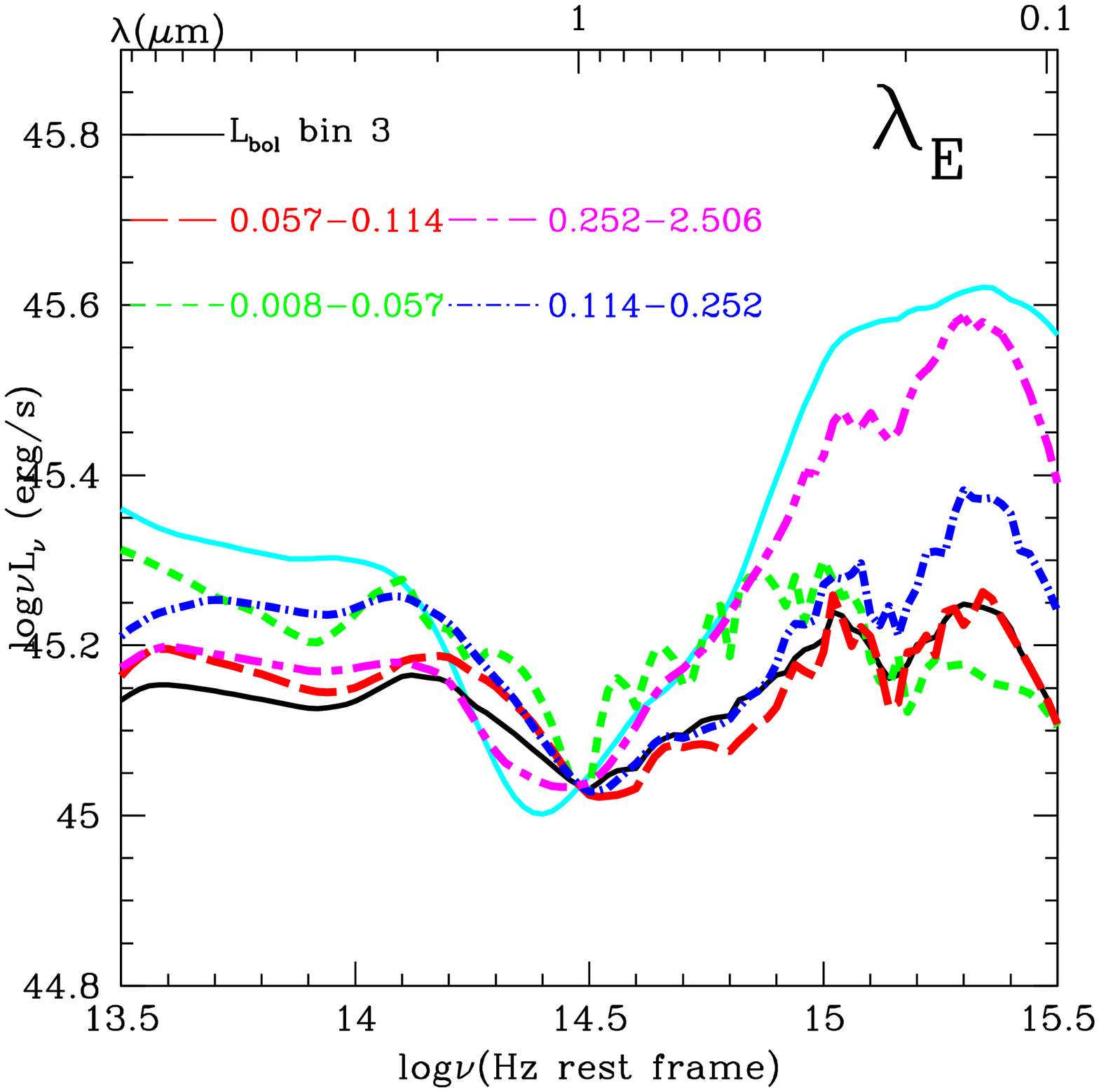}
\includegraphics[angle=0,width=0.32\textwidth]{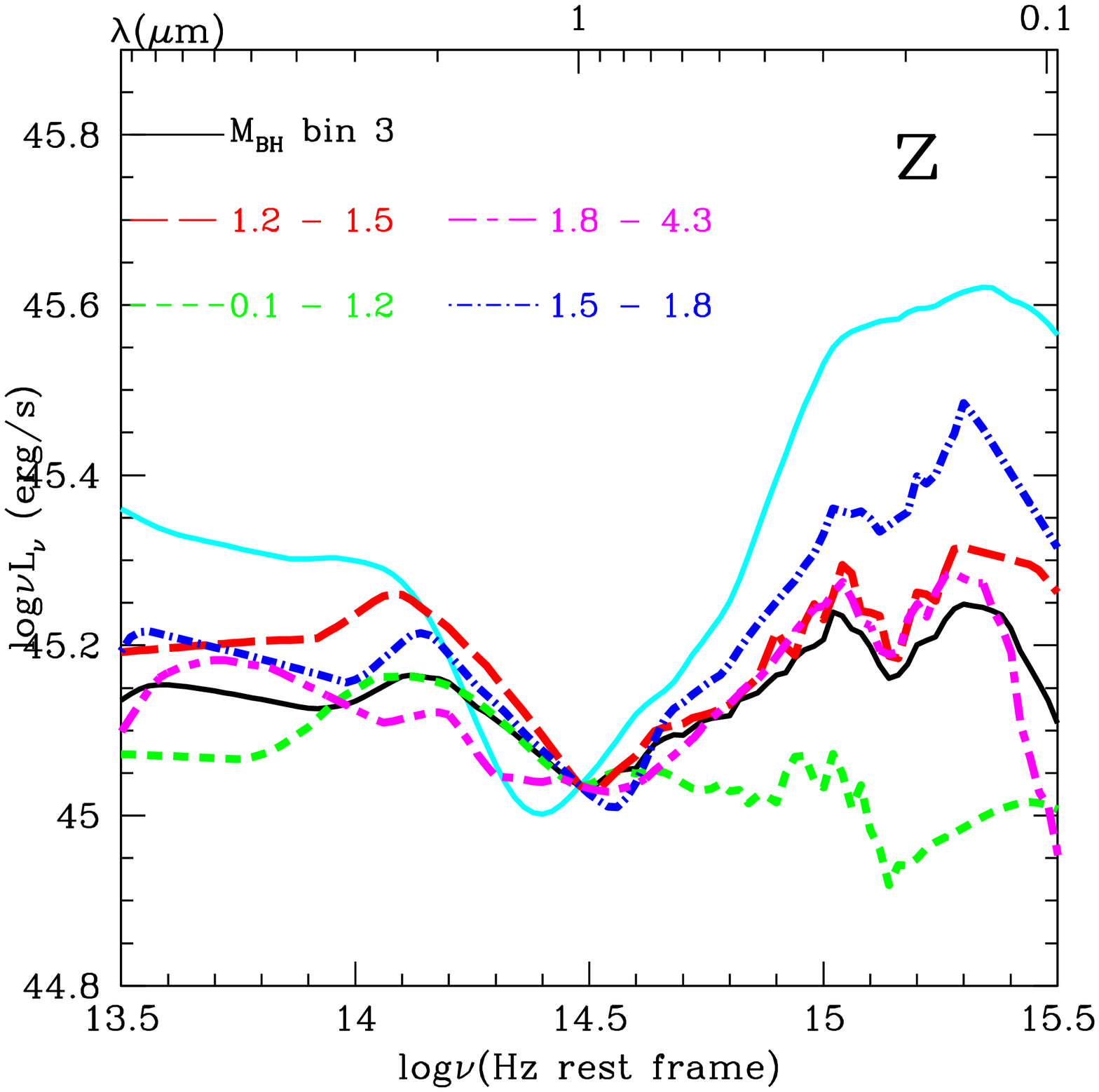}
\includegraphics[angle=0,width=0.32\textwidth]{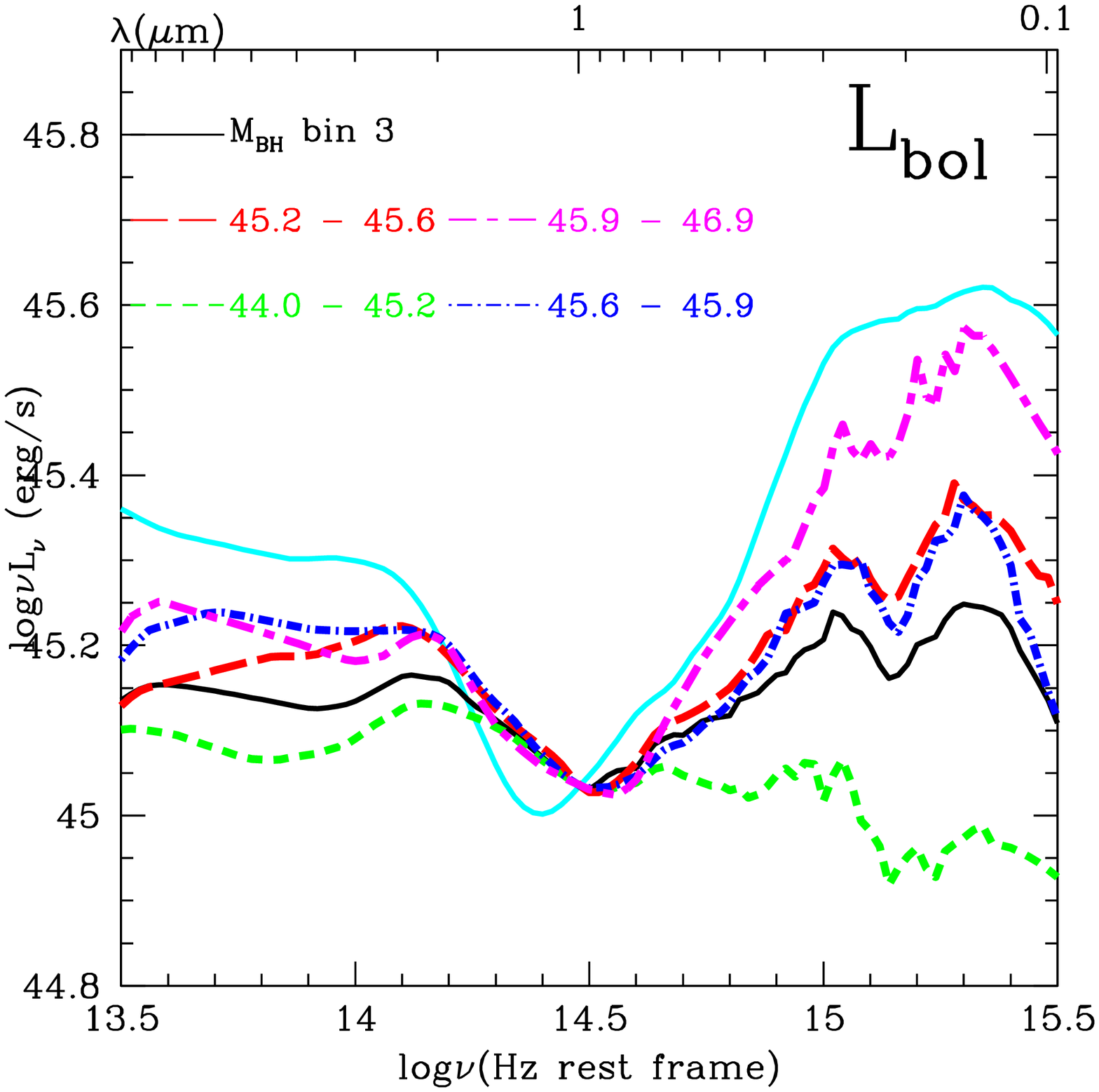}
\includegraphics[angle=0,width=0.32\textwidth]{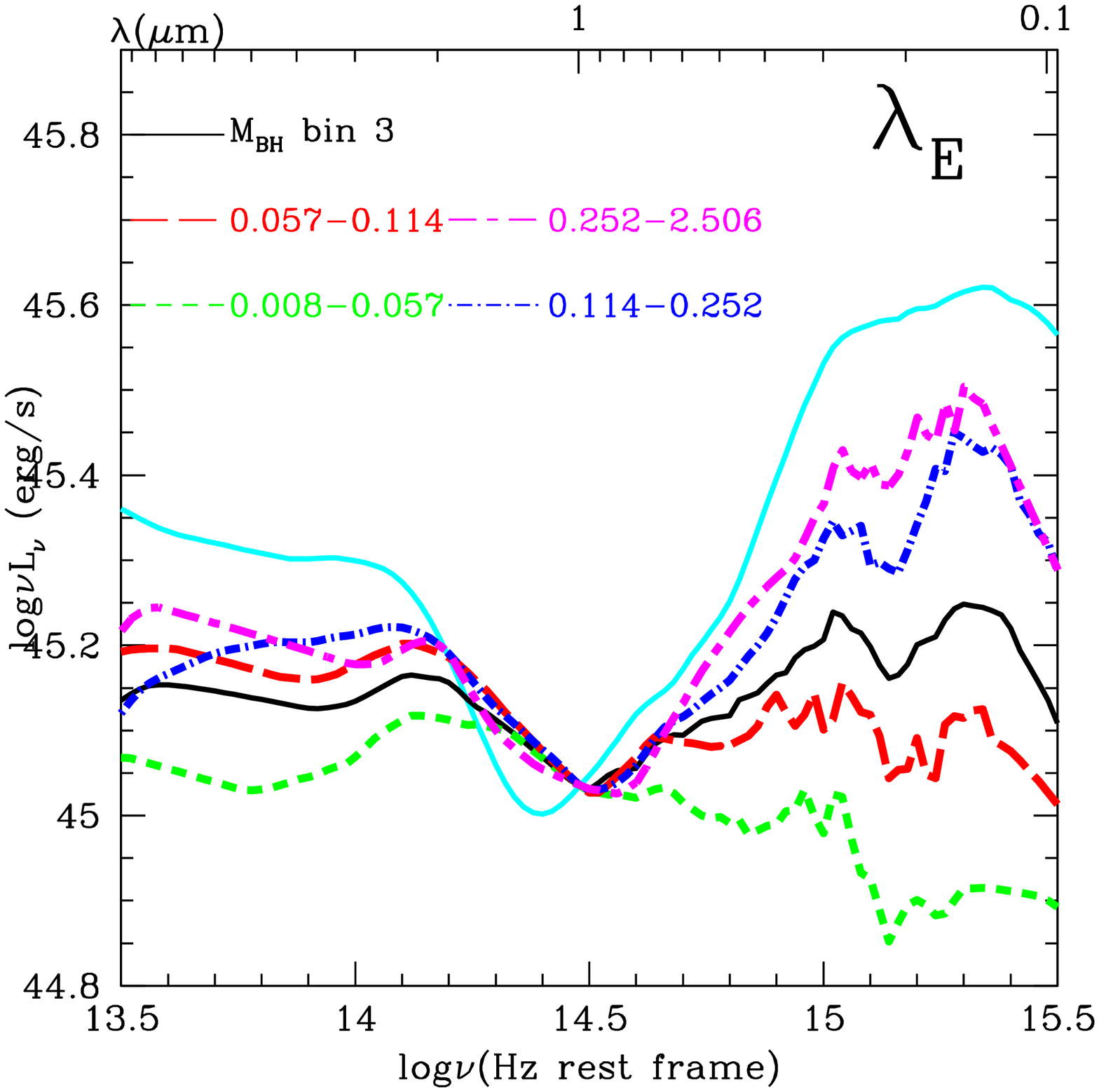}
\includegraphics[angle=0,width=0.32\textwidth]{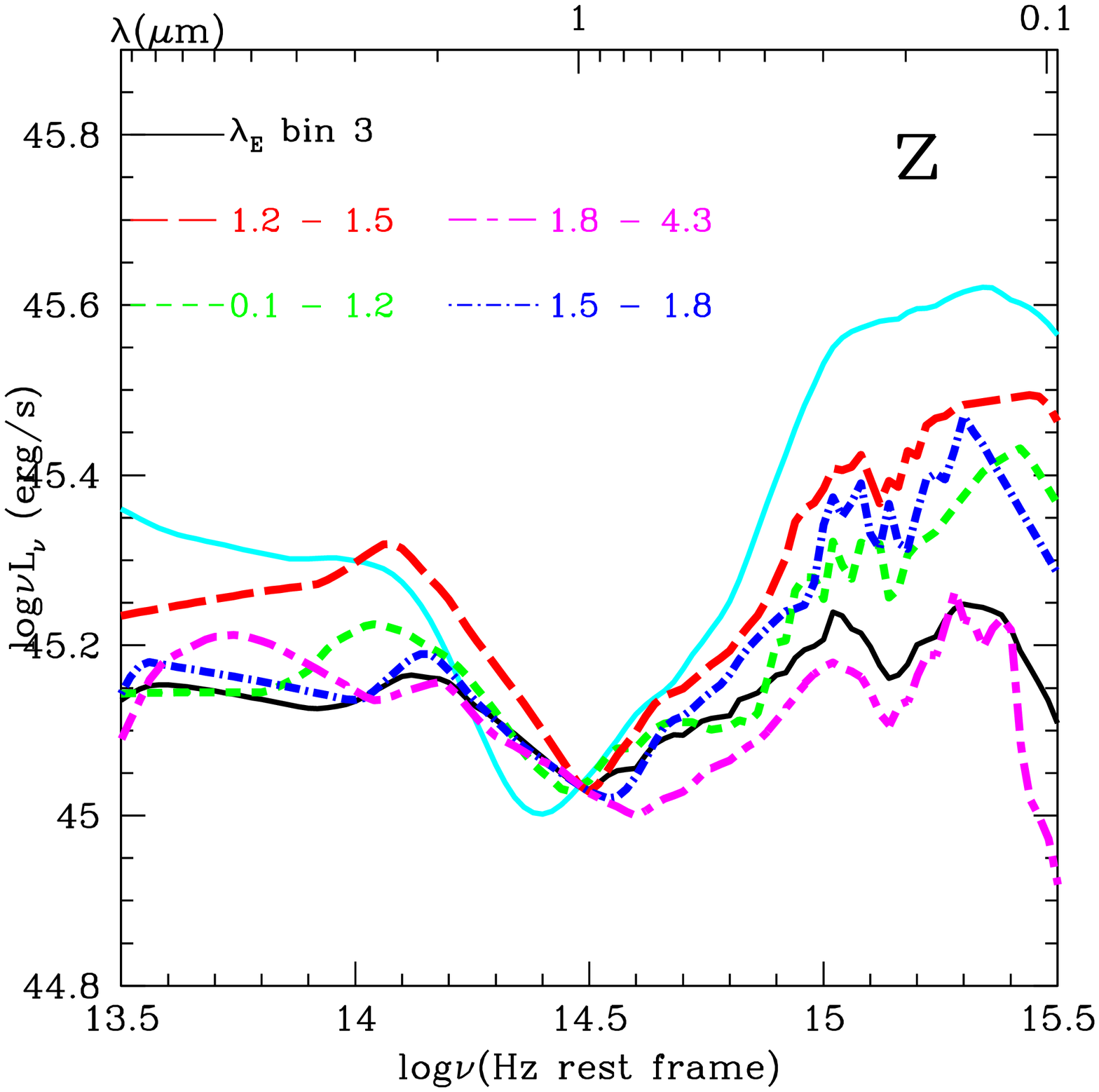}
\includegraphics[angle=0,width=0.32\textwidth]{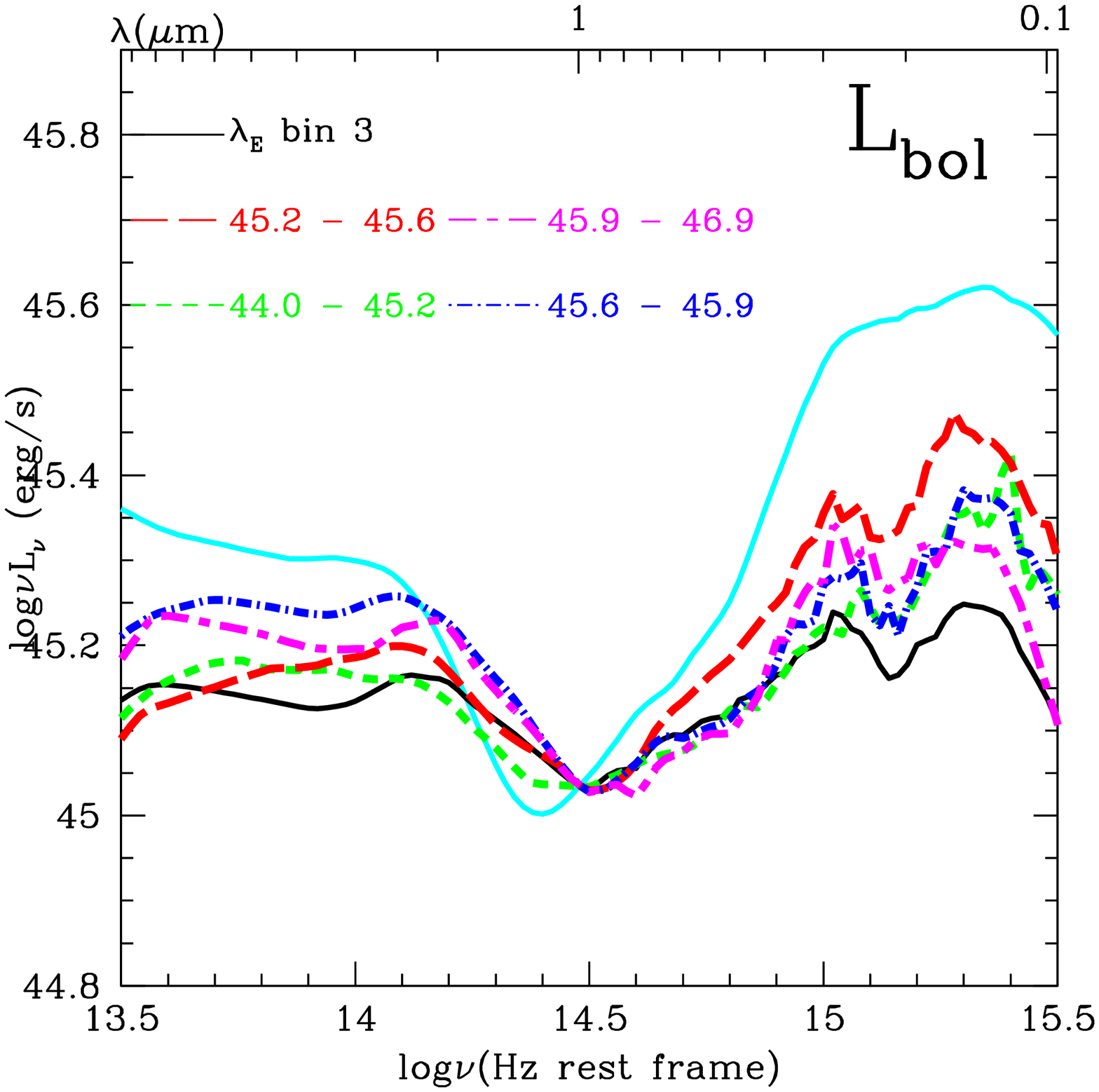}
\includegraphics[angle=0,width=0.32\textwidth]{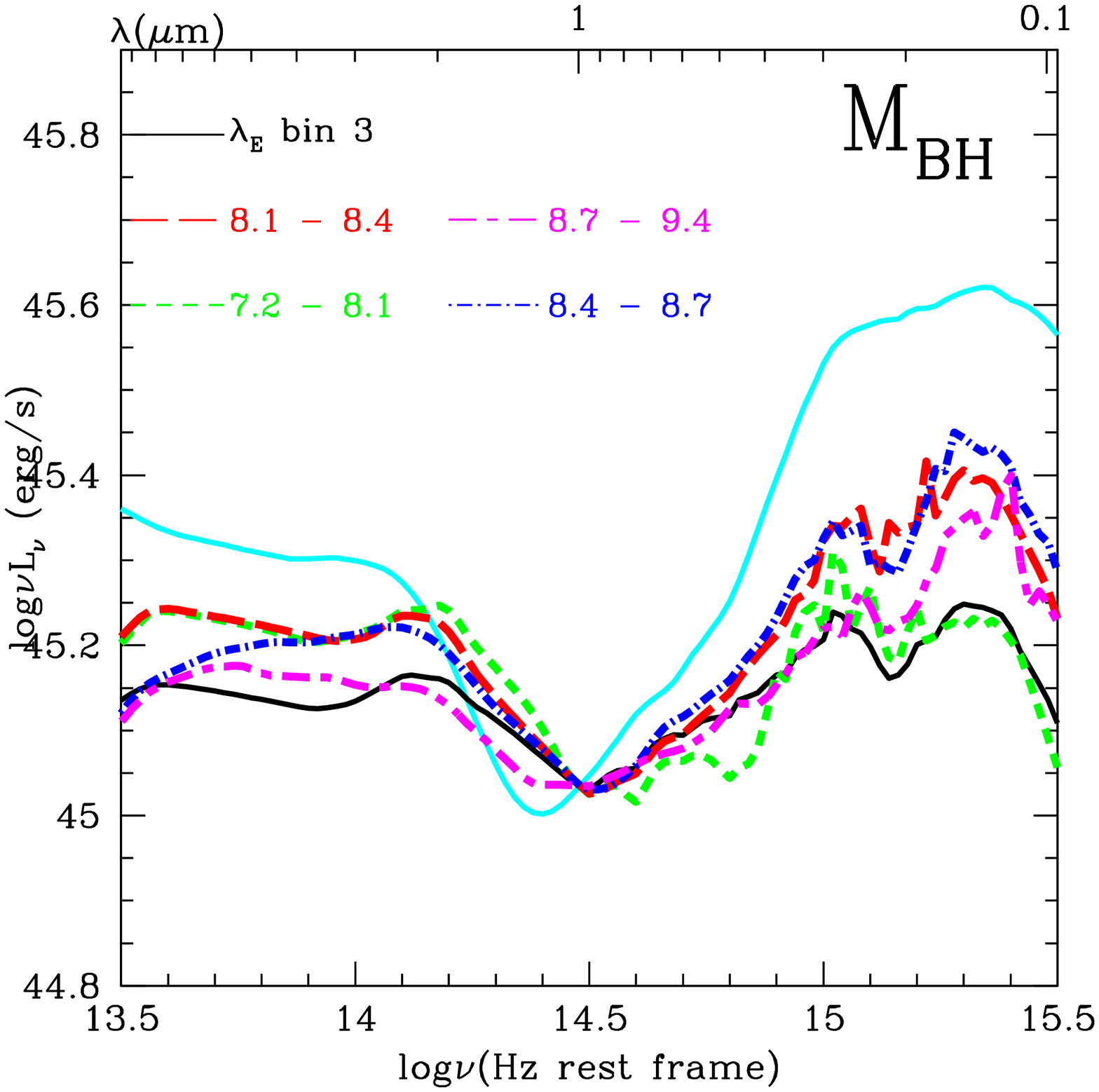}
\caption{The mean host-subtracted SED normalized at 1$\mu m$
compared to E94 mean radio-quiet SED (red solid line). Different
rows are for quasars in specific bins: (1) $z$ bin 3 ($1.5<z<1.8$);
(2) $\log L_{bol}$ bin 3 ($45.6<\log L_{bol}<45.9$); (3)
$\log(M_{BH}/M_{\bigodot})$ bin 3
($8.4<\log(M_{BH}/M_{\bigodot})<8.7$); (4) $\log\lambda_E$ bin 3
(right half, $0.114<\lambda_E<0.252$). The symbol on the upper right
corner of the each plot shows which sub-bin is considered. The lines
are color coded as in Figure~\ref{msedbin1}~\&~\ref{msedbin2}.
\label{msedbinpart3}}
\end{figure*}

We can look for trends in the SEDs by checking the mean SED shape
diversity in different bins. For each of the $z$, $L_{bol}$,
$M_{BH}$ and $\lambda_{E}$ bins the mean SED are calculated as in
Paper I. Briefly we: (1) converted the flux densities at each
frequency for each object to luminosity, using a $\Lambda$CDM
Concordance Cosmology (Komatsu et al. 2009); (2) shifted them to the
rest frame for each source; (3) corrected for the small Galactic
extinction (E(B-V)$\simeq$0.017); (4) limited the variability by
restricting the photometry data in use to 2004 - 2007; (5) corrected
for broad emission line contributions, which can be significant in
the intermediate width Subaru bands; (6) linearly interpolated the
SED to a uniform frequency grid ($\Delta\log\nu=0.02$); (7)
calculated the mean SED at each $\log \nu$ grid point. To avoid the
SED shape being dominated by the few luminous quasars in each bin,
we also calculated the mean of the SED after normalizing the SED at
1$\mu m$.

As the COSMOS optical and infrared data used here were taken over a
4 year interval, from 2004 to 2007, and the SDSS data for the field
were taken as early as 2001, variability is common in the XC413
sample (Paper I). Salvato et al. (2009) defined a convenient
variability parameter $\Upsilon$ (the rms of the magnitude offsets
at the sampled epochs) to quantify the variability of the sources.
Salvato et al. (2009) found that $\Upsilon > 0.25$ efficiently
separates out variable XMM-COSMOS sources (including both point-like
and extended sources). We plot the $\Upsilon$ histogram in the left
panel of Figure~\ref{varsel} for both the XC413 (blue dashed line)
and SSRQ200 (black solid line) sample. Half of the XC413 AGN
(199/413) and SSRQ200 AGN (94/200) have $\Upsilon > 0.25$. As in
Paper I, we do not use the Salvato et al. (2009) method to correct
the SED, to avoid the modification to the SED shape. Alternatively,
we restrict the data set to a shorter time period. Using $\chi^2$
fits to the continuum (using quadratic functions to fit the observed
data from rest frame 9000~\AA\ to 912\AA), we find that using only
the data in the interval from 2004 to 2007 reduces the variability
issue (right panel of Figure~\ref{varsel}). The right panel of
Figure~\ref{varsel} shows how the reduced time span improves
$\chi^2$ after applying these restrictions.

The resulting mean SEDs in different $z$ and $L_{bol}$ bins for
XCRQ407 are shown in Figure~\ref{rqmsedbin}. For ease of comparison,
we also plot the E94 mean SED and a galaxy template from SWIRE
(Polletta et al. 2007) normalized to the value of $L_{*}$ from the
UKIDSS Ultra Deep Survey (Cirasuolo et al. 2007, $M_K^*=-23$). The
galaxy SED shown is an Elliptical galaxy with an age of 5~Gyr
(hereafter E5). Different galaxy templates have similar shapes at
around 1$\mu m$, so here we just choose E5 as a representative case.

The SEDs in our sample have much less pronounced $1\mu m$ inflection
point than in E94. In Paper I we concluded that this shape is
probably due to the host galaxy contribution. It is clear from
Figure~\ref{rqmsedbin} that for low $z$ and low $L_{bol}$ sources
the galaxy component around $1\mu m$ strongly affects the shape of
the SEDs. Indeed, at higher $z$ and $L_{bol}$, the $1\mu m$
inflection becomes more obvious, as the quasar component becomes
relatively stronger. However, even for high $z$ and $L_{bol}$
sources, the optical big blue bump is not as strong as in E94. This
is probably because E94 considered a UV selected sample which picks
out the bluest quasars (Schmidt \& Green 1983). The mean SED shapes
in the central $z$ and $L_{bol}$ bins are quite similar to the mean
SEDs of all the quasars in XCRQ407.

For the SSRQ200 sample, we calculate the mean host-corrected SED for
each bin, as shown in Figure~\ref{msedbin1}~\&~\ref{msedbin2}. Even
with the host-correction, the mean SEDs for the low $z$ and
$L_{bol}$ sources are still relatively flat and different from E94.
This may be due to insufficient host galaxy correction in some cases
because of the scatter in the scaling relationship. The mean SEDs
show that the top three bins of $z$ and $L_{bol}$ have similar
shapes, showing no sign of dependency on any of these parameters.

The $1~\mu m$-normalized mean SED shapes of three $M_{BH}$ bins are
also similar to each other except for the second bin
($8.1<\log(M_{BH}/M_{\bigodot})<8.4$). The optical slopes of the
mean SEDs of different $\lambda_E$ bins increase as $\lambda_E$
increases, that is the mean SED is redder when the Eddington ratio
is smaller. This trend is in agreement with the difference seen in
typical quasar with $\lambda_E>0.01$ and extremely low Eddington
ratio AGN with $\lambda_E<10^{-4}$ (Ho et al. 2008, Trump et al.
2011).

The $z$ bins have the tightest distribution compared to other
parameters (e.g., at $3\mu m$ and 3000\AA\ the difference of the
mean SEDs in different bins is 0.1 dex and 0.15 dex respectively,
smaller than corresponding values of the other parameters). The
difference of the mean SEDs in all the bins (except for the lowest
bin) of each parameter are less than a factor of 2 even at the
infared/UV end.

To check for partial dependencies of the SED shape on the physical
parameters, we checked the SED shape difference with one parameter
when fixing another. In order to have a large enough number of
quasars in the mean SED calculation in each sub-bin, we consider the
central two bins (Figure~\ref{msedbinpart2} \& \ref{msedbinpart3}).
The number of quasars in each sub-bin is listed in
Table~\ref{t:npart}. As the number of quasars in each sub-bin is
relatively small, these mean SEDs are more affected by particular
SED shapes. In all these plots, the bottom~/~first bins are
sometimes still affected by the host contribution, and the
top~/~last bins are sometimes affected strongly by several specific
SEDs (e.g. hot-dust-poor quasars, Hao et al. 2010, 2011). However,
in general, the mean SEDs shapes in different sub-bins are very
similar to each other and to the mean SED in that bin. For instance,
when fixing $z$, the top~/~last two $L_{bol}$, $M_{BH}$ or
$\log\lambda_E$ sub-bins have a similar mean SED shape respectively.

As shown in the following section (\S~\ref{s:seddisp}), the
dispersion of the SED in different bins ranges from 0.3 to 0.6,
larger than the difference between the mean SED of adjacent bins;
and the normalized SED dispersion can be as high as $\sim 0.5$ much
larger than the difference between the mean normalized SED of
adjacent bins (for details see \S~\ref{s:seddisp}). The lack of SED
shape dependency on $z$, $L_{bol}$, $M_{BH}$ and $\lambda_E$ we
observed in the XCRQ407 and SSRQ200 suggests that neither the
emission mechanism nor the accretion disk and torus structure alters
systematically or dramatically with these parameters. In other
words, a single intrinsic quasar SED in the optical/UV to
near-infrared range is a meaningful concept. There is clearly still
difference between individual SEDs, for example in the normalization
due to $L_{bol}$. However, most of the differences in shapes are
caused by different host contributions or reddening in quasars. The
SED shape does not show a statistically significant dependence on
the parameters we investigated.

\section{SED Dispersion Dependency on Physical Parameters}
\label{s:seddisp}
\begin{figure*}
\includegraphics[angle=0,width=0.45\textwidth]{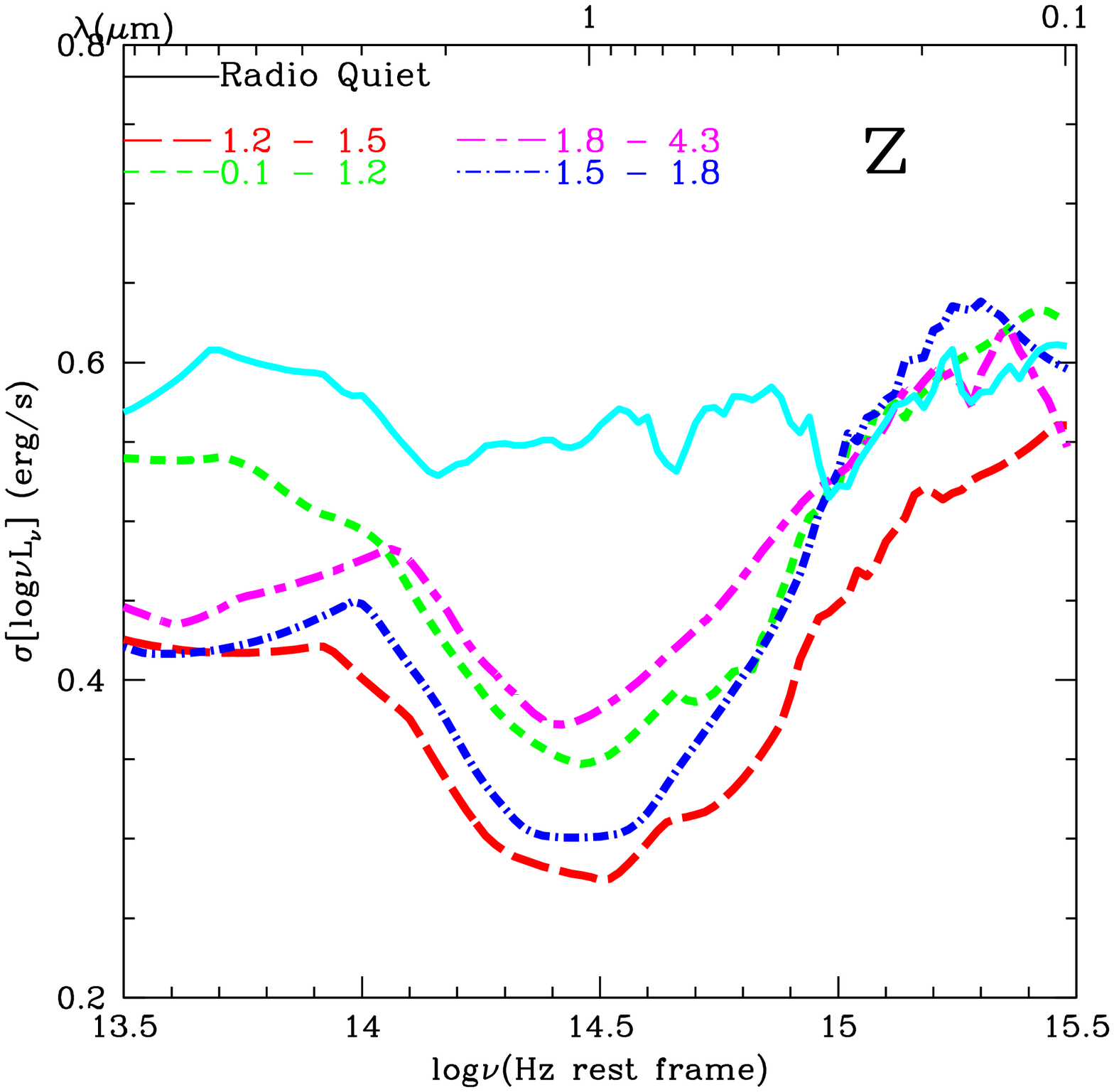}
\includegraphics[angle=0,width=0.45\textwidth]{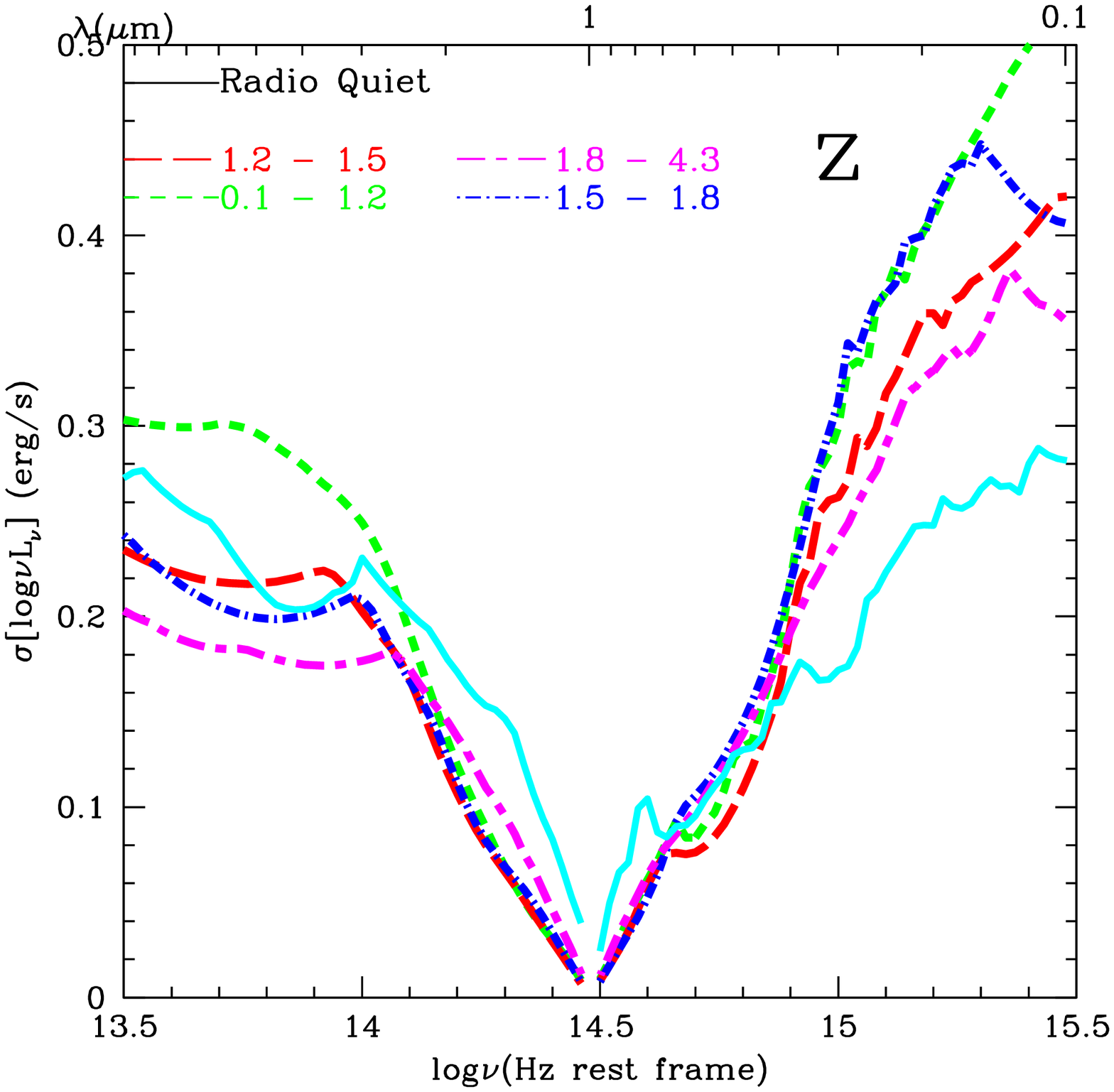}
\includegraphics[angle=0,width=0.45\textwidth]{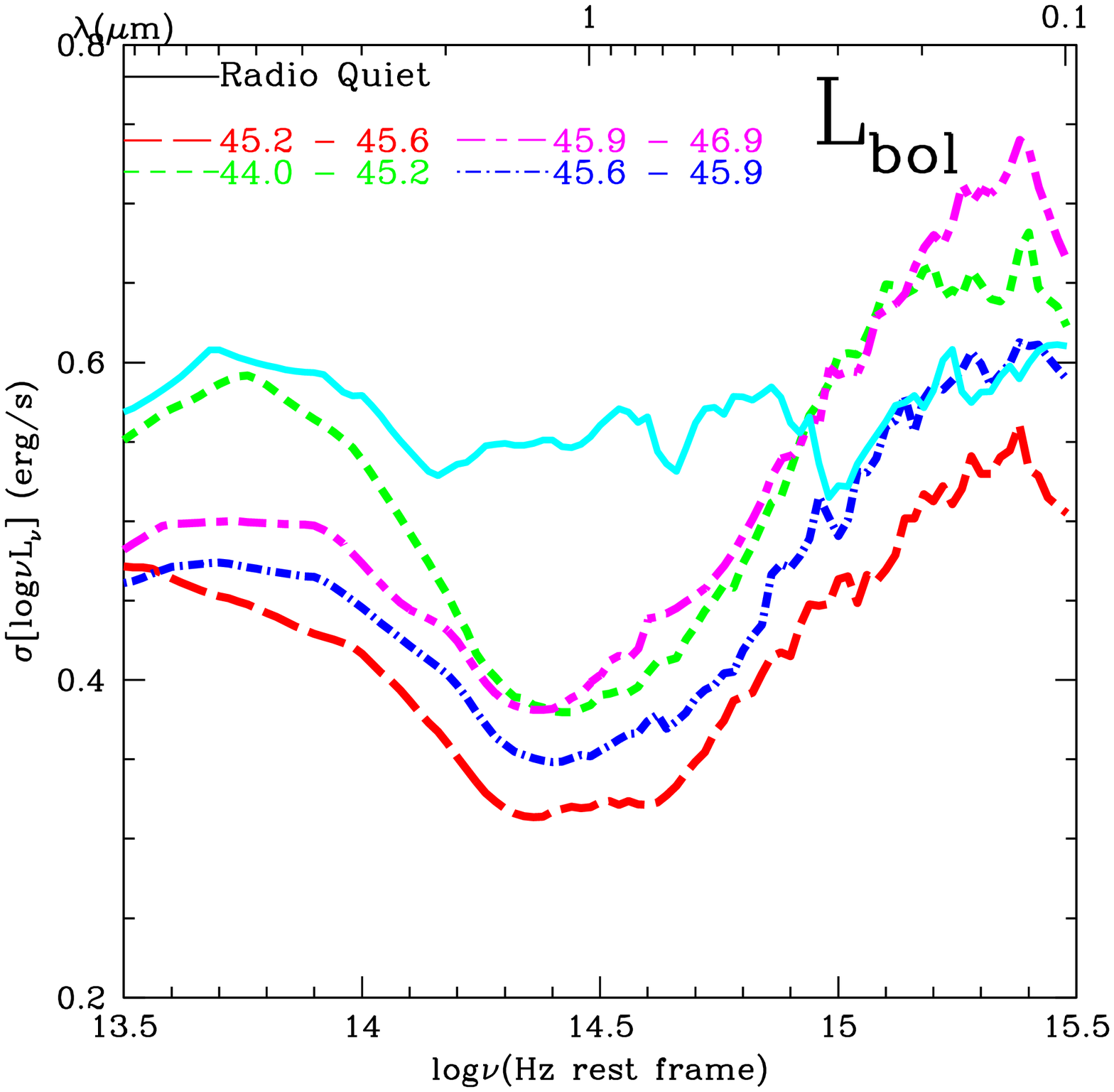}
\includegraphics[angle=0,width=0.45\textwidth]{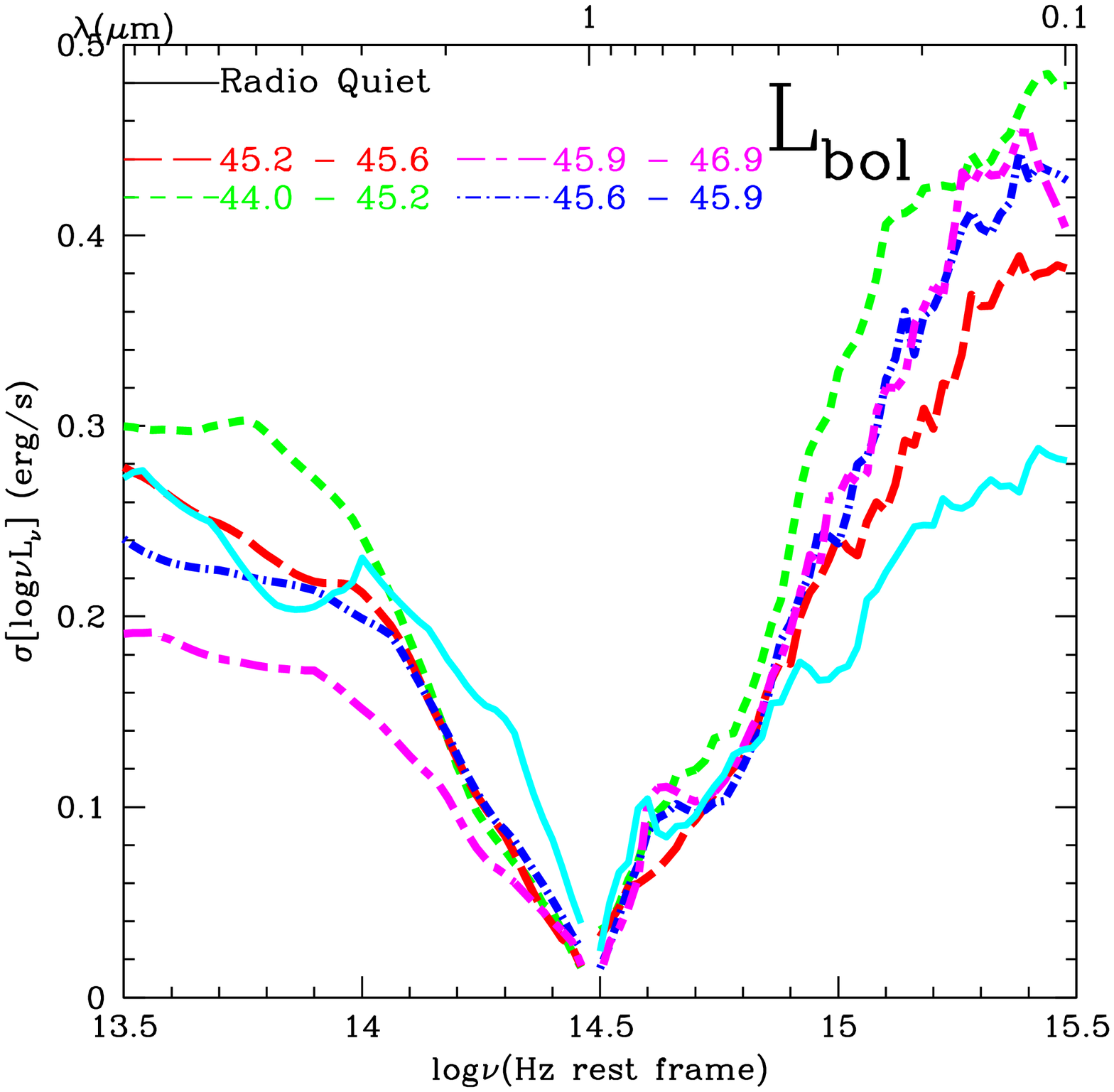}
\caption{The dispersion of the SEDs for XCRQ407 in bins of $z$ and
$L_{bol}$ before (left) and after (right) normalization at 1$\mu m$
compared to corresponding E94 radio-quiet SED dispersion (cyan solid
curve). \label{seddisprqbin}}
\end{figure*}

\begin{figure*}
\includegraphics[angle=0,width=0.45\textwidth]{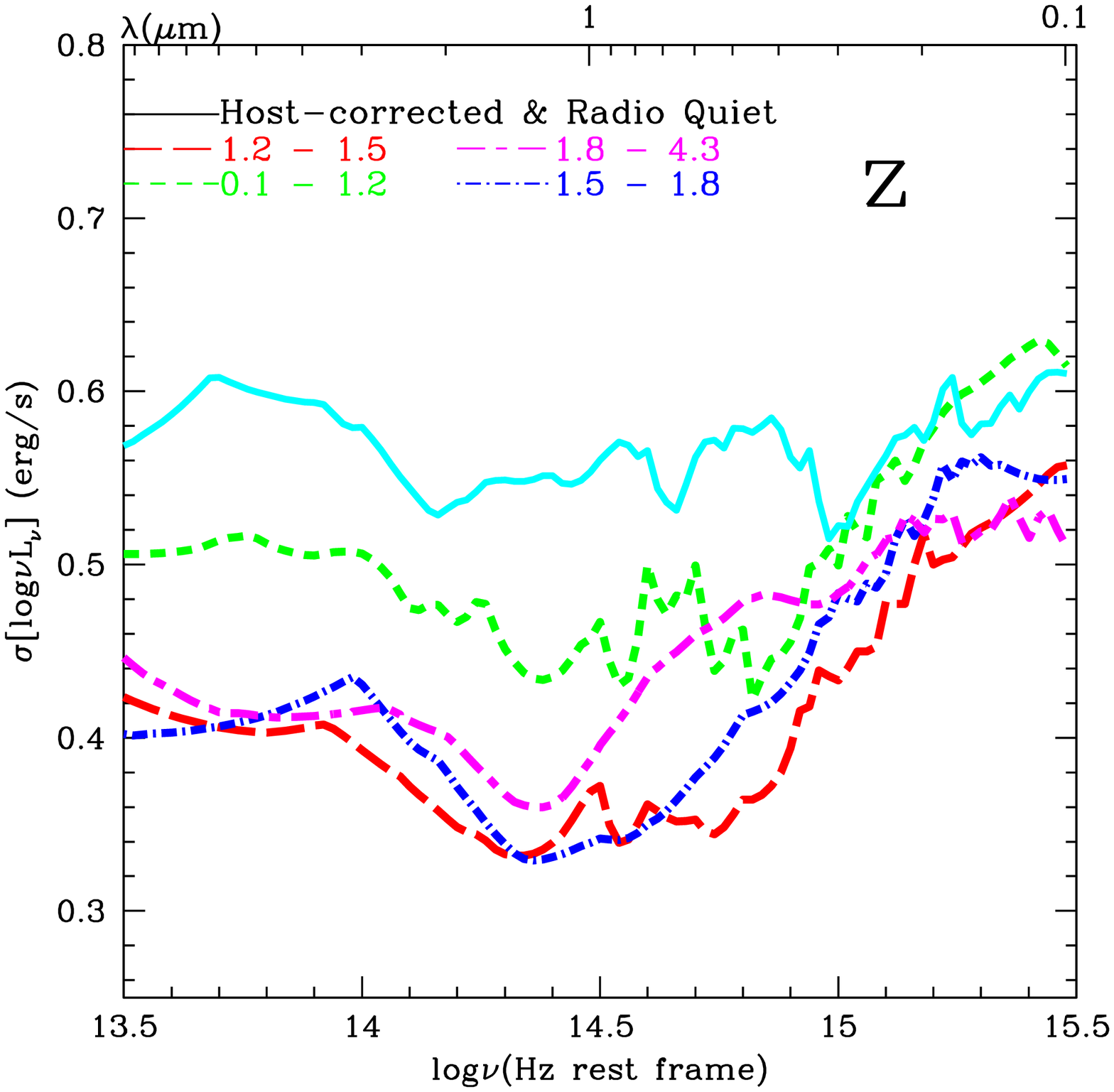}
\includegraphics[angle=0,width=0.45\textwidth]{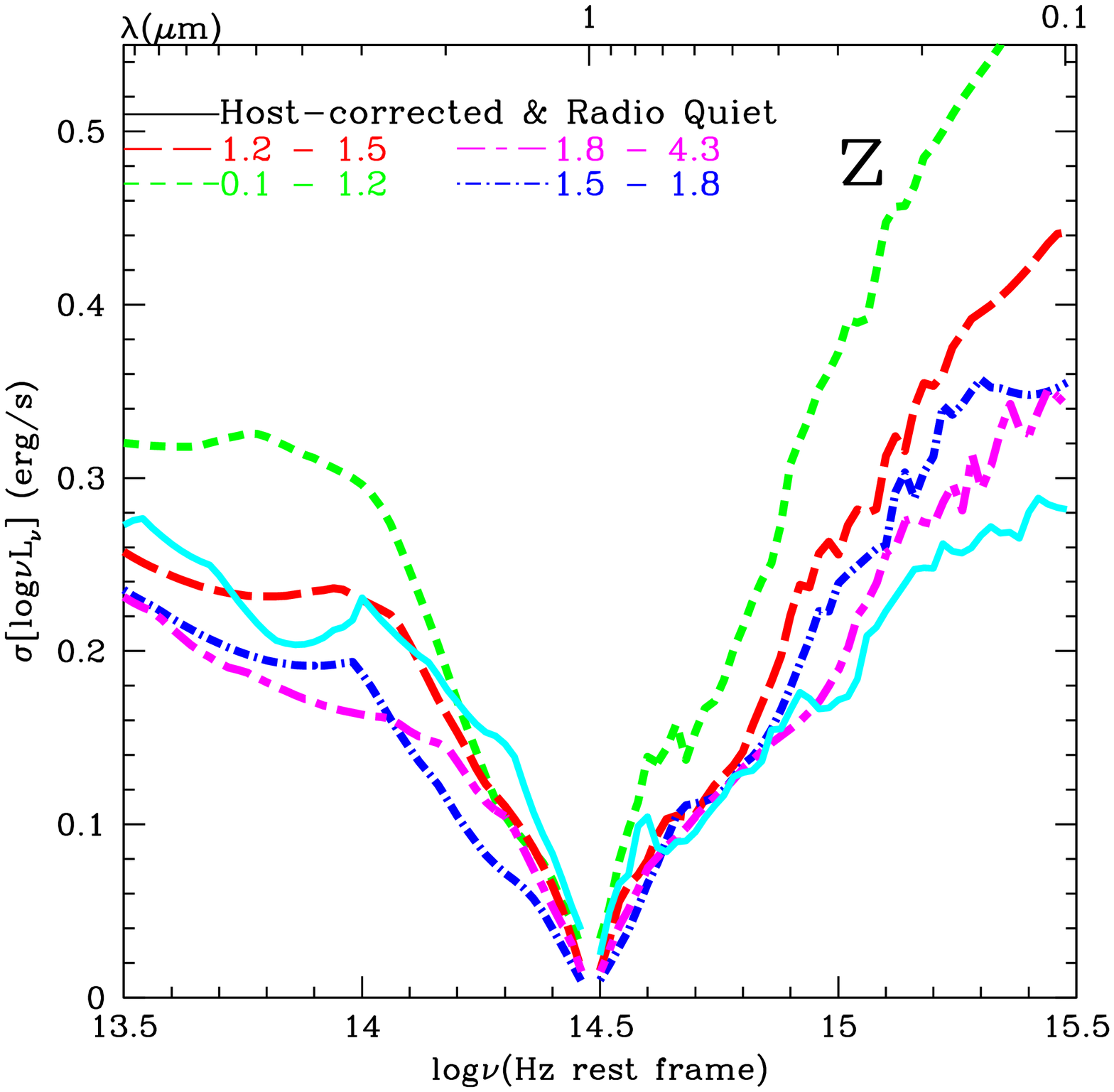}
\includegraphics[angle=0,width=0.45\textwidth]{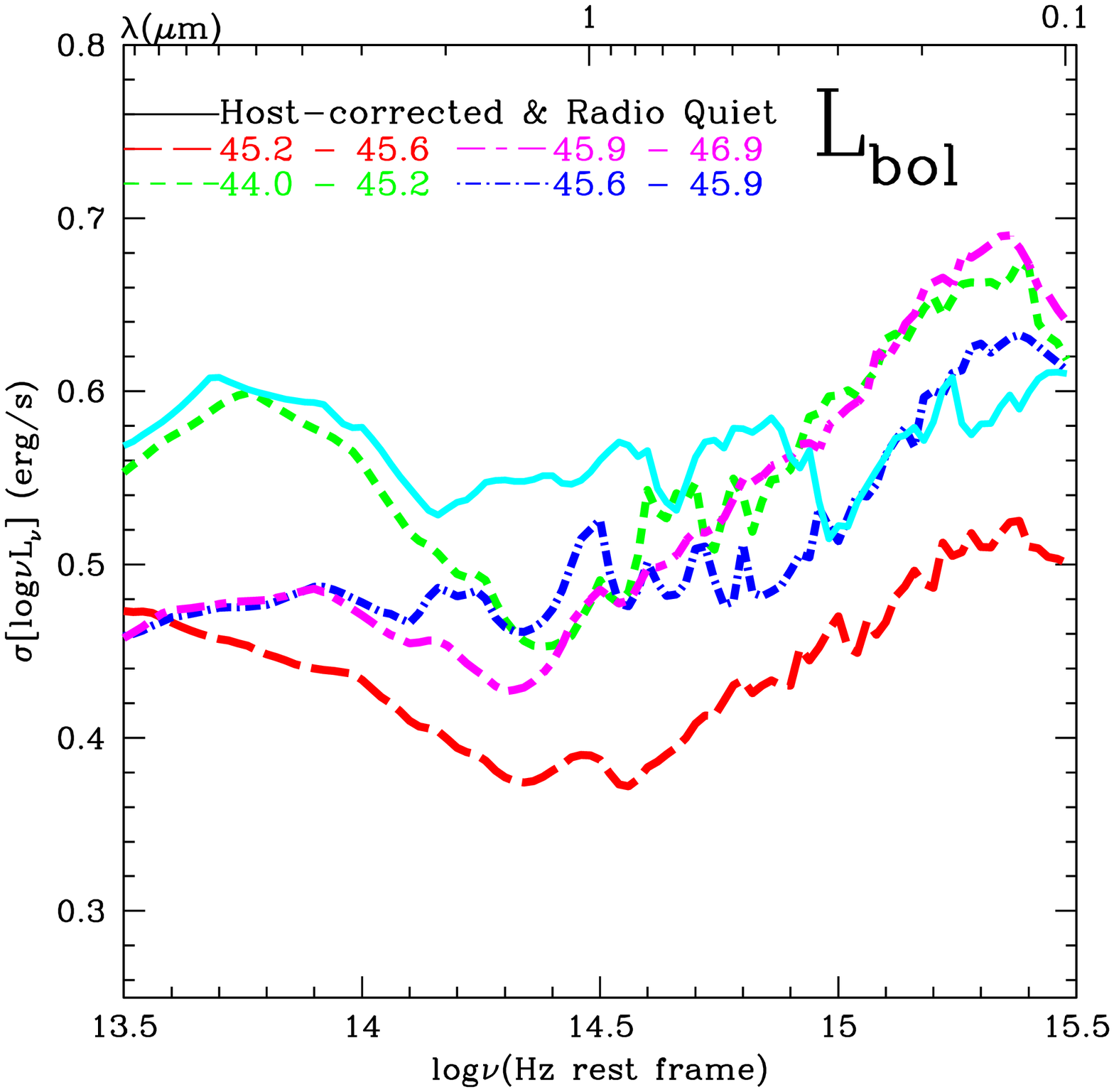}
\includegraphics[angle=0,width=0.45\textwidth]{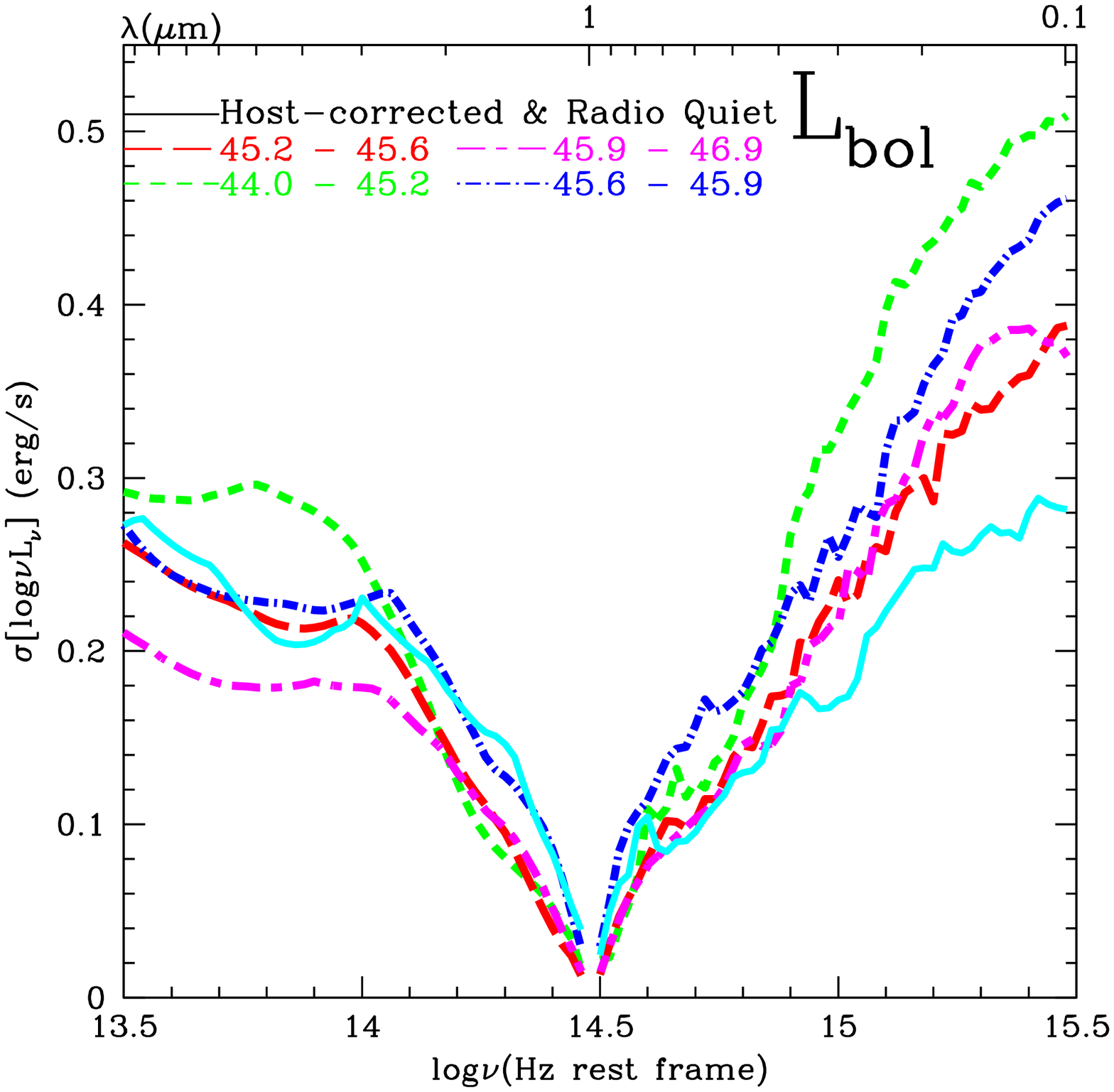}
\caption{The dispersion of host-subtracted SEDs for SSRQ200 in bins
of $z$ and $L_{bol}$ before (left) and after (right) normalized at
1$\mu m$ compared to corresponding E94 radio-quiet SED dispersion
(cyan solid curve). \label{seddisphcbin1}}
\end{figure*}

\begin{figure*}
\includegraphics[angle=0,width=0.45\textwidth]{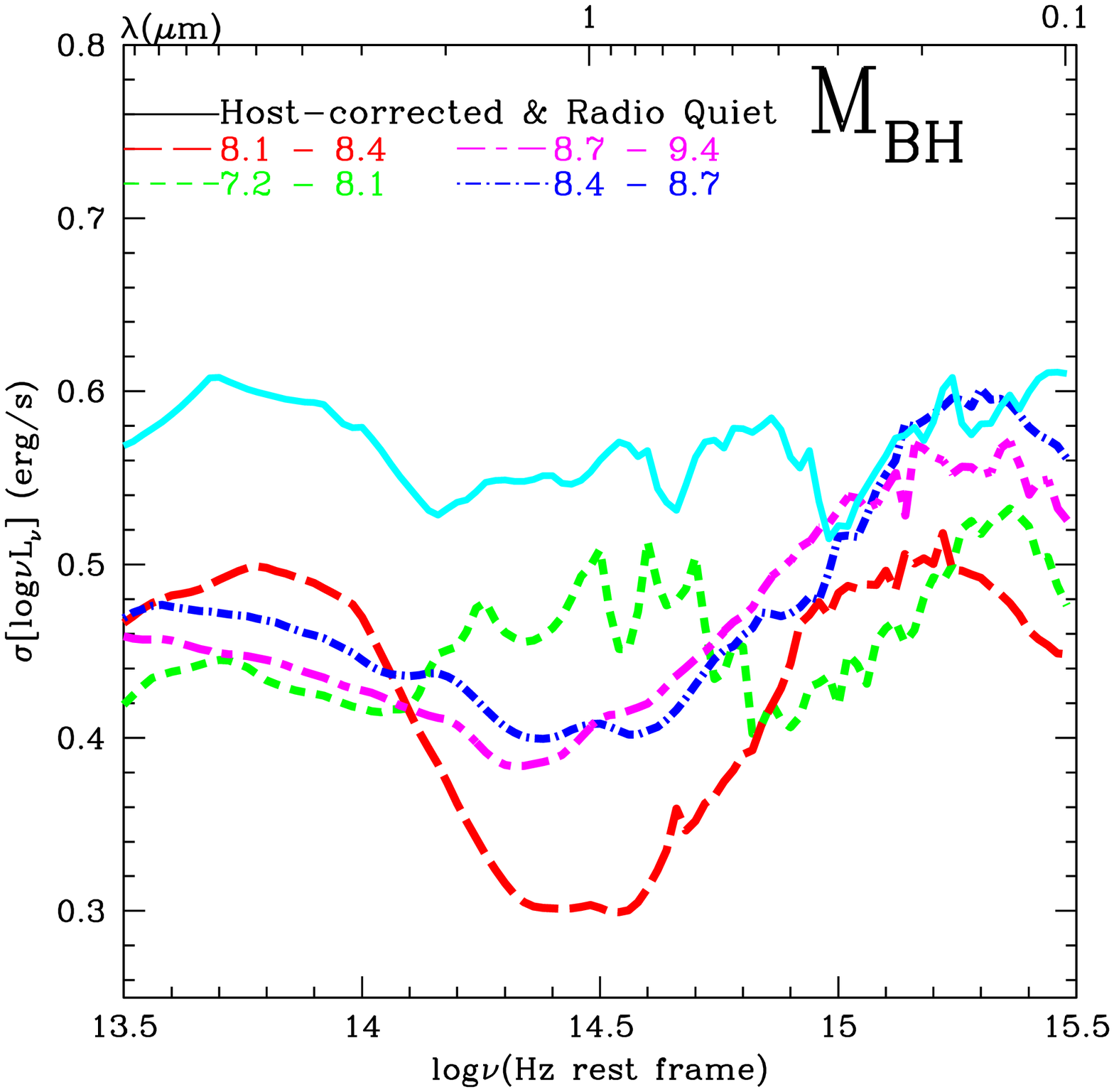}
\includegraphics[angle=0,width=0.45\textwidth]{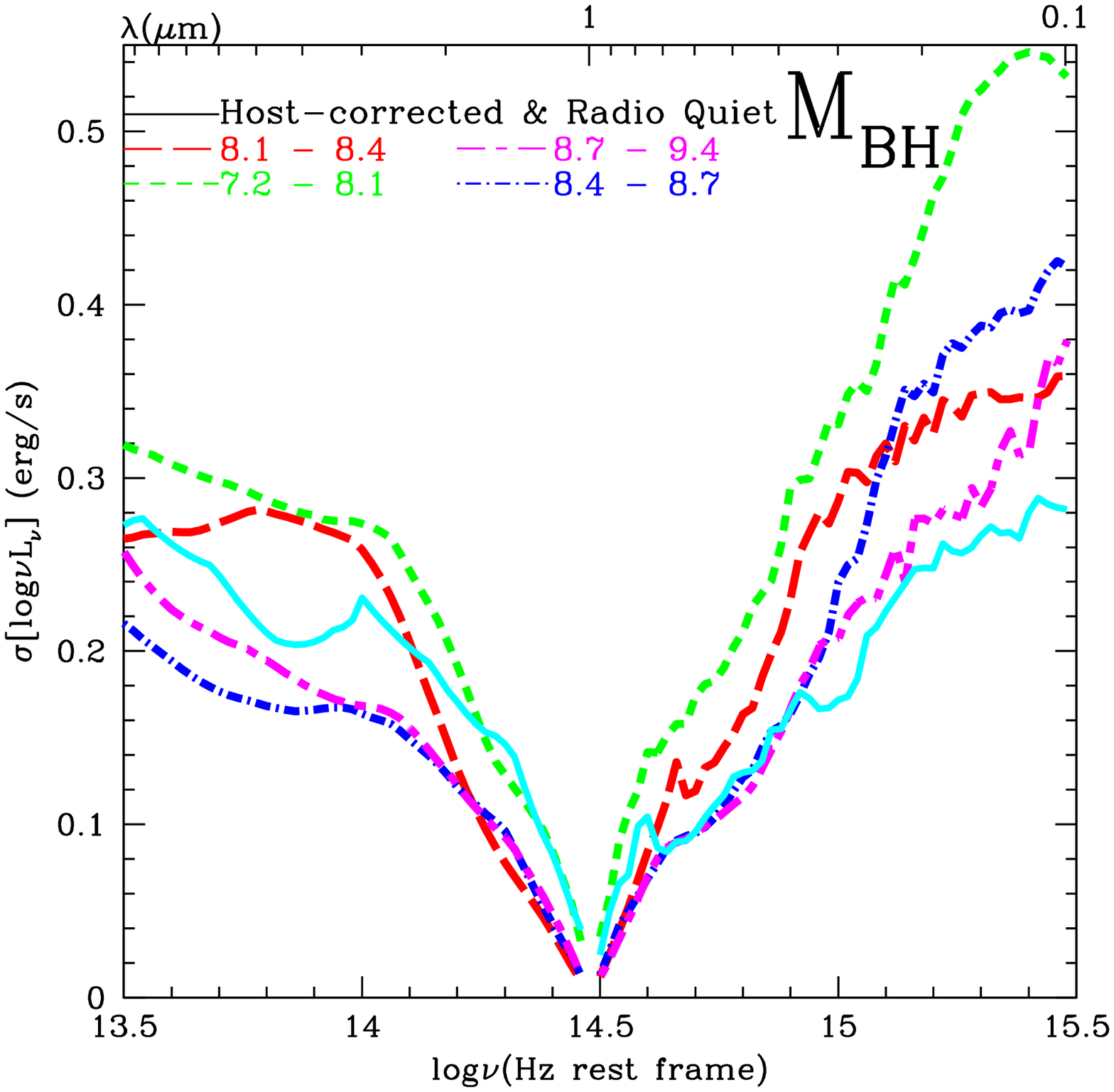}
\includegraphics[angle=0,width=0.45\textwidth]{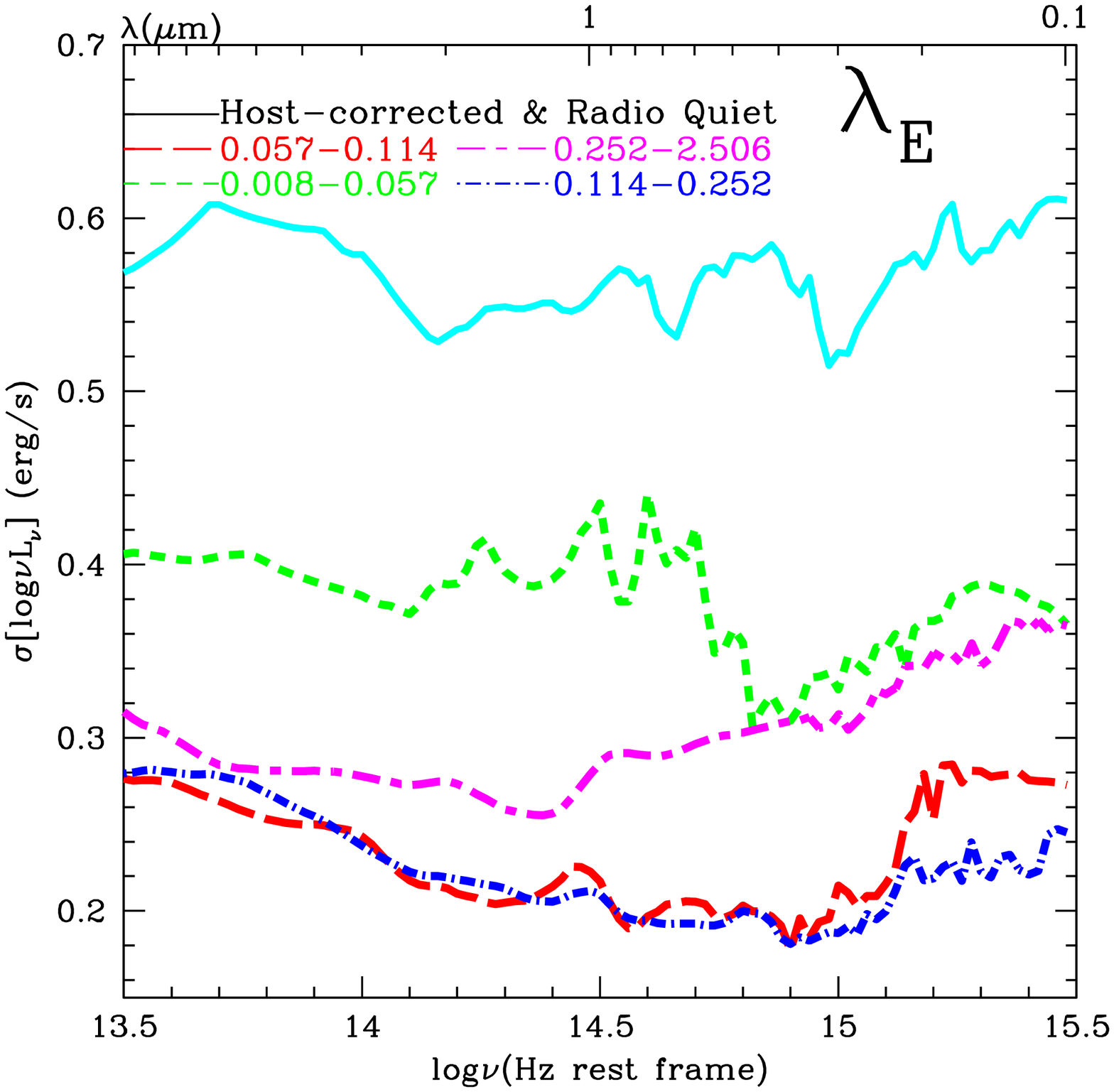}
\includegraphics[angle=0,width=0.45\textwidth]{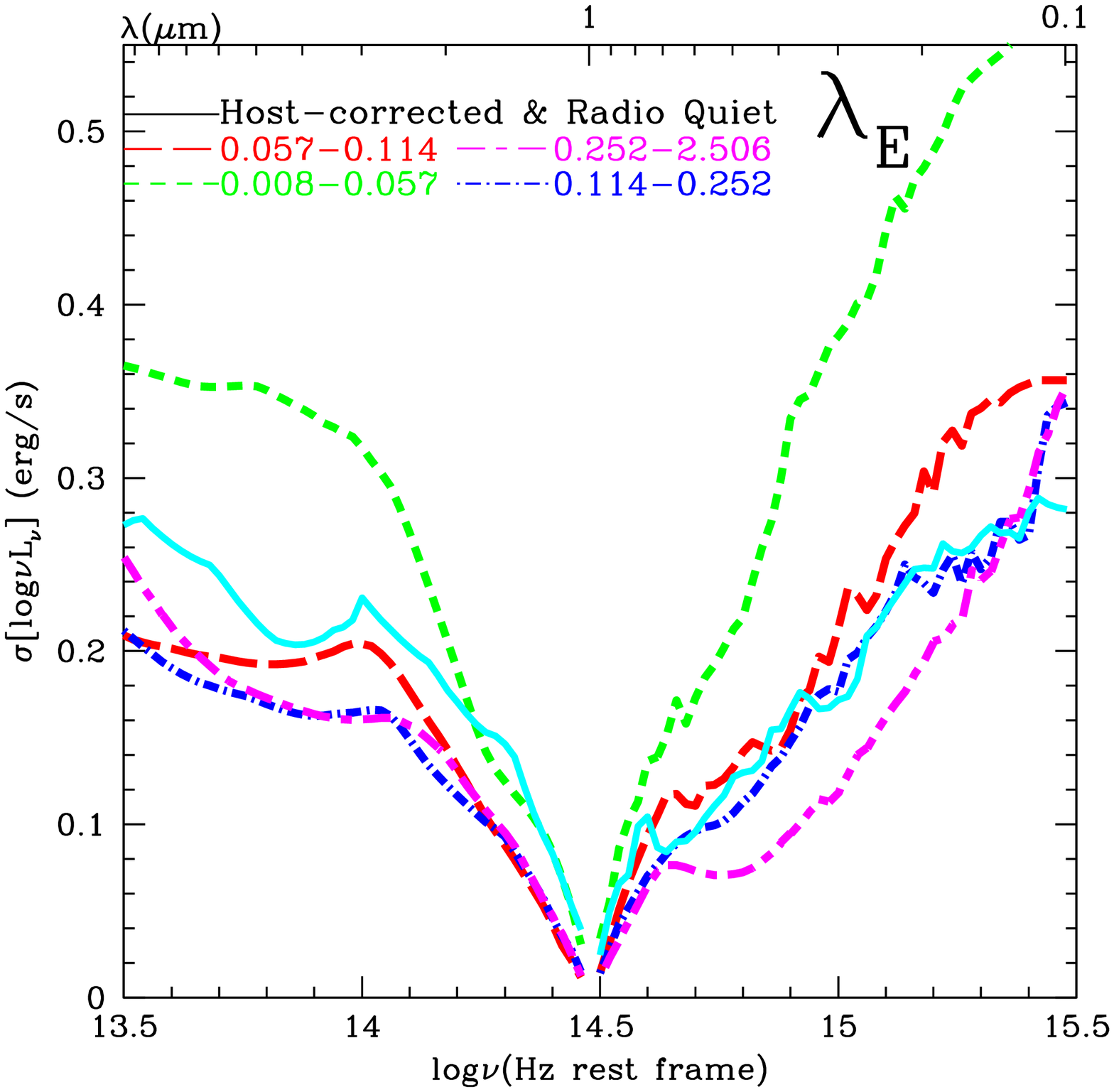}
\caption{The dispersion of host-subtracted SEDs for SSRQ200 in bins
of $M_{BH}$ and $\lambda_{E}$ before (left) and after (right)
normalized at 1$\mu m$ compared to corresponding E94 radio-quiet SED
dispersion (cyan solid curve). \label{seddisphcbin2}}
\end{figure*}

\begin{table}
\begin{minipage}{\columnwidth}
\centering \caption{Bolometric Luminosity Dispersion in Each bin
\label{t:lsig}}
\begin{tabular}{@{}c|cc|cccc@{}}\hline
sample & \multicolumn{2}{c}{XCRQ407} & \multicolumn{4}{c}{SSRQ200} \\
bin & z & $L_{bol}$ & z & $L_{bol}$ & $M_{BH}$ & $\lambda_{E}$ \\
 & $\sigma_{L_{bol}}$ & $\sigma_{L_{bol}}$ & $\sigma_{L_{bol}}$ & $\sigma_{L_{bol}}$ & $\sigma_{L_{bol}}$ &
 $\sigma_{L_{bol}}$\\
\hline
      1 & 0.46 & 0.50 & 0.45 & 0.50 & 0.36 & 0.26 \\
      2 & 0.38 & 0.38 & 0.35 & 0.38 & 0.38 & 0.10 \\
      3 & 0.42 & 0.43 & 0.41 & 0.43 & 0.43 & 0.09 \\
      4 & 0.42 & 0.49 & 0.39 & 0.48 & 0.39 & 0.25 \\
\hline
\end{tabular}
\end{minipage}
\end{table}

\begin{table}
\begin{minipage}{\columnwidth}
\centering \caption{Normalized SED Dispersions at Certain
Wavelength\label{t:seddisp}}
\begin{tabular}{@{}cc|ccccc@{}}
\hline\multicolumn{2}{c}{bin} & \multicolumn{5}{c}{Wavelength}\\
& sub-bin & 10~$\mu$m & 3~$\mu$m & 3000\AA & 2500\AA & 1000\AA \\
\hline
                    & 1 & 0.30 & 0.25 & 0.30 & 0.36 & 0.50 \\
XCRQ407             & 2 & 0.24 & 0.20 & 0.26 & 0.30 & 0.42 \\
  $z$               & 3 & 0.24 & 0.21 & 0.31 & 0.37 & 0.41 \\
                    & 4 & 0.20 & 0.18 & 0.24 & 0.28 & 0.35 \\
                    \hline
                    & 1 & 0.30 & 0.24 & 0.33 & 0.38 & 0.48 \\
XCRQ407             & 2 & 0.28 & 0.21 & 0.24 & 0.26 & 0.38 \\
$\log L_{bol}$      & 3 & 0.24 & 0.20 & 0.24 & 0.30 & 0.43 \\
                    & 4 & 0.19 & 0.15 & 0.26 & 0.31 & 0.40 \\
                    \hline
                    & 1 & 0.32 & 0.30 & 0.37 & 0.42 & 0.57 \\
SSRQ200             & 2 & 0.26 & 0.23 & 0.26 & 0.28 & 0.44 \\
  $z$               & 3 & 0.24 & 0.19 & 0.24 & 0.26 & 0.36 \\
                    & 4 & 0.23 & 0.16 & 0.19 & 0.24 & 0.34 \\
                    \hline
                    & 1 & 0.29 & 0.25 & 0.33 & 0.37 & 0.51 \\
SSRQ200             & 2 & 0.26 & 0.22 & 0.24 & 0.26 & 0.39 \\
$\log L_{bol}$      & 3 & 0.27 & 0.23 & 0.25 & 0.28 & 0.46 \\
                    & 4 & 0.21 & 0.18 & 0.22 & 0.27 & 0.37 \\
                    \hline
                    & 1 & 0.32 & 0.27 & 0.33 & 0.37 & 0.53 \\
SSRQ200             & 2 & 0.26 & 0.26 & 0.29 & 0.31 & 0.36 \\
$\log M_{BH}$       & 3 & 0.22 & 0.16 & 0.24 & 0.30 & 0.42 \\
                    & 4 & 0.26 & 0.17 & 0.21 & 0.23 & 0.38 \\
                    \hline
                    & 1 & 0.36 & 0.32 & 0.38 & 0.42 & 0.56 \\
SSRQ200             & 2 & 0.21 & 0.20 & 0.21 & 0.23 & 0.36 \\
$\log \lambda_{E}$  & 3 & 0.21 & 0.17 & 0.18 & 0.21 & 0.34 \\
                    & 4 & 0.25 & 0.16 & 0.12 & 0.15 & 0.35 \\
\hline
\end{tabular}
\end{minipage}
\end{table}

Even if the mean SEDs show little or no dependency on physical
parameters, the SED dispersion may change with them. We checked the
SED dispersion in different bins for both the XCRQ407 and SSRQ200 by
calculating the dispersion of $\nu L_{\nu}$ at each frequency in
each bin. The dispersion of the SEDs could purely~/~mostly caused by
the luminosity difference among quasars in each bin. In order to
distinguish the dispersion caused by different brightness of quasars
and the SED shape dispersion, we consider the dispersion of the SEDs
before and after normalizing all the SEDs at 1$\mu m$
(Figure~\ref{seddisprqbin}-\ref{seddisphcbin2}). The bolometric
luminosity dispersion in each bin is listed in Table~\ref{t:lsig}.
From the left panels of
Figure~\ref{seddisprqbin}-\ref{seddisphcbin2} and
Table~\ref{t:lsig}, we can see majority of the SED dispersion is
caused by the luminosity dispersion within bins. For the $\lambda_E$
bins, the quasar brightness effect is the least. So we will
concentrate on the dispersion of the normalized SEDs (right panel of
Figure~\ref{seddisprqbin}-\ref{seddisphcbin2}) to see if there is
any SED dispersion dependency on physical parameters. The normalized
SED dispersion in different bins at certain specific wavelengths are
listed as examples in Table~\ref{t:seddisp}.

The resulting dispersion of normalized SEDs in $z$ and $L_{bol}$
bins for XCRQ407 compared to E94 radio-quiet SED dispersion are
shown in right panel Figure~\ref{seddisprqbin}. The XMM-COSMOS SEDs
generally have a larger dispersion in the optical to UV range
compared to E94 radio-quiet sample. This is probably because E94 is
biased toward blue quasars, unlike the XMM selected XCRQ407, thus it
does not include quasars with various optical shapes/colors. The
1--10 $\mu m$ dispersion of XCRQ407 is closely similar to E94,
except at $\sim 2 \mu m$, where it is somewhat lower. The XMM-COSMOS
SED dispersion is quite similar in the 3~$\mu m$ to 3000~\AA\ range
for different $z$ or $L_{bol}$ bins. The SED dispersions in this
range are all below a factor of 2.

For the host-corrected SSRQ200 sample, the dispersion of the SED for
each bin is shown in Figure~\ref{seddisphcbin1} \&
\ref{seddisphcbin2}. 
The SED dispersion in the 3~$\mu m$ to 3000~\AA\ range is not as
tight as the un-corrected sample, implying extra dispersion induced
by the host-correction process depends on $z$, $L_{bol}$, $M_{BH}$
or $\lambda_E$. The lowest bin for each parameter always has the
largest dispersion in most frequency range in all the dispersion
plots. 

In all these SED dispersion plots, the UV (at $\sim0.1\mu m$)
dispersion is generally larger than the near-IR (at $\sim10\mu m$)
dispersion. This could mean that the reprocessing of the hot dust
component has slightly smoothed the discrepancy of the accretion
disk emission. Alternatively, this could be completely caused by the
variability of the quasar which affects the UV SED most, or by
reddening.

We can compare the difference between mean SEDs in
\S~\ref{s:msedevl} and the SED dispersion discussed above. Before
normalization, the SED dispersion is comparable and slightly larger
than the mean SED difference between adjacent bins. As majority of
the difference and dispersion is caused by the bolometric luminosity
distribution, we will focus our comparison for the normalized SEDs.
The normalized SED dispersion can reach up to $\sim0.5$. At each
frequency, the dispersion is much larger than the normalized mean
SED difference. We could thus state that the mean SEDs are invariant
within $1\sigma$ (at most frequencies $>2\sigma$).

\section{Discussion and Conclusion}
We analyzed the dependence of both the mean and dispersion of the
SED shapes in the optical-UV to IR range on the parameters $z$,
$L_{bol}$, $M_{BH}$ and $\lambda_E$ for the 407 XMM-COSMOS
radio-quiet type 1 AGN sample. We also calculated the bolometric
correction at UV-optical to near-IR for the host-corrected SSRQ200,
and in four different redshift bins. The bolometric correction for
different redshift bins are quite similar to each other.

As the XMM-COSMOS quasar sample is an X-ray-selected sample, it
includes also sources with low nucleus to host contrast. The mean
SED of the whole sample is greatly affected by the host galaxy
emission for low redshifts and luminosities. Therefore, we mainly
studied the mean and dispersion SEDs of a host-corrected sub-sample
of 200 radio-quiet quasars, SSRQ200, in four quartile bins of $z$,
$L_{bol}$, $M_{BH}$ and $\lambda_E$. The mean SED shapes in the
different bins are quite similar to each other. These SEDs are also
generally similar to the E94-like mean SED shape, but they tend to
have less pronounced optical to UV bump emission than in E94, as E94
is biased towards blue quasars. Even if we fix one parameter (e.g.
redshift), the mean SEDs show no statistical significant evolution
with the others.

We checked the dispersion of SEDs in different parameter bins. The
near-IR SED dispersion is generally smaller compared to the UV SED
dispersion, which might be due to the variability of the quasar that
mainly affects UV SED. The SEDs before and after host correction
have very similar dispersion in the 3~$\mu m$ to 3000~\AA\ range for
different bins, which implies an invariant intrinsic dispersion of
SED shapes in this wavelength range. However, in this wavelength
range, the host-corrected SED dispersion is not as tight as the
uncorrected sample, probably due to the extra dispersion induced by
the scaling relationship depends on the physical parameters. When we
compare the SED dispersion with the difference of the mean SEDs, we
could conclude that the mean SED is invariant within $1\sigma$ (at
most frequencies more than $2\sigma$).

There is no statistical significant dependency of the mean SED with
$z$, $L_{bol}$, $M_{BH}$ or $\log\lambda_E$. This result implies
that the global quasar structures (the accretion disk and the torus)
are invariant with respect to these parameters. Despite the strong
expectation that feeding of the AGN and host galaxy is likely to
change (e.g. Merloni et al 2010), as well as the X-ray corona is
likely to change ($\alpha_{OX}$ varies with $\lambda_E$, e.g. Lusso
et al. 2010), the accretion disk and the torus seem to not depend on
this mechanism. Thus an intrinsic SED in UV-optical to near-IR
exists.

The differences among the SEDs show there might be diversity only at
the second order. Host contamination, reddening and quasar
variability should lead to a dispersion in the UV and near-IR SEDs.
The lack of SED dispersion evolution indicates that these effects
themselves do not show much systematic dependence on physical
parameters.

All the above conclusions depend on the reliability of the host
correction process. This assumes the scaling relationship between
black hole mass and bulge mass (e.g. Marconi \& Hunt 2003) with an
evolutionary term (Bennert et al. 2010, 2011). The black hole mass
estimates can lead to a dispersion as large as 0.4 (e.g.,
Vestergaard \& Peterson 2006) besides dispersion of the relationship
itself. The evolution of this relationship is still under debate
(see e.g., Schramm \& Silverman 2013). Uncertainties in this
correlation might limit our ability to detect SED shape dependency
on physical parameters and it is hard to estimate the exact amount
of dispersion associated with the host correction process.

A less model dependent analysis will be described in a following
paper (Hao et al. 2013a), where we introduce a new mixing diagram to
readily distinguish different SED shape, i.e. galaxy-dominated,
quasar-dominated and reddening dominated SEDs.

So far, we have only studied the SED shape one decade on either side
of $1\mu m$. To study the SED shape evolution in UV or FIR is less
easy. The UV is greatly affected by both reddening and variability
that is different for different sources. The FIR is greatly affected
by host galaxy star formation, which also varies from case to case.
We will discuss these two regions of the quasar SED in following
papers.

\section*{Acknowledgments}

HH thanks Belinda Wilkes and Martin J. Ward for discussion. This
work was supported in part by NASA {\em Chandra} grant number
G07-8136A (HH, ME, FC) and the Smithsonian Scholarly Studies (FC).
Support from the Italian Space Agency (ASI) under the contracts
ASI-INAF I/088/06/0 and I/009/10/0 is acknowledged (AC and CV). MS
acknowledges support by the German Deutsche Forschungsgemeinschaft,
DFG Leibniz Prize (FKZ HA 1850/28-1). KS gratefully acknowledges
support from Swiss National Science Foundation Grant
PP00P2\_138979/1.

\label{lastpage}

\end{document}